\documentclass[twocolumn,apj,dvipdfm]{emulateapj}
\usepackage{color}
\usepackage{hhline}
\usepackage{CJK}
\usepackage{morefloats}

\def\mbf#1{\mbox{\boldmath ${#1}$}}
\def\Alfven{Alfv\'{e}n~}
\def\Alfvenic{Alfv\'{e}nic~}

\shorttitle{MHD simulations of Global Disks with $B_z$}
\shortauthors{Suzuki \& Inutsuka}

\begin{document}
\begin{CJK*}{UTF8}{gbsn}
\title{MHD Simulations of Global Accretion Disks with Vertical Magnetic Fields}

\author{Takeru K. Suzuki$^{1}$ \& Shu-ichiro Inutsuka$^{1}$}
\email{stakeru@nagoya-u.jp}
\altaffiltext{1}{Department of Physics, Nagoya University,
Nagoya, Aichi 464-8602, Japan}

\begin{abstract}
We report results of three dimensional mangetohydrodynamical (MHD) 
simulations of global accretion disks threaded with weak vertical magnetic 
fields. 
We perform the simulations in the spherical coordinates with different 
temperature profiles and accordingly different rotation profiles.
In the cases with a spatially constant temperature, because the rotation 
frequency is vertically constant in the equilibrium condition, general 
properties of the turbulence excited by magnetorotational instability (MRI) 
are quantitatively similar to those obtained in local shearing box simulations.
On the other hand, in the cases with a radially variable temperature profile,
the vertical differential rotation, which is inevitable in the equilibrium 
condition, winds up the magnetic field lines, in addition to the usual 
radial differential rotation. 
As a result, the coherent wound magnetic fields contribute to the Maxwell 
stress in the surface regions. 
Our global simulations give somewhat larger density fluctuation, $\delta \rho/
\rho=0.1-0.2$, near the midplane than 
the values obtained in previous local shearing box simulations and global 
simulations without net vertical magnetic field. 
The velocity fluctuations, dominated by the radial component, are 
$\approx 0.1-0.2$ of the local sound speed. 
The azimuthal power spectra of the magnetic fields show shallow slopes, 
$\propto m^0\sim m^{-1}$, where $m$ is an azimuthal mode number, which might 
be related to the energy injection by MRI from small scales. 
On the other hand, the power spectra 
of the velocities and density show steeper slopes, $\propto m^{-1}\sim m^{-2}$.
We observe intermittent and structured disk winds driven by the Poynting flux 
associated with the MHD turbulence, with the slightly smaller mass fluxes 
than that obtained in our local simulations.
The Poynting flux originating from magnetic tension is injected from the 
regions above a scale height toward both the midplane and the surfaces. 
Related to this, 
sound waves are directed to the midplane from the surface regions. 
The mass accretion mainly occurs near the surfaces and the gas near the 
midplane slowly moves outward in the time domain of the present simulations. 
The vertical magnetic fields are also dragged 
inward in the surface regions, while they stochastically move outward and 
inward around the midplane. 
The difference of the velocities 
at the midplane and the surfaces might cause large-scale meridional 
circulations. Applying to protoplanetary disks, these waves and circulation 
are supposed to play an important role in the dynamics of solid particles.
We also discuss an observational implication of induced spiral structure in 
the simulated turbulent disks.
 
\end{abstract}

\section{Introduction}
Magnetohydrodynamical (MHD) turbulence is believed to play a central role 
in the transport of the angular momentum and the mass in accretion disks 
\citep[][and reference therein]{bh98}. 
Magnetorotational instability \citep[MRI;][]{vel59,cha61,bh91} is a 
promising mechanism that drives MHD turbulence in accretion disks. 
To date various attempts have been carried out to understand fundamental 
properties of MRI-driven turbulence. MHD simulations with local shearing boxes 
have been extensively performed 
\citep{hgb95,mt95,bra95,sto96,tur03,san04,si09,hir09,shi10,dav10,suz10}.  
One of the important findings by local MHD simulations is that the net vertical 
magnetic field controls the saturation level of the turbulence \citep{hgb95,
san04,pes07,si09,oh11}, which essentially determines the strength of the 
transport of angular momentum and resulting mass accretion.

On the other hand, the shearing box approximation is somewhat too idealistic 
since various effects that are important in realistic accretion 
disks are neglected. For instance, mass accretion cannot be handled in 
the shearing box treatment; instead the accretion rate is simply estimated 
from the transported angular momentum under the time-steady condition. 
In order to study realistic accretion disks, global MHD simulations have also 
been performed recently 
\citep{mac00,haw00,pn03,mm03,nis06,fn06,bec09,flo11,flo12,fro11,fro13,haw11,haw13,
pb13a,pb13b}. 
However, global simulations of accretion disks threaded with vertical magnetic 
fields have not been extensively performed except for a limited number 
of attempts \citep[e.g.][]{bec09}. 

In this paper, we investigate properties of MHD turbulence in  
accretion disks threaded with weak vertical magnetic fields by global MHD 
simulations.  In global disks the rotation profile in the equilibrium 
condition is determined by the distribution of density and temperature. 
In general, the rotation frequency changes in the vertical direction, 
in addition to the radial direction, unless the gas pressure satisfies 
a barotropic equation of state, $p=p(\rho)$
\citep[generalization of von Zeipel (1924) theorem; 
{\it e.g.},][; see \S \ref{sec:incd}]{koz78,mae99}. 
Then, the vertical magnetic fields are 
wound by the vertical differential rotation. In order to study this 
effect, we simulate disks with different temperature profiles. 

This paper is organized as follows. In \S \ref{sec:setup}, we describe 
the setups of the global simulations. After presenting overall time evolutions 
(\S \ref{sec:ote}) and some snapshots of the disks (\S \ref{sec:snp}), 
in \S \ref{sec:trb} we inspect details of MHD turbulence in the global 
accretion disks.

\section{Simulation Setups}
\label{sec:setup}
We simulate the time evolution of global accretion disks threaded 
by weak net vertical magnetic fields. 
Our simulations are performed in spherical coordinates, $(r,\theta,\phi)$ 
and solve a following set of ideal MHD equations:
\begin{equation}
\frac{d \rho}{d t} + \rho \mbf{\nabla}\cdot \mbf{v} = 0,
\end{equation}
\begin{equation}
\rho\frac{d\mbf{v}}{dt} = -\mbf{\nabla}\left(p + \frac{B^2}{8\pi}\right) 
+ \left(\frac{\mbf{B}}{4\pi}\cdot \mbf{\nabla}\right)\mbf{B}
-\rho \mbf{\nabla}\Phi
\label{eq:mom}
\end{equation}
and
\begin{equation}
\frac{\partial \mbf{B}}{\partial t} = \mbf{\nabla \times (v\times B)}, 
\end{equation}
where the variables have the conventional meanings, and we take into 
account Newtonian gravity, $\Phi=-GM/r$, by a central object with 
mass $M$ in Equation (\ref{eq:mom}), but neglect the self-gravity in 
the accretion disks. We consider different
temperature distributions described in \S \ref{sec:incd}. 
At each location we assume locally isothermal gas and do not solve an 
energy equation. Gas pressure, $p$, and density, 
$\rho$, are connected by sound speed, $c_{\rm s}$, which is spatially variable 
but constant with time as 
\begin{equation}
p = \rho c_{\rm s}^2. 
\end{equation}
For the data analyses, we mainly use cylindrical coordinates, $(R,\phi,z)$. 
To do so, we convert data in the $(r,\theta,\phi)$ coordinates to 
the $(R,\phi,z)$ coordinates. 

We modify the simulation code used for the simulations in local shearing 
boxes \citep{si09,suz10} to handle global disks in the spherical coordinates.
The adopted scheme is 2nd order Godunov-CMoCCT method \citep{san99}, in which 
we solve nonlinear Riemann problems with magnetic pressure at cell boundaries 
for compressive waves \citep{ii11} and adopt the consistent method of 
characteristics (CMoC) for the evolution of magnetic fields \citep{cl96,sn92} 
under the constrained transport (CT) scheme \citep{eh88} 
for the conservation of magnetic flux. We use the CFL condition of 0.3 for 
the time update in all the cases.

The accretion disks are set up in the simulation box that extends in 
$\theta=\frac{\pi}{2}\pm 0.5$. The radial and azimuthal sizes of each 
model are summarized in Table \ref{tab:models}.  
The difference between Cases I and II is the temperature profile (see 
\S \ref{sec:incd}); in Case I the radial box is only in $< 25 r_{\rm in}$ 
because the equilibrium rotation profile does not exist in the outer region; 
in Case II we use a very large box size, $>400r_{\rm in}$, where $r_{\rm in}$ 
is the inner radius of the simulation box, to avoid effects of 
the unphysical reflection at the outer boundary. 
We use homogeneous grid spacing, $\Delta \theta$ and $\Delta \phi$, 
in the $\theta$ and $\phi$ directions.
The radial grid size, $\Delta r$, is set up in proportion to $\propto r$. 
Then, the ratio of the grid sizes in the $r$, $\theta$, and $\phi$ directions, 
$(\Delta l_r,\Delta l_{\theta},\Delta l_{\phi})\equiv (\Delta r,r\Delta \theta,
r\sin\theta\Delta \phi)$, is constant with $r$ at the midplane, which is 
$(1:1:\pi)$ for low-resolution runs and $\approx (1:1.25:2)$ for 
high-resolution runs (Table \ref{tab:models}). 

\subsection{Initial Condition}
\label{sec:incd}

\begin{figure*}[t]
\begin{center}
\includegraphics[height=0.25\textheight,width=0.4\textwidth]{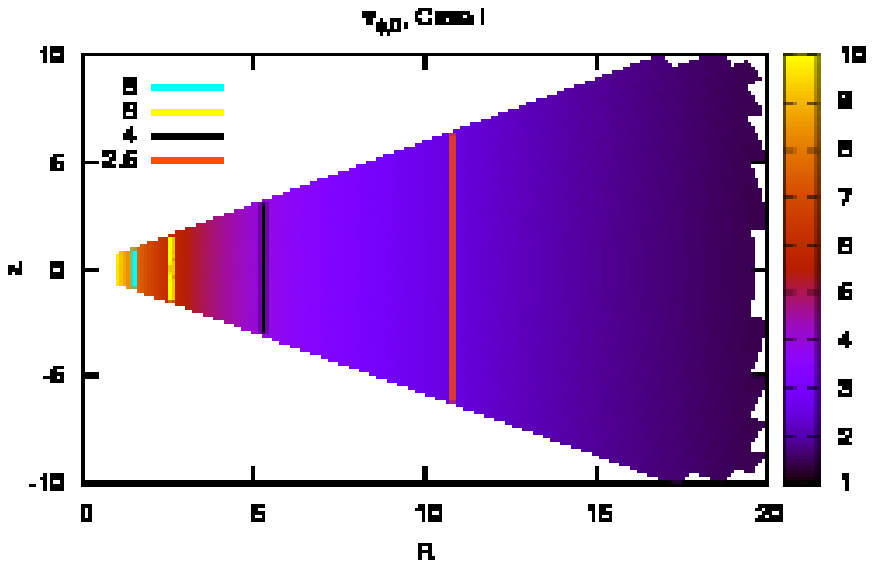}
\includegraphics[height=0.25\textheight,width=0.4\textwidth]{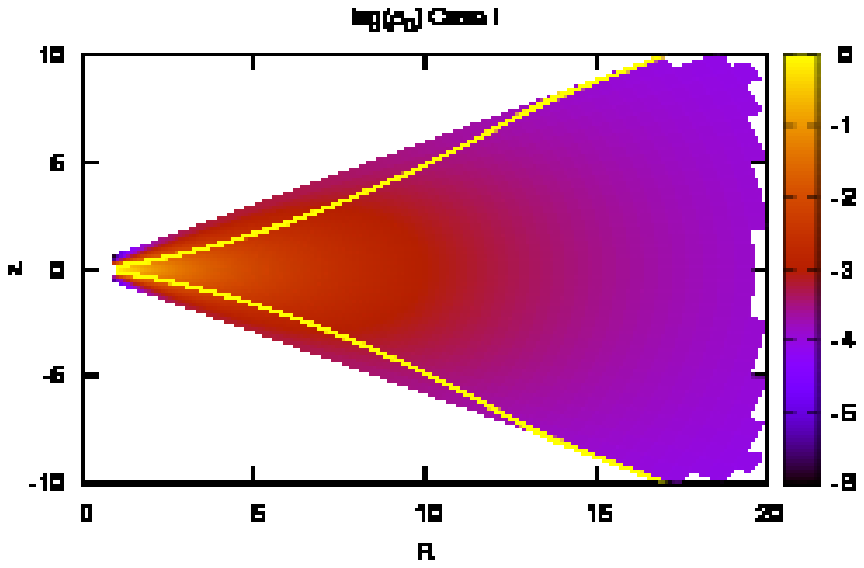}\\
\includegraphics[height=0.25\textheight,width=0.4\textwidth]{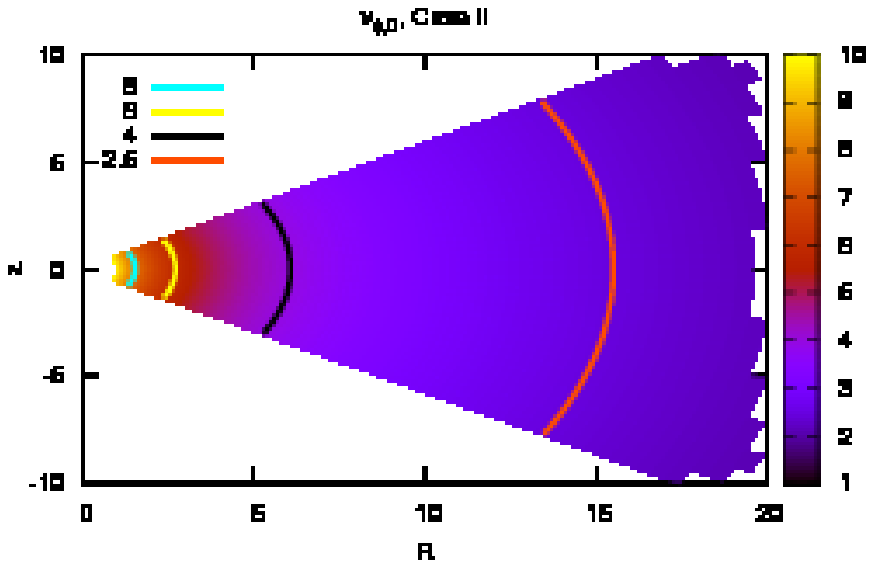}
\includegraphics[height=0.25\textheight,width=0.4\textwidth]{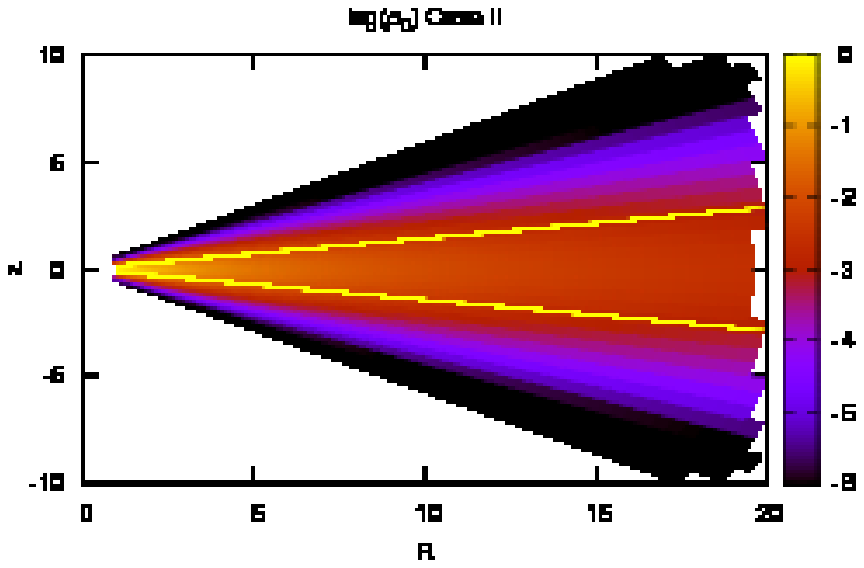}
\end{center}
\caption{Initial condition of Case I ($T=$const.; {\it upper}) and Case II 
($T\propto 1/r$; {\it lower}) in an $(R,z)$ plane. 
The left panels compare the rotation velocities 
of the two cases. The right panels compare the densities, where the yellow 
lines correspond to the locations of 1 scale height. In these figures, 
only the region inside $<20r_{\rm in}$ is shown. }
\label{fig:intcond_col}
\end{figure*}

\begin{deluxetable*}{cccccccc}[h]
\tabletypesize{\scriptsize}
\tablecaption{Simulation setups. \label{tab:models}
}
\tablehead{
 & \colhead{Simulation box} 
& \colhead{Resolution} 
& \colhead{Grid Size} 
& \colhead{$\nu$} &  
& \colhead{$t_{\rm end}$} 
& \colhead{$\Delta t_{\rm ave}$}\\
\colhead{Model}& \colhead{$(r,\theta,\phi)$} &\colhead{$(r,\theta,\phi)$} 
& \colhead{at midplane} & \colhead{($T\propto r^{-\nu}$)} 
& \colhead{$\beta_{z,{\rm mid}}$} & \colhead{$(t_{\rm rot,in})$} 
& \colhead{$(t_{\rm rot,in})$}\\
 & & & \colhead{$\Delta l_r:\Delta l_{\theta}:\Delta l_{\phi}$}& & & &
}
\startdata
I--high & (1-25,$\frac{\pi}{2}\pm 0.5$,0--$\pi$) & (512,128,256) & 
$\approx 1:1.25:2$ & 0 & $10^5$ & 600 & 200--300\\
II--high & (1-640,$\frac{\pi}{2}\pm 0.5$,0--$\pi$) & (1024,128,256) &
$\approx 1:1.25:2$ & 1 & $10^5$ & 1830 & 1200--1800\\
I-low & (1-22,$\frac{\pi}{2}\pm 0.5$,0--$2\pi$) & (192,64,128) & 
$=1:1:\pi$ & 0 & $10^5$ & 1000 & 300--500\\
II--low & (1-470,$\frac{\pi}{2}\pm 0.5$,0--$2\pi$) & (384,64,128) & 
$=1:1:\pi$ & 1 & $10^5$ & 2000 & 1200--2000
\enddata
\tablecomments{From left to right, tabulated are the name of models, 
a simulation box size, the numbers of grid points, the ratio of grid spacing 
at the midplane, the power law index of the temperature profile, 
the initial plasma $\beta$ value at the midplane, the simulation time 
in unit of inner rotation time, and the duration for the time average.}
\end{deluxetable*}

The gas pressure is initially distributed with a power-law dependence on 
$r$ at the midplane, $\theta=\pi/2$, of the disks, 
\begin{equation}
p_{\rm mid} = p_{\rm in,mid}\left(\frac{r}{r_{\rm in}}\right)^{-\mu}, 
\end{equation}
where the subscript ``in'' denotes the inner boundary of the simulation box and 
the subscript ``mid'' denotes the midplane of the disks. 
In this paper we only consider cases with $\mu=3$. 
According to the adopted temperature distributions, the initial density and 
rotation profiles are determined to satisfy the equilibrium configuration
as explained in \S \ref{sec:caseI} \& \S \ref{sec:caseII}. 
As a seed for MRI we include small perturbations in the three components 
of $v$ with the amplitude equal to $5\times 10^{-3}$ of the local sound speed.

We set up the initial vertical magnetic fields,   
\begin{equation}
B_z = B_{z,{\rm in}} \left(\frac{R}{r_{\rm in}}\right)^{-\mu/2}, 
\label{eq:Bzinit}
\end{equation}
to give a constant plasma $\beta_{z,{\rm mid}}=8\pi p_{\rm mid}/B_z^2$, 
a ratio of gas 
pressure to magnetic pressure, at the midplane of the accretion disks. 
In this paper, we simulate cases with initial $\beta_{z,{\rm mid}}=10^5$.  
In order to set up the initial straight vertical field lines in the spherical 
coordinates exactly satisfying $\mbf{\nabla\cdot B}=0$ within the accuracy 
of round-off errors, we use the $\phi$ component of the vector potential 
(see the Appendix).

We initially set up the equilibrium configurations of the accretion disks 
by taking into account the force balance of the gas component while neglecting 
the magnetic pressure since the initial $B_z$ is quite small. 
We perform simulations with two types of temperature distributions: 
$T=$ const., 
which we call Case I, and $T\propto 1/r$, which we call Case II. 
The most important 
difference between the two cases is the difference of the rotation profiles 
of the equilibrium configurations.
If we neglect the effect of the magnetic fields, an equilibrium rotation 
profile of a differentially rotating accretion disk is derived from the 
force balance among the gas pressure, the centrifugal force, and the gravity 
of Equation (\ref{eq:mom}) in the $(R-z)$ plane of the cylindrical coordinates, 
\begin{equation}
-\frac{1}{\rho}\frac{\partial p}{\partial R} + R \Omega^2 -\frac{\partial \Phi} 
{\partial R} = 0,
\label{eq:hydstr}
\end{equation}
and
\begin{equation}
-\frac{1}{\rho}\frac{\partial p}{\partial z} -\frac{\partial \Phi} 
{\partial z} = 0, 
\label{eq:hydstz}
\end{equation}
where $\Omega$ is rotation frequency. 
Differentiating Equation (\ref{eq:hydstr}) with $z$ and Equation 
(\ref{eq:hydstz}) with $R$, and subtracting them from each other, we have
\begin{equation}
-\frac{\partial}{\partial z}\left(\frac{1}{\rho}\frac{\partial p}{\partial R}
\right) + \frac{\partial}{\partial R}\left(\frac{1}{\rho}
\frac{\partial p}{\partial z} \right) 
+ \frac{\partial}{\partial z} (R \Omega^2) = 0.
\end{equation}
If gas pressure is globally barotropic, $p=p(\rho)$, the first and second 
terms are canceled out, hence, 
\begin{equation}
\frac{\partial \Omega}{\partial z} = 0 
\end{equation}
\citep[von Zeipel (1924) theorem, e.g.][]{koz78}.
In Case I the gas pressure satisfied a barotropic equation of state, and 
the rotation frequency is constant along the vertical direction, while 
in Case II the disks rotate 
differentially along the {\it vertical} direction in addition to the radial
direction. 

\begin{figure}[h]
\begin{center}
\includegraphics[height=0.4\textheight,width=0.5\textwidth]
{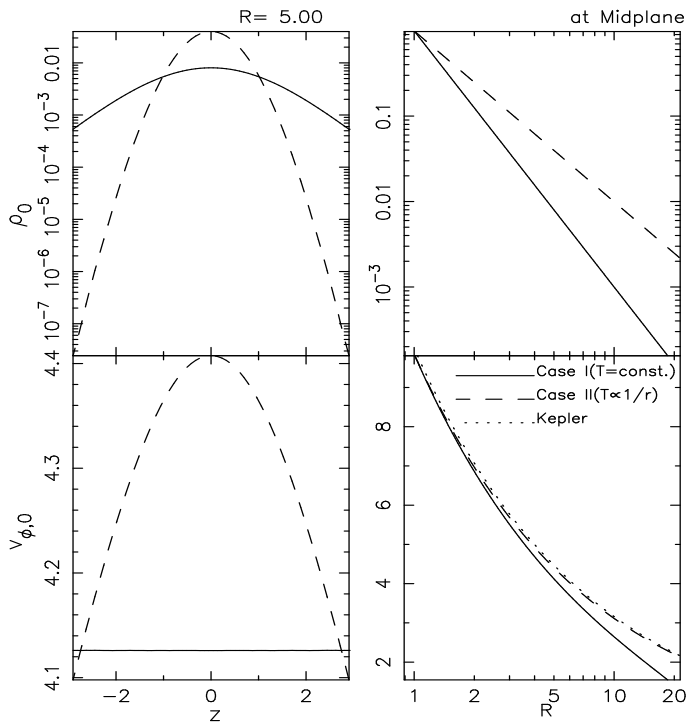}
\end{center}
\caption{Comparison of the initial equilibrium structures of Case I (solid) 
and II (dashed). The density structures and the rotation velocities are 
plotted in the upper and lower panels, respectively. 
On the left the vertical ($z$) structures at $R=5$ are compared, 
and on the right the radial structures at the midplane are compared. 
In the bottom right panel, the Keplerian rotation speed is also 
compared ({\it dotted}).}
\label{fig:intcond_lin}
\end{figure}

\subsubsection{Case I}
\label{sec:caseI}
In Case I, we adopt constant temperature in the simulation box, namely 
the sound speed is constant everywhere:  
\begin{equation}
c_{\rm s} = c_{\rm s,in}=0.1\sqrt{\frac{GM}{r_{\rm in}}} = {\rm const.} ,
\label{eq:cItp}
\end{equation}
which denotes that we set the ratio of the Keplerian rotation speed to 
the sound speed as 10 at the inner boundary, $r=r_{\rm in}$. 
The ratio decreases with $R$ since the Keplerian rotation speed decreases. 
The initial density structure that satisfies the equilibrium configuration
is given as 
\begin{eqnarray}
\rho &=& \rho_{\rm in,mid} 
\left(\frac{R}{r_{\rm in}}\right)^{-\mu}
\exp\left[\frac{GM}{c_{\rm s}^2}\left(\frac{1}{r}
-\frac{1}{R}\right)\right] \nonumber \\
&\approx& \rho_{\rm in,mid}\left(\frac{R}{r_{\rm in}}\right)^{-\mu}\exp
\left[-\frac{GMz^2}{2R^3 c_{\rm s}^2}\right] \nonumber \\
&\equiv& \rho_{\rm in,mid}
\left(\frac{R}{r_{\rm in}}\right)^{-\mu}\exp\left[-\frac{z^2}{H^2}\right], 
\label{eq:cIdnsap}
\end{eqnarray}
where note that $r=\sqrt{R^2+z^2}$.
This profile gives a familiar expression of 
the scale height\footnote{Note that instead of $H$ in 
Equation (\ref{eq:cIdnsap}), $H^{'} = c_{\rm s}/\Omega_{\rm K} (=H/\sqrt{2})$, 
is often used \citep[e.g.,][]{fn06,flo11}, which gives 
$\rho \propto \exp\left[-\frac{z^2}{2({H^{'})}^2}\right]$}, 
$H \approx \sqrt{2}c_{\rm s}/\Omega_{\rm K}$, where 
$\Omega_{\rm K}$ is the Keplerian frequency, 
$\Omega_{\rm K}=\sqrt{\frac{GM}{r^3}}$. Therefore, $H\propto R^{3/2}$, or 
\begin{equation}
\frac{H}{R}= \sqrt{2} c_{\rm s,in}\left(\frac{R}{r_{\rm in}}\right)^{1/2} 
= 0.14\left(\frac{R}{r_{\rm in}}\right)^{1/2}. 
\label{eq:cIhr}
\end{equation}
Our simulation box covers the spherical coordinates of $\theta=\pi/2 \pm 0.5$. 
Therefore, the vertical extent of the simulation box, $R\tan(\theta
-\frac{\pi}{2})$, 
measured in $H$ decreases with increasing $R$. At the inner boundary 
$R=r_{\rm in}$, the simulation 
box covers $\pm 4H$, but at $R=10r_{\rm in}$ it covers 
$\approx \pm 1.3 H$ (Figure \ref{fig:intcond_col}). 
When analyzing the simulation data, we need to carefully take into account this 
effect. First, properties of the disk winds depend on the vertical box 
size in scale height \citep{suz10,fro13}. Second, saturation levels of 
MRI-driven turbulence could depend on $r$ because one scale height is 
resolved by larger numbers of $\theta$ ($\approx$ vertical) grids for 
larger $r$.

For a given $\mu$($=3$ throughout this paper)  the rotation velocity, 
$v_{\phi}$, is self-consistently determined as 
\begin{equation}
v_{\phi}^2 = \frac{GM}{R} - \mu c_{\rm s}^2. 
\label{eq:rotI}
\end{equation}
Here, this equation shows that because of the second term on the 
right-hand side the disks rotate with sub-Keplerian velocities. 
An important aspect of the rotation profile is that the rotation speed, 
$v_{\phi}$, is a constant along the vertical direction and the direction 
of the differential rotation is exactly along the cylindrically radial 
direction, $\mbf{R}$ (Figures \ref{fig:intcond_col} and 
\ref{fig:intcond_lin}). 
Therefore, vertical field lines are not wound up by shearing motions of 
the differential rotation, which is the most important difference from 
Case II described below.

Equation (\ref{eq:rotI}) indicates that the rotation velocity becomes 
0 for a large $R$, because the radial force balance is satisfied between 
the gravity and the gas pressure without the contribution from the 
centrifugal force. For the adopted parameters, $\mu=3$ and Equation 
(\ref{eq:cItp}), $v_{\phi}=0$ at $R=33r_{\rm in}$, because the gravity 
is solely supported by the gas pressure gradient. Outside of this radius, 
no equilibrium profile is achieved. 
Therefore, we set the simulation boxes for Cases I-high and I-low 
in $r<30r_{\rm in}$ (Table \ref{tab:models}). 

\subsubsection{Case II}
\label{sec:caseII}
In Case II, we consider the temperature distribution in proportion to $1/r$, 
and then the sound speed has a dependence \footnote{In previous studies, 
a similar temperature profile, $c_{\rm s}\propto R^{-1/2}$, which 
depends on cylindrical $R$ instead of spherical $r$, 
is often adopted \citep[e.g.][]{fn06,flo11,flo12}. 
We do not believe that the difference significantly affects our 
simulation results. }, 
\begin{equation}
c_{\rm s}^2 = c_{\rm s,{\rm in}}^2\left(\frac{r}{r_{\rm in}}\right)^{-1}, 
\end{equation}
where $c_{\rm s,in}$ is set to be 0.1 of the Keplerian rotation speed 
at $r_{\rm in}$, which is the same as in Case I (Equation \ref{eq:cItp}). 
In Case II the ratio of $c_{\rm s}$ to the Keplerian rotation speed is 
kept constant $=0.1$ with $r$ owing to the radial temperature gradient. 
We can derive a density structure and a rotation profile that satisfy 
the force balance: 
\begin{equation}
\rho = \rho_{\rm in,mid} \left(\frac{r}{r_{\rm in}}\right)^{-\mu+1} 
\sin^{\nu}\theta, 
\label{eq:dnstrII}
\end{equation}
and 
\begin{equation}
v_{\phi}^2 = \nu c_{\rm s}^2, 
\label{eq:rotII}
\end{equation}
where $\mu$ and $\nu$ satisfy  
\begin{equation}
\mu + \nu = \frac{GM}{r c_{\rm s}^2} = 100 (={\rm const.}), 
\label{eq:asprt}
\end{equation}
The comparison between Equations (\ref{eq:rotII}) and (\ref{eq:asprt}) 
shows that the rotation speed $v_{\phi}=\sqrt{\nu}c_{\rm s}$ is smaller than 
the Keplerian velocity, $r \Omega_{\rm K}$ 
because of the contribution from the gas pressure
gradient to the force balance.
In this paper, we adopt $\mu=3$ and accordingly $\nu =97$.
A vertical scale height, $H$, can be approximately derived from Equation 
(\ref{eq:dnstrII}). We expand $\theta$ around $\theta = \frac{\pi}{2}$ 
assuming $\theta - \pi/2 (\approx z/r) \ll 1$: 
\begin{eqnarray}
\sin^{\nu} \theta \approx 1-\frac{\nu}{2}\left(\frac{z}{r}\right)^2
&\approx& \exp\left(-\frac{\nu}{2}\left(\frac{z}{r}\right)^2\right) 
\nonumber \\
&\equiv& \exp\left(-\frac{z^2}{H^2}\right), 
\end{eqnarray}
where $H$ is further transformed by using Equation (\ref{eq:asprt}) as
\begin{equation}
H^2=\frac{2r^2}{\nu}=\frac{2 r^2 c_{\rm s}^2}{GM}\frac{\mu+\nu}{\nu} 
\approx \frac{2 r^2 c_{\rm s}^2}{GM} = \frac{2 c_{\rm s}^2}{\Omega_{\rm K}^2},
\label{eq:cIIsch}
\end{equation}
which gives an asymptotic expression for $H$ in Case II, similar to 
Equation (\ref{eq:cIdnsap}) for Case I. 

In contrast to Case I, in Case II $H\propto R$ from Equation 
(\ref{eq:cIIsch}), or 
\begin{equation}
\frac{H}{R} = \sqrt{2}\frac{c_{\rm s}}{R \Omega_{\rm K}}=0.14
\label{eq:cIIhr}
\end{equation}
Thus, the vertical size, $R\tan(\frac{\pi}{2}\pm 0.5)=\pm 0.55R$, of 
the simulation box covers the constant scale height $\approx\pm 4H$ 
and $H$ is resolved by the same number of $\theta$ grid points, 
which is independent of $R$. 
When analyzing the disk winds and the saturation levels 
of MRI turbulence, the setup for Case II is supposed to be more 
straightforward. 
Also in contrast to Case I, the equilibrium rotation profile exists 
even in the outer region. 
Thus, we take sufficiently large radial box sizes 
(470 for the low-resolution run and 640 for the high-resolution run; Table 
\ref{tab:models}) to avoid unphysical wave reflection at the outer boundary, 
$r=r_{\rm out}$. 
However, in this paper we mainly study the region inside $<20r_{\rm in}$ because 
in the outer region the growth of the magnetic field, which is scaled by 
the rotation frequency, is slow and the saturated state is not achieved 
in the simulation time. In \S \ref{sec:disII} we briefly discuss the time 
evolution in the entire region of Case II-high. 

\subsection{Boundary Condition}
\label{sec:bc}
The boundary condition for the $\phi$ direction is straightforward. In the 
low-resolution runs, we treat the full ($2\pi$) disks and connect one 
edge to the other edge (technically, this is the same as the periodic 
boundary). In the high-resolution runs for the half ($\pi$) 
disks, we adopt a simple periodic boundary condition.

In the $\theta$ direction ($\sim$ the vertical direction), we prescribe 
the outgoing boundary condition for mass and MHD waves by using the seven 
MHD characteristics \citep{si06}, in order to handle disk winds 
\citep[see also][]{si09,suz10}.

We use a viscous accretion condition for the $r$ direction, which is a method 
adopted in \citet{fn06}\footnote{\citet{fn06} also adopted a method with 
resistive buffer zones for both inner and outer radial boundaries for most 
of their simulations, which is different from the viscous accretion condition 
we use for our simulations. 
A reason why we adopt the viscous accretion condition 
is to avoid pileups of masses in the buffer zones for long-time simulations}. 
At both the inner and outer radial boundaries, 
we fix small $v_r$ estimated from the $\alpha$ prescription for standard 
accretion disks \citep{ss73}, 
$v_r=-\frac{3}{2}\frac{\alpha c_{\rm s}^2}{r\Omega}$ 
with $\alpha=5\times 10^{-3}$. 
We fix $v_{\theta}$ and $v_{\phi}$ to the initial values, {\it i.e.}, 
$v_{\theta}=0$ and $v_{\phi}=$ sub-Keplerian rotation speed under 
equilibrium (Equations \ref{eq:rotI} \& \ref{eq:rotII}).
The densities at the inner and outer radial boundaries are also fixed to 
the initial values. 
As for the magnetic fields, we assume the zero-gradient for 
$r^2(B_{\theta}^2+B_{\phi}^2)$ across the boundaries, which {\it nearly}
(not exactly) corresponds to the equilibrium from the magnetic pressure. 
The condition for the magnetic fields also allows flows of magnetic fluxes 
across the boundaries. 
Limitations for the prescribed boundary conditions will be discussed in 
\S \ref{sec:dwd} \& \ref{sec:rf}.

\subsection{Averaged Quantities}
In order to quantitatively analyze numerical results, we take various 
kinds of averages of the obtained physical quantities. 
While we perform the simulations in the 
spherical coordinates $(r,\theta,\phi)$, the data are often analyzed in 
the cylindrical coordinates, $(R,\phi,z)$. We take averages of a physical 
quantity, $A(t,R,\phi,z)$ in the following ways. 

As an example, we explain how to derive a time- and $\phi$-averaged vertical 
structure at a certain $R$ (\S \ref{sec:vst} \& \S \ref{sec:vsvrho}). 
The simple average of $A$ is 
\begin{equation}
\langle A \rangle_{t,\phi}(R,z)
= \frac{\int_{\Delta t_{\rm ave}}dt \int_{\phi_{\rm min}}^{\phi_{\rm max}} d\phi A} 
{\Delta t_{\rm ave} (\phi_{\rm max}-\phi_{\rm min})},
\label{eq:qtavtp}
\end{equation}
where the subscripts ($t,\phi$ in this case) of the bracket correspond 
to the independent variables over which the average is taken and the rest 
($R,z$ in this case) of the independent variables are written in the arguments. 
We integrate $A$ over $\Delta t_{\rm ave}$ (Table \ref{tab:models}) 
and in the entire region with respect to $\phi$ from $\phi_{\rm min}(=0)$ to 
$\phi_{\rm max}$($=\pi$ for the high-resolution runs and $2\pi$ 
for the low-resolution runs). In the denominator, the integration 
($\int_{\Delta t_{\rm ave}}\int_{\phi_{\rm min}}^{\phi_{\rm max}}d\phi = \Delta t_{\rm ave}
(\phi_{\rm max}-\phi_{\rm min})$) is already carried out. 
In contrast to the simple average in Equation (\ref{eq:qtavtp}) 
the density-weighted average is derived as  
\begin{equation}
\frac{\langle \rho A \rangle_{t,\phi}(R,z)}{\langle \rho \rangle_{t,\phi}(R,z)}
=\frac{\int_{\Delta t_{\rm ave}}dt \int_{\phi_{\rm min}}^{\phi_{\rm max}} d\phi \rho A}
{\int_{\Delta t_{\rm ave}}dt \int_{\phi_{\rm min}}^{\phi_{\rm max}} d\phi \rho}.
\label{eq:denweiave}
\end{equation}
For variables concerning magnetic field (\S \ref{sec:vst}), {\it e.g.,} 
magnetic energy, $B^2/8\pi$, we take the simple average, Equation 
(\ref{eq:qtavtp}), and for variables concerning velocity (\S \ref{sec:vsvrho}), 
{\it e.g.} flow speed, $v$, and kinetic energy per mass, $v^2/2$, 
we take the density-weighted average, Equation (\ref{eq:denweiave}); 
in principle we take the average of a variable in units of energy density 
or momentum density.

To study the time evolution of overall trends in the disks (\S \ref{sec:ote}), 
we examine a box average, 
\begin{equation}
\langle A \rangle_{R,\phi,z}(t)
= \frac{ \int_{\Delta R}RdR \int_{\Delta z}dz \int_{\phi_{\rm min}}^{\phi_{\rm max}} 
d\phi A}{\int_{\Delta R}RdR \int_{\Delta z}dz (\phi_{\rm max}-\phi_{\rm min})
},
\label{eq:qtavrpz}
\end{equation}
where we integrate $A$ in the entire region with respect to $\phi$, but 
the regions for the $R$ and $z$ integrations, 
$\Delta R$ and $\Delta z$, are case-dependent and are explained later;
in the denominator 
the integration with $\phi$ 
is carried out, but the integrations with $R$ and $z$ are left 
as they stand because 
$\Delta z$ depends on $R$. 
As explained above we use variables in units per volume for $A$; 
for instance, to check a plasma $\beta(=8\pi p/B^2)$ value, 
we see $8\pi\langle p \rangle/\langle B^2 \rangle$, 
{\it which essentially corresponds to the density-weighted average}, 
rather than the simple volumetric average, $8\pi\langle p /B^2 \rangle$.

When we examine face-on snapshots of the disks (\S \ref{sec:fov}), we take 
a vertical average,
\begin{equation}
\langle A \rangle_{z_{\rm tot}}(t,R,\phi) = \frac{\int_{\Delta z_{\rm tot}} dz A}
{z_{\rm top} - z_{\rm bot}}, 
\end{equation}
over the entire region, 
\begin{equation}
\Delta z_{\rm tot} : z_{\rm bot} \Rightarrow z_{\rm top}
\label{eq:dztot}
\end{equation}
from the lower surface, 
$z_{\rm bot}=-R\tan(\theta_{\rm max}-\frac{\pi}{2})(<0)$, 
to the upper surface, $z_{\rm top}=-R\tan
(\theta_{\rm min}-\frac{\pi}{2})$$(>0$; note that the upper (lower) surface 
corresponds to 
the location at $\theta=\theta_{\rm min\;(max)}=\frac{\pi}{2}\mp 0.5$). 
For the analysis of time-averaged radial dependences (\S \ref{sec:rpB} \& 
\S \ref{sec:rpvrho}), we take the following average,
\begin{equation}
\langle A \rangle_{t,\phi,z}(R)
= \frac{\int_{\Delta t}dt \int_{\Delta z} dz \int_{\phi_{\rm min}}^{\phi_{\rm max}} 
d\phi A} {\Delta t \Delta z(R) (\phi_{\rm max}-\phi_{\rm min})},
\label{eq:qtavtpz}
\end{equation}
where we do not only consider the whole region of $\Delta z_{\rm tot}$, 
but we also take the averages in regions near 
the midplane and surfaces. 
As for the midplane region, we consider the region in $z=\pm H$, 
\begin{equation}
\Delta z_{\rm mid}: -H \Rightarrow H, 
\label{eq:dzmid}
\end{equation}
and for the surface regions, we take averages over
\begin{equation}
\Delta z_{\rm sfc}: (z_{\rm bot} \Rightarrow \frac{3}{4}z_{\rm bot}) + 
(\frac{3}{4}z_{\rm top} \Rightarrow z_{\rm top}),  
\label{eq:dzsfc}
\end{equation}
where see \S \ref{sec:vsvrho} for an example.

In \S \ref{sec:apsb} \& \ref{sec:apsv} we evaluate azimuthal power spectra 
of magnetic fields, velocities, and density fluctuations (see \citet{pb13b} 
for 3D spectra from simulations in the spherical coordinates). 
We take the usual Fourier transformation of a variable $A$ in the azimuthal 
direction,  
\begin{equation}
A(t,R,m,z) = \frac{1}{\sqrt{2\pi}}\int A(t,R,\phi,z)e^{-im\phi}d\phi,  
\label{eq:aps}
\end{equation}
where $m$ has the relation of $m=R k_{\phi}$ to the wave number, $k_{\phi}$, in 
the $\phi$ direction. 
Then, we derive an azimuthal power spectrum after taking a simple 
average over time and $R$--$z$ space, 
\begin{equation}
\langle |A(m)^2| \rangle_{t,R,z} = \frac{\int_{\Delta t} dt\int_{\Delta R} dR 
\int_{\Delta z}dz |A^2(t,R,m,z)|}{\Delta t \int_{\Delta R} dR \int_{\Delta z} dz}.
\label{eq:apsave}
\end{equation}
Here $|A^2|=AA^{\ast}$, where $A^{\ast}$ is the complex conjugate of $A$.

\section{Overview of Time Evolution}
\label{sec:ote}
\begin{figure*}[h]
\begin{center}
\includegraphics[height=0.25\textheight,width=0.33\textwidth]{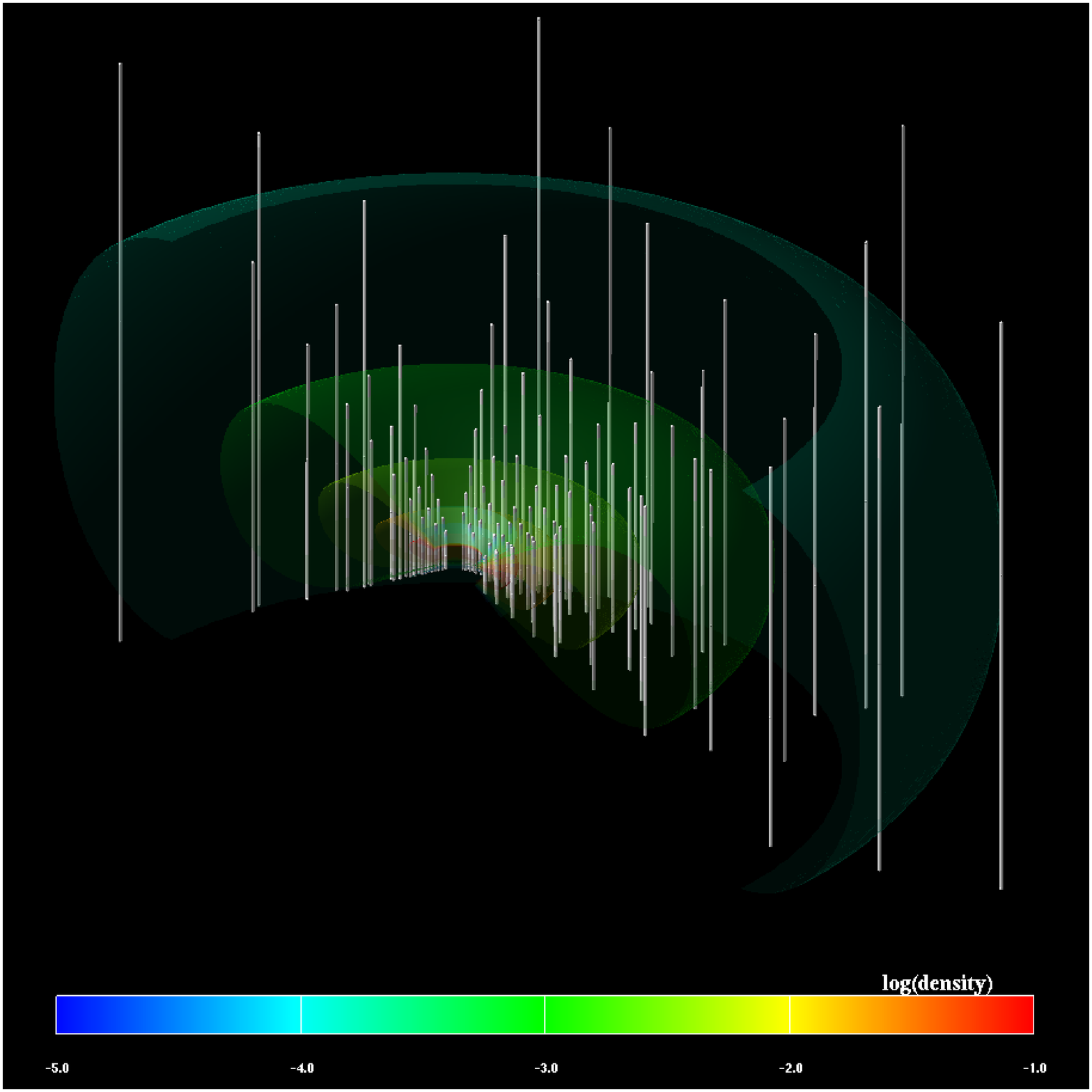}
\includegraphics[height=0.25\textheight,width=0.33\textwidth]{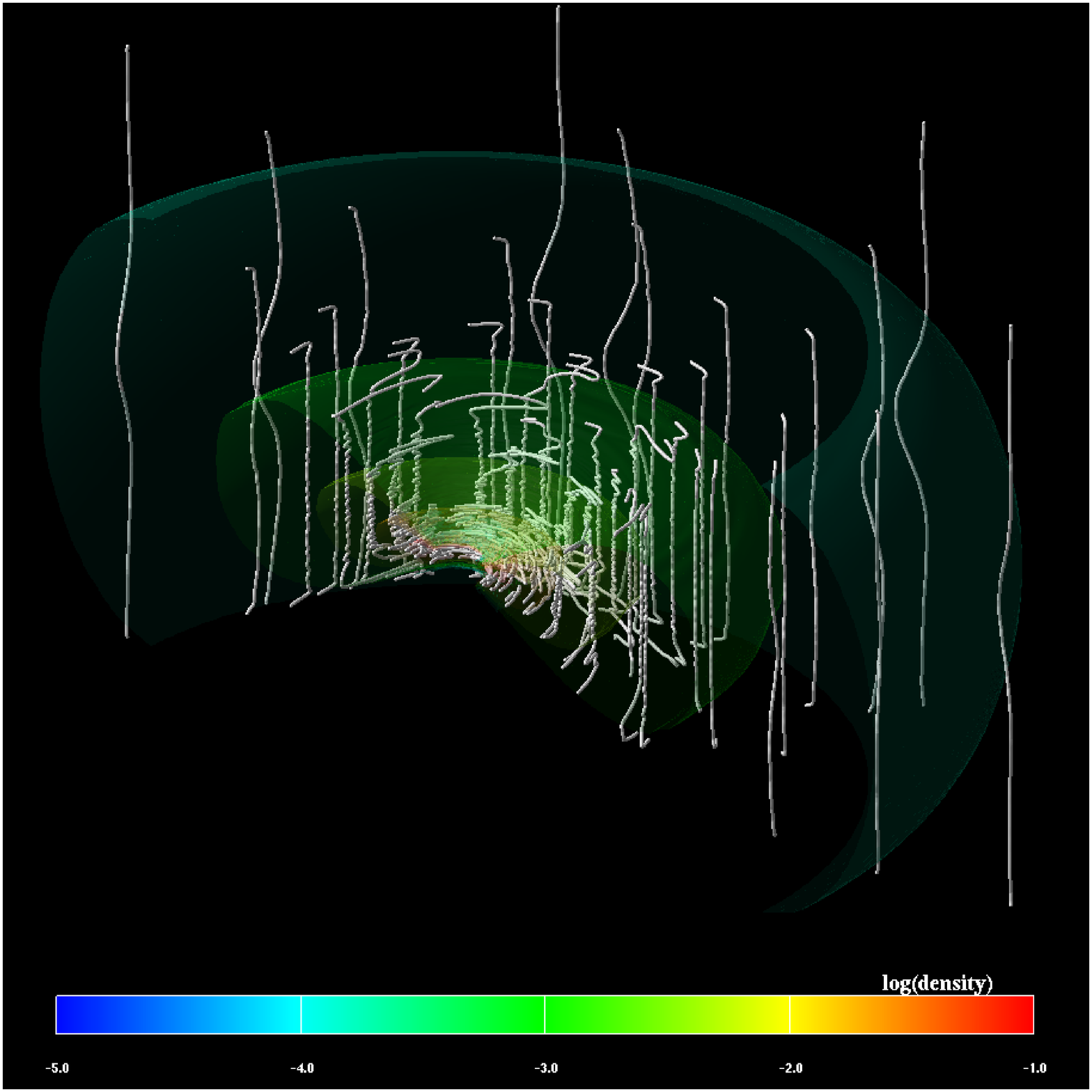}
\includegraphics[height=0.25\textheight,width=0.33\textwidth]{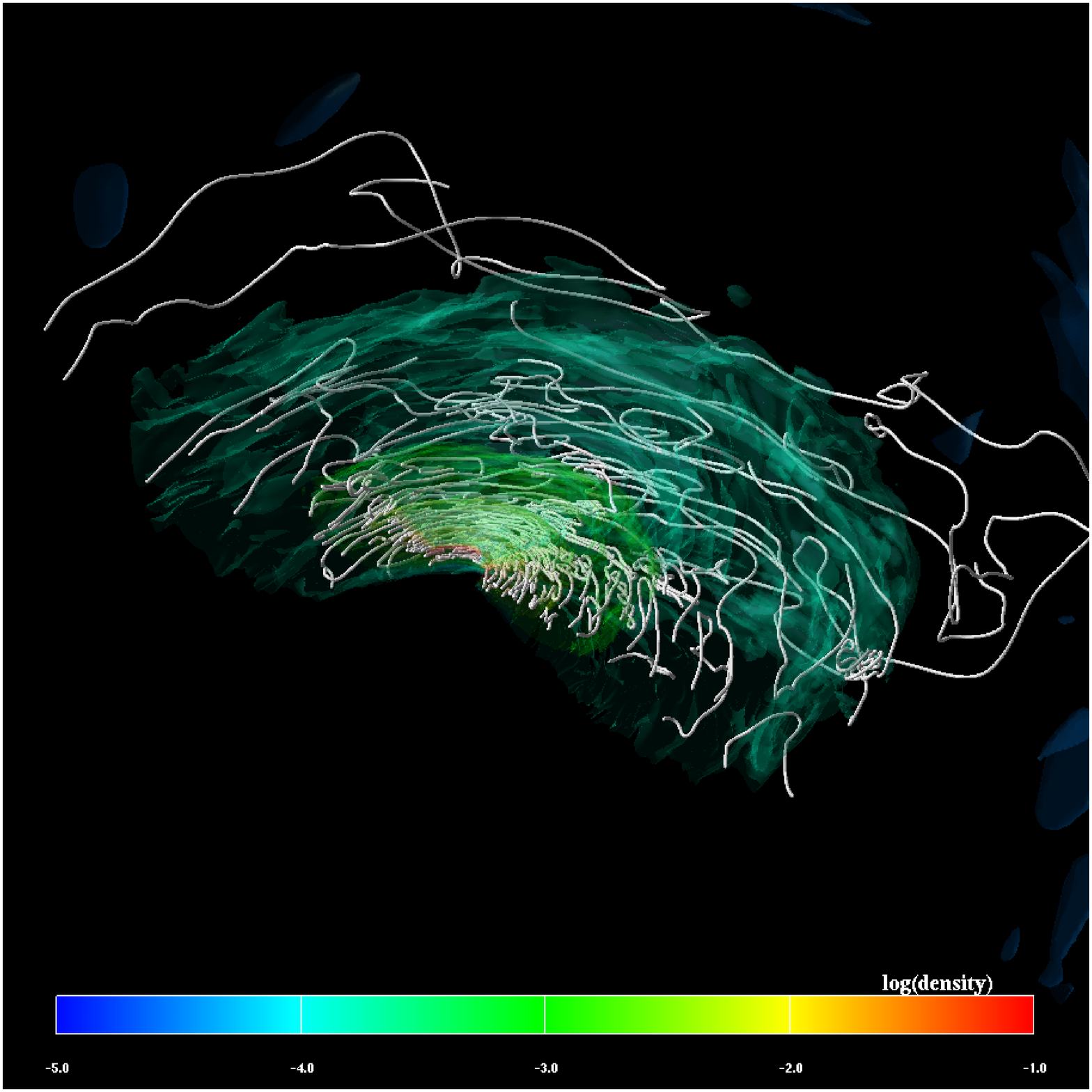}\\
\includegraphics[height=0.25\textheight,width=0.33\textwidth]{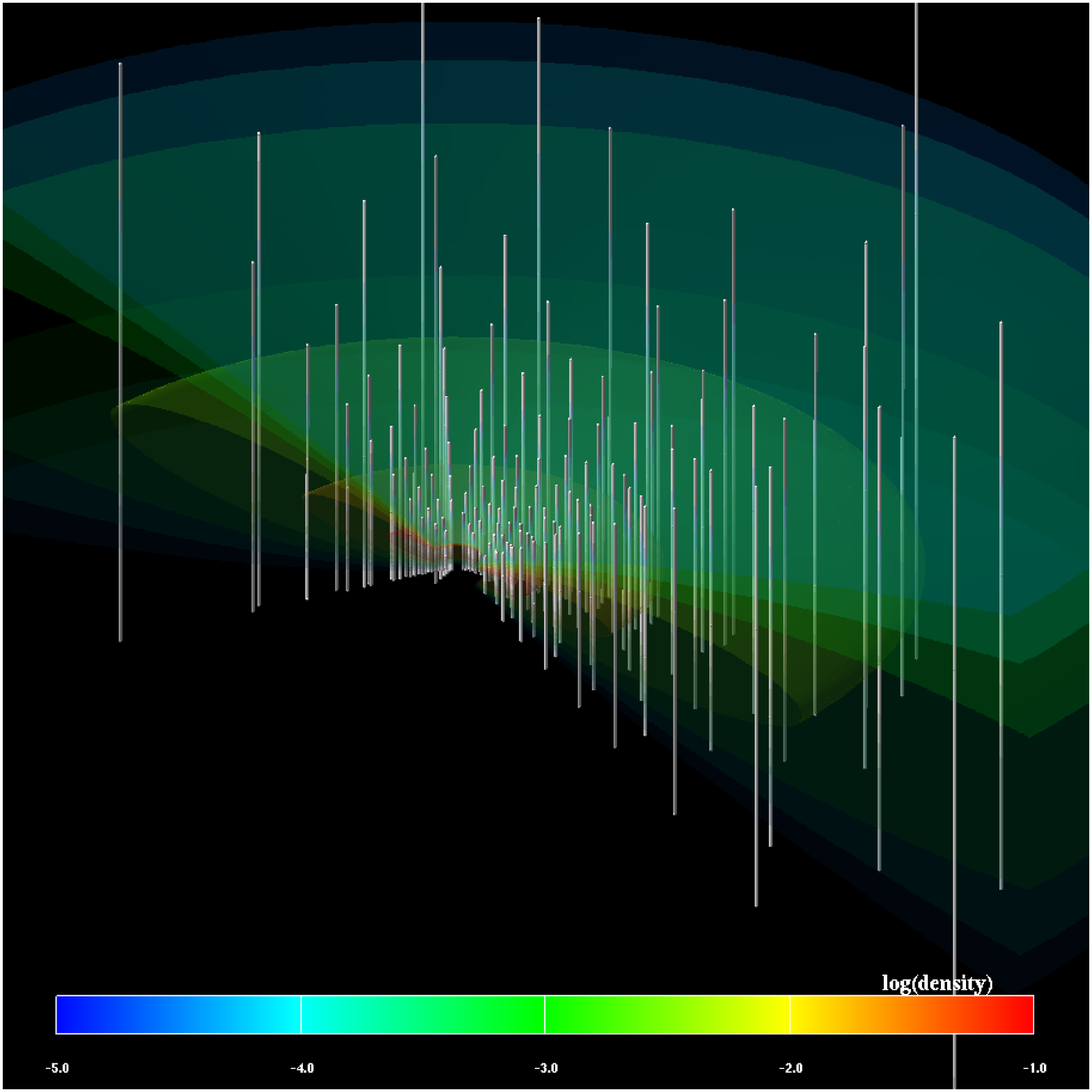}
\includegraphics[height=0.25\textheight,width=0.33\textwidth]{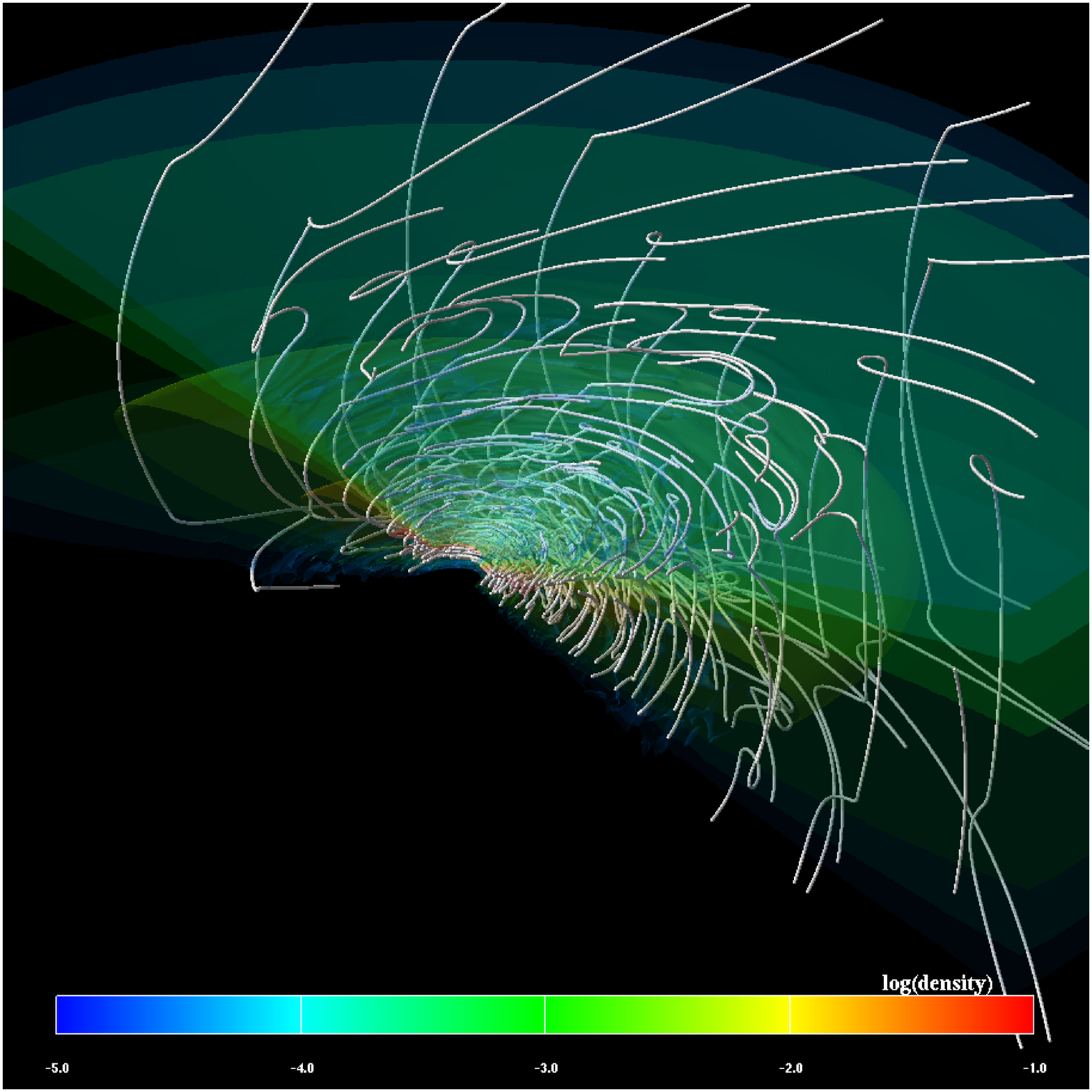}
\includegraphics[height=0.25\textheight,width=0.33\textwidth]{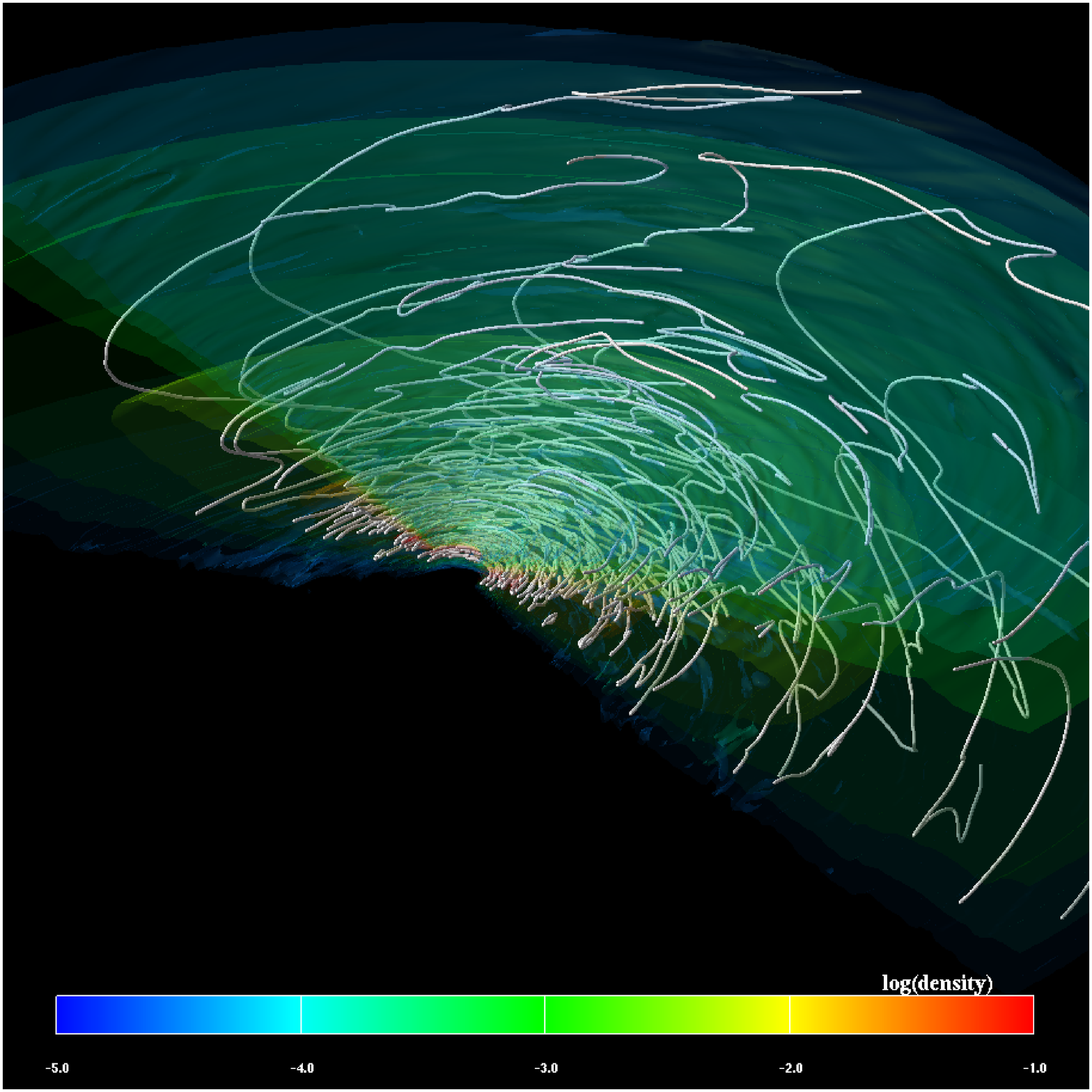}\\
\includegraphics[height=0.25\textheight,width=0.33\textwidth]{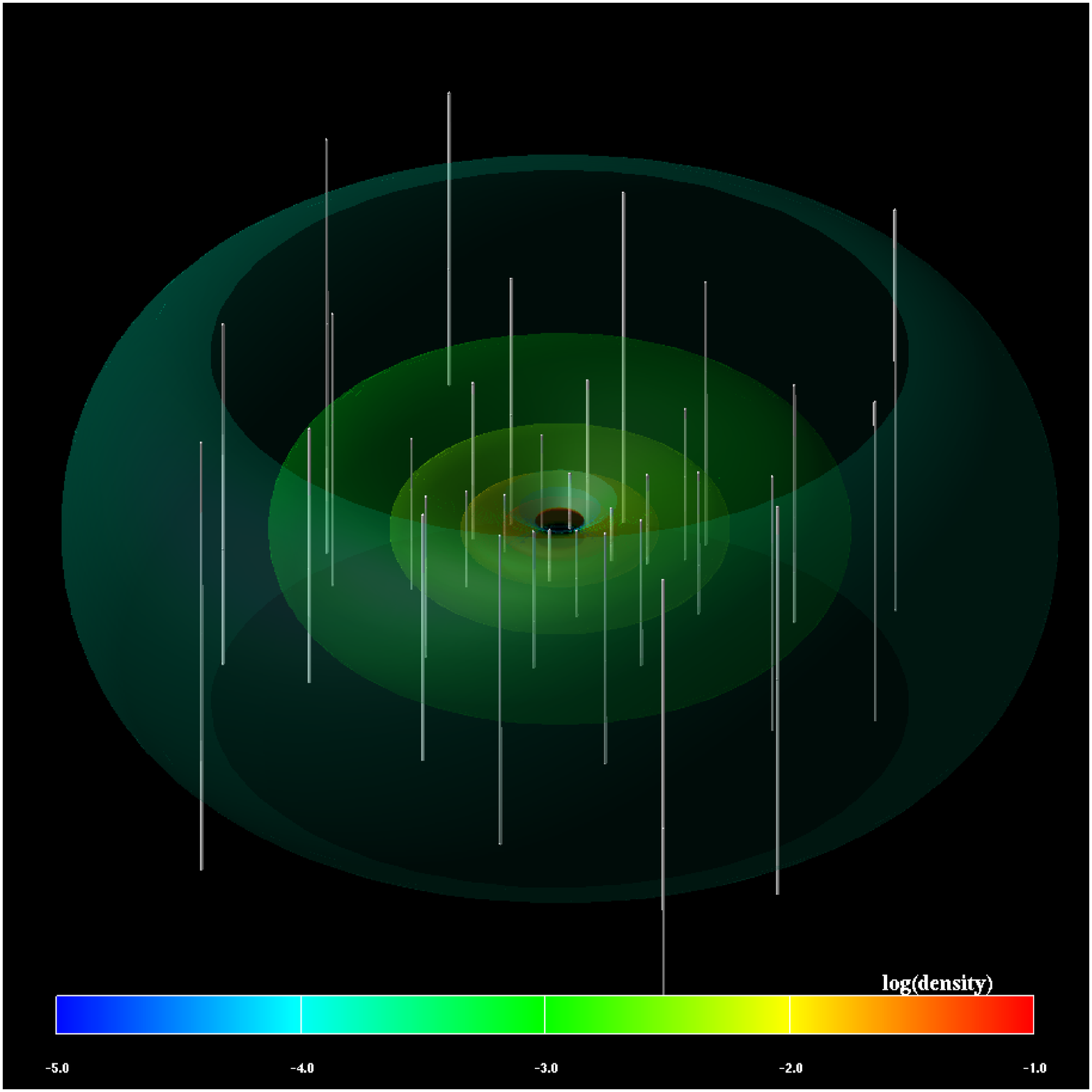}
\includegraphics[height=0.25\textheight,width=0.33\textwidth]{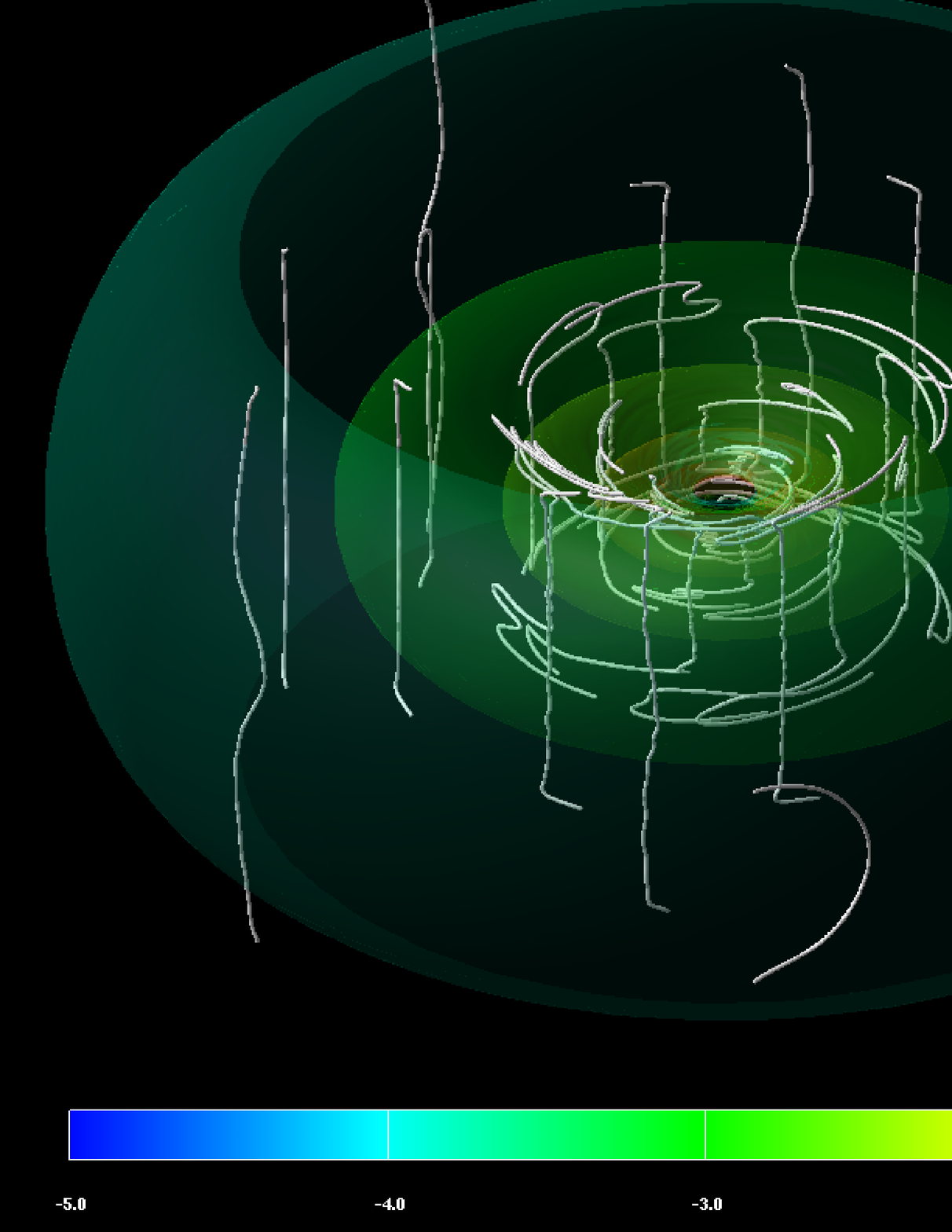}
\includegraphics[height=0.25\textheight,width=0.33\textwidth]{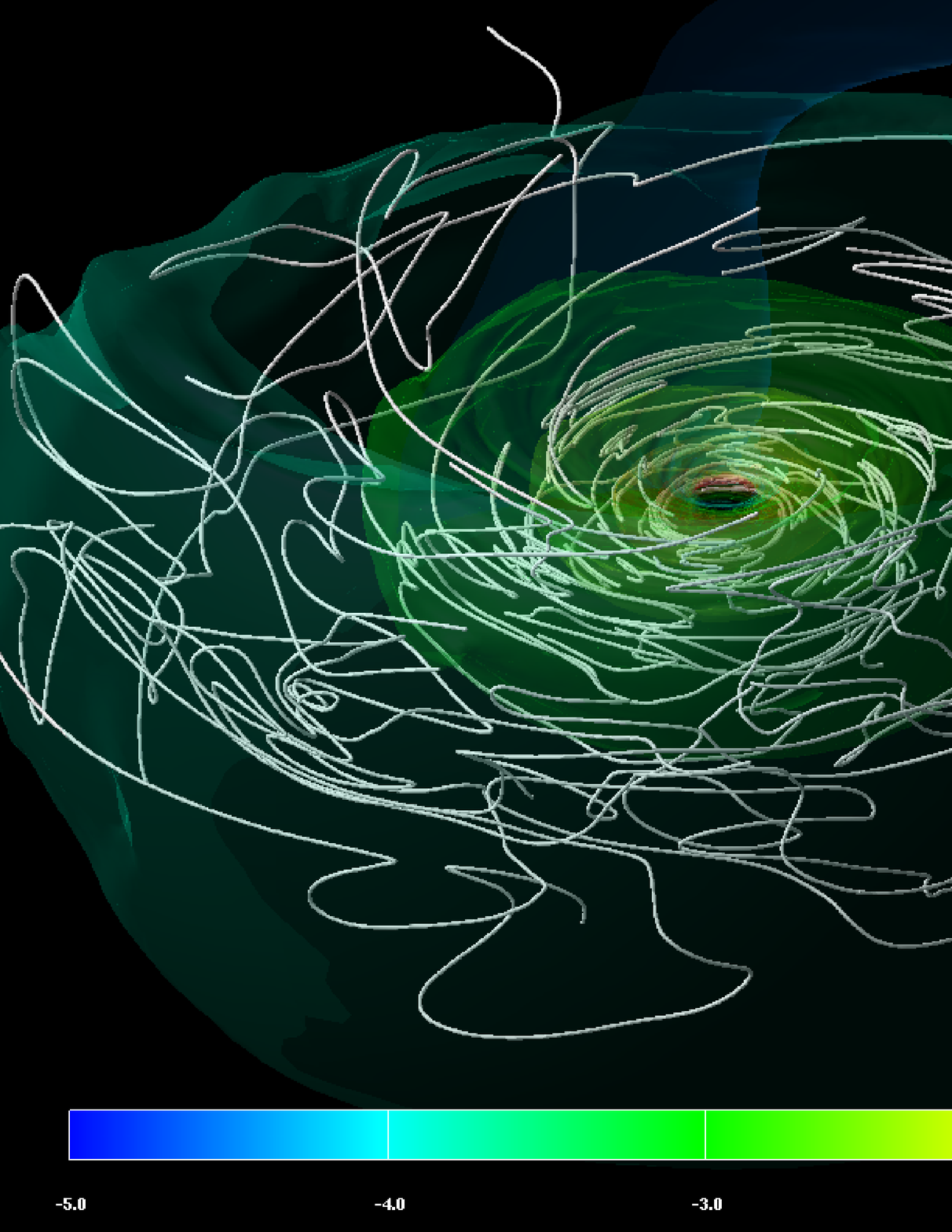}\\
\includegraphics[height=0.25\textheight,width=0.33\textwidth]{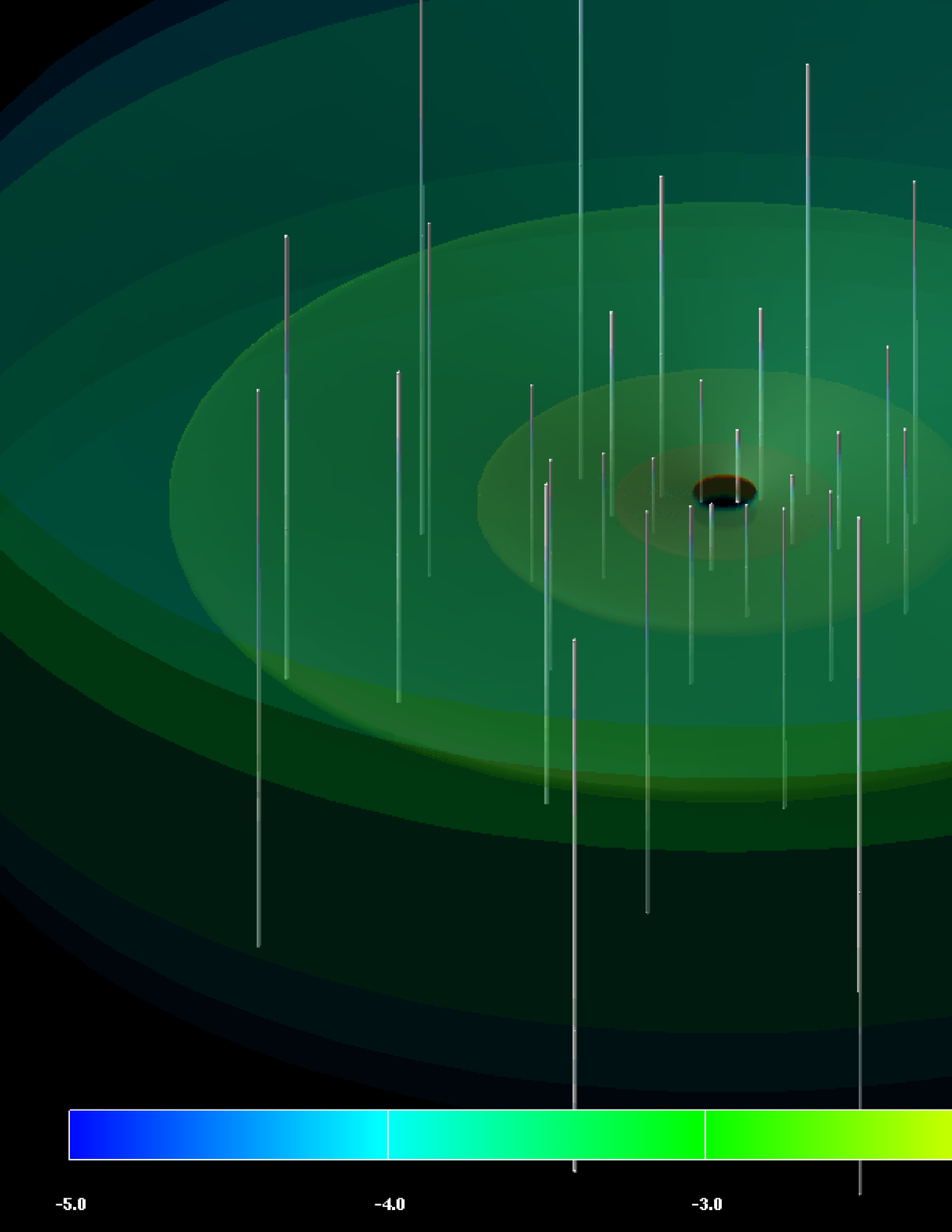}
\includegraphics[height=0.25\textheight,width=0.33\textwidth]{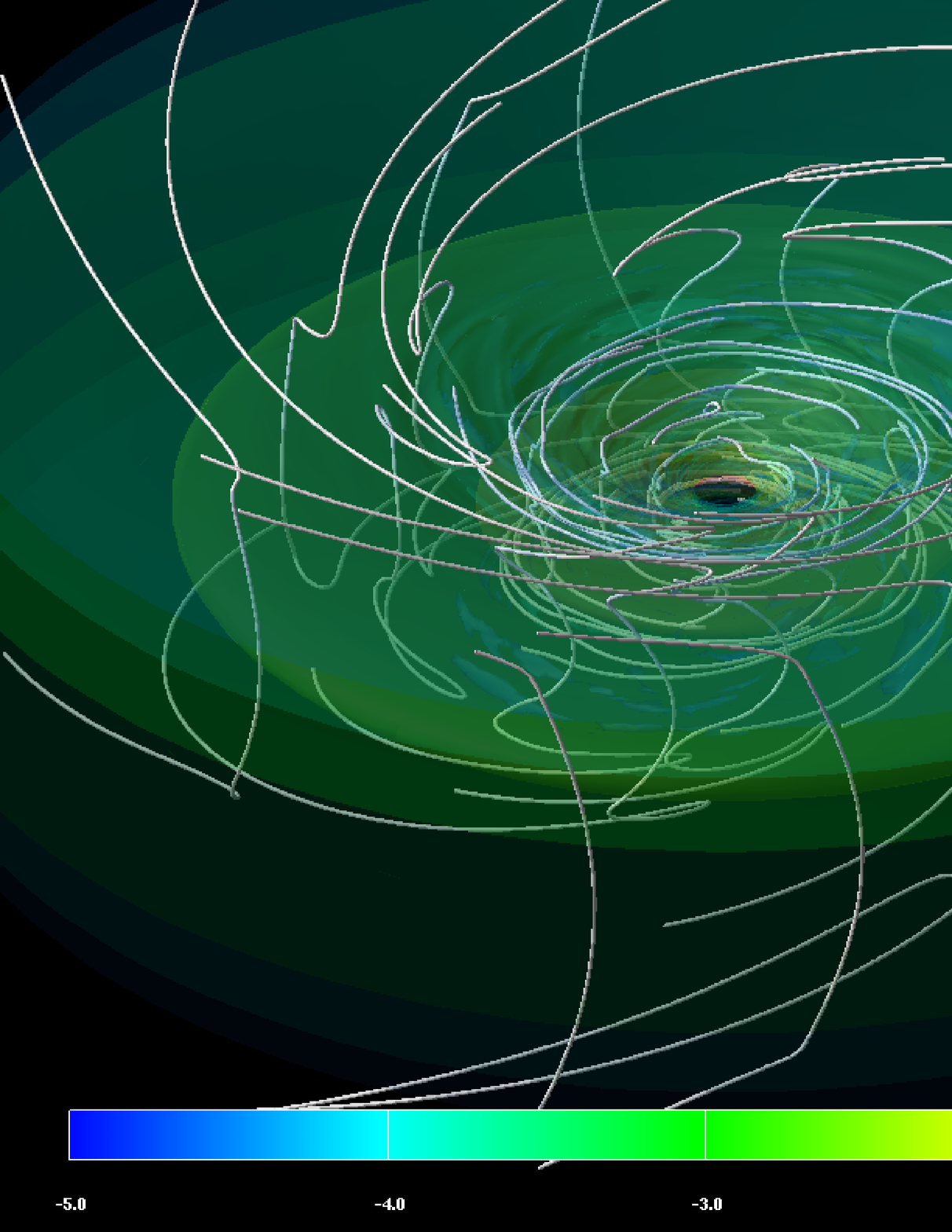}
\includegraphics[height=0.25\textheight,width=0.33\textwidth]{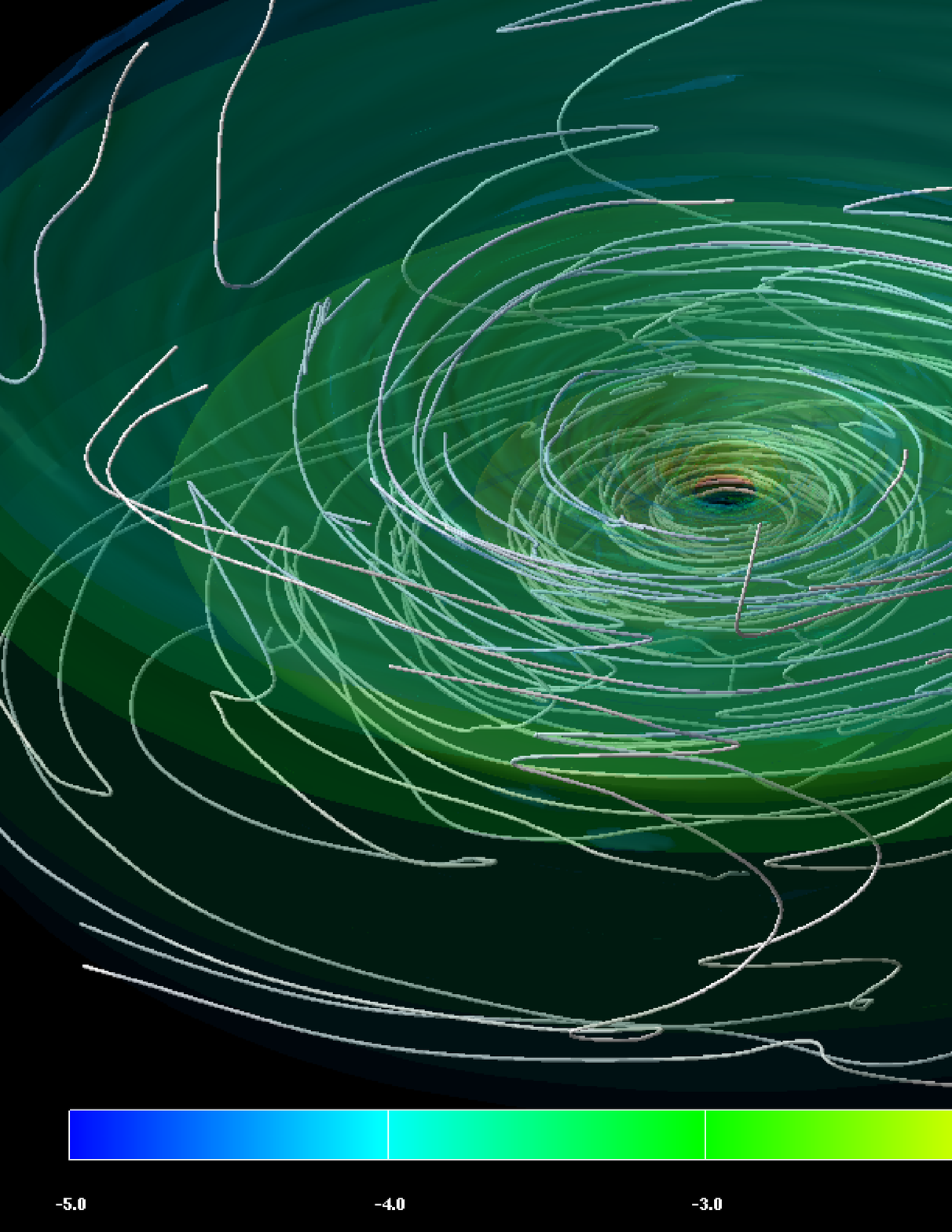}
\end{center}
\caption{Time evolution of the 4 cases. 
From top to bottom displayed are Cases I-high, II-high, I-low, and II-low. 
From left to right, the results at $t=0,50$, \& $500$ inner rotations. White 
lines illustrate magnetic fields and colors indicate iso-density surfaces.
Animations are available as {\it online materials} (also at 
www.ta.phys.nagoya-u.ac.jp/stakeru/research/glbdsk).}
\label{fig:3Dov}
\end{figure*}

Figure \ref{fig:3Dov} shows 3D views of time evolution of the 4 cases. 
The animations for these 4 cases up to $t_{\rm end}$ can be downloaded as 
{\it online materials}\footnote{Animation files are also available at www.ta.phys.nagoya-u.ac.jp/stakeru/research/glbdsk}.
Cases I-low and I-high exhibit typical evolutions of MRI, initiated 
by the development of channel-mode flows, which are clearly seen in the 
panels at $t=50$ inner rotations (the middle column). 
On the other hand, in Cases II-low and II-high the initial vertical 
magnetic field lines are strongly wound particularly in the surface regions 
by the vertical differential rotation. 
As a result, the configurations of the magnetic fields of Cases I \& II 
look different at later times (right column); Cases I-high and I-low show 
more turbulent magnetic fields, while Cases II-high and II-low appear to be
dominated by coherent magnetic fields wound by both radial and vertical 
differential rotation. 
From now on, we inspect the difference between the magnetic 
field structures as a result of the different temperature profiles
and extensively discuss their outcomes in the following sections.

\begin{figure}[h]
\begin{center}
\includegraphics[height=0.45\textheight,width=0.48\textwidth]{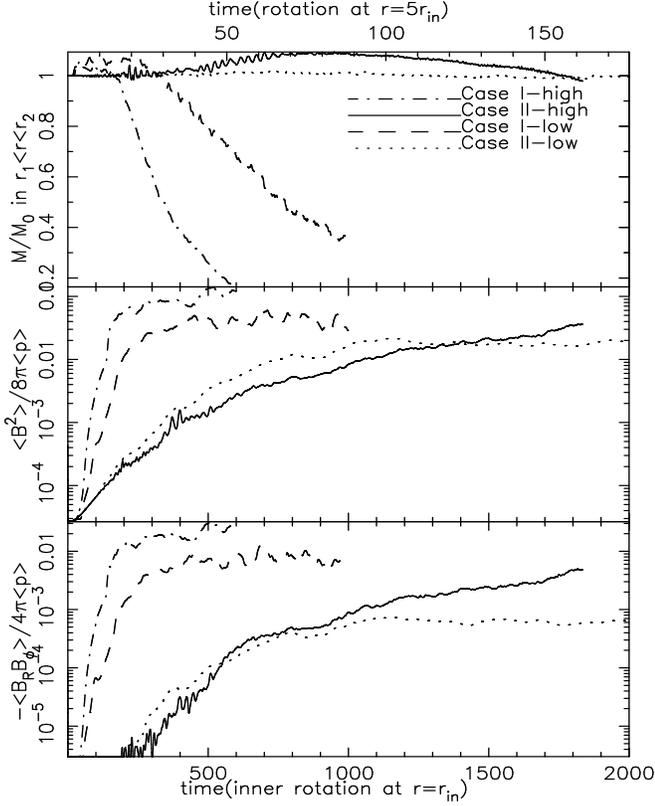}
\end{center}
\caption{Time evolution of characteristic quantities 
of Case I-high (dash-dotted), Case II-high (solid), 
Case I-low (dashed), and Case II-low (dotted). 
On the horizontal axis, time is measured in units of inner rotation time at 
$r=r_{\rm in}$ on the bottom and in units of rotation time at $r=5r_{\rm in}$ 
on the top.
{\it Top}: Mass in $r_1(=5r_{\rm in})<
r<r_2(=10r_{\rm in})$ normalized by the initial 
value. 
{\it Middle}: Volume-integrated magnetic energy in the midplane region, 
$\Delta z_{\rm mid}$, of $r_1<r<R_2$, normalized 
by the volume-integrated gas pressure in the same region, 
$\frac{\langle B^2\rangle_{R,\phi,z_{\rm mid}}(t)}{8\pi\langle p 
\rangle_{R,\phi,z_{\rm mid}}(t)}$, which essentially 
corresponds to the inverse of density weighted plasma $\beta$.
{\it Bottom}: Volume-integrated Maxwell stress normalized 
by volume-integrated gas pressure, corresponding to density weighted 
nondimensional Maxwell stress. The integration is done in the same 
region as in the middle panel. (see text for the detail)}
\label{fig:tevvint}
\end{figure}

Figure \ref{fig:tevvint} presents some representative quantities for 
the evolution of the disks. Here we focus on the physical quantities 
between $r_1=5r_{\rm in}$ and $r_2=10r_{\rm in}$; the region with $r<r_{1}$ 
is influenced by the inner boundary, particularly the decrease of 
the surface density by accretion and disk wind (\S \ref{sec:trb}); 
in the outer region, 
$r>r_{2}$, the magnetic fields are still in growth phases at the end of the 
simulations for Cases II-low and II-high because the dynamical time 
($\propto r^{-3/2}$) is long there. 
From top to bottom in Figure \ref{fig:tevvint}, we compare the 
following quantities of the four cases: the mass in the entire 
region of $r_1 < r < r_2$ in spherical coordinates, 
\begin{equation}
M(r_1<r<r_2) = \int_{\phi_{\rm min}}^{\phi_{\rm max}} d\phi 
\int_{\theta_{\rm min}}^{\theta_{\rm max}}\sin \theta d\theta \int_{r_1}^{r_2}
r^2 dr \rho,
\end{equation}
the inverse of the plasma $\beta$ value integrated in the midplane region 
of $\Delta z_{\rm mid}$ (Equation \ref{eq:dzmid}) in cylindrical 
coordinates, 
\begin{equation}
\frac{1}{\langle \beta \rangle_{R,\phi,z_{\rm mid}}(t)} 
\equiv \frac{\langle B^2\rangle_{R,\phi,z_{\rm mid}}(t)}{8\pi\langle p 
\rangle_{R,\phi,z_{\rm mid}}(t)} 
\label{eq:volbeta}
\end{equation}
and the Maxwell stress normalized by gas pressure, which is integrated 
in the same region,  $\Delta z_{\rm mid}$, 
\begin{equation}
-\frac{\langle B_R B_z\rangle_{R,\phi,z_{\rm mid}}(t)}
{4\pi \langle p \rangle_{R,\phi,z_{\rm mid}}(t)} 
\label{eq:volmaxw}
\end{equation}
Note that they essentially correspond to the inverse of 
the density-weighted plasma $\beta$ and the density-weighted nondimensional 
Maxwell stress. 

\begin{figure}[h]
\includegraphics[height=0.25\textheight,width=0.45\textwidth]{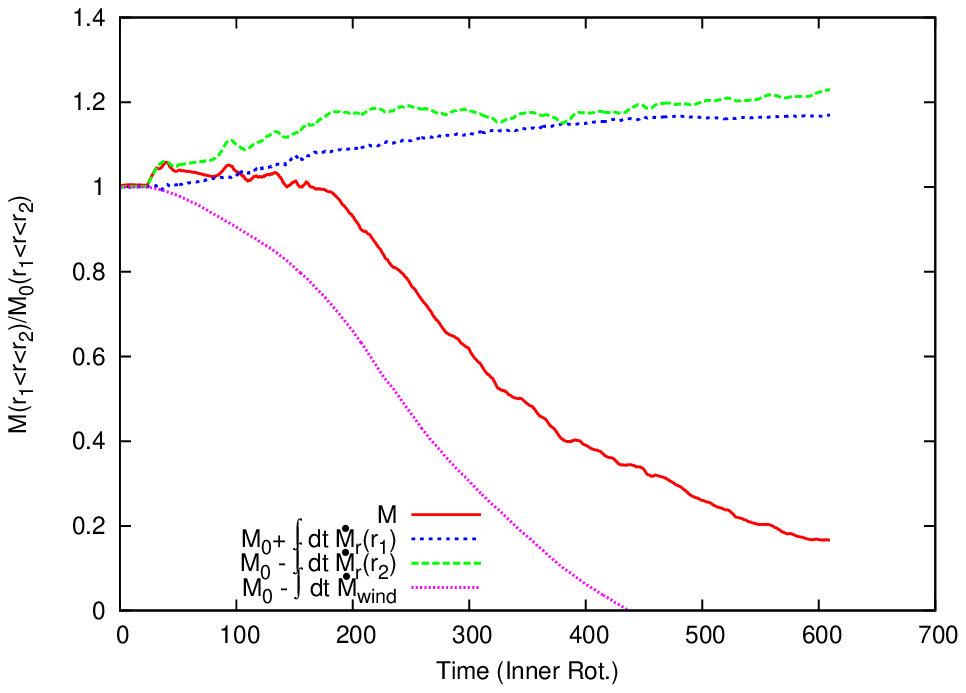}
\includegraphics[height=0.25\textheight,width=0.45\textwidth]{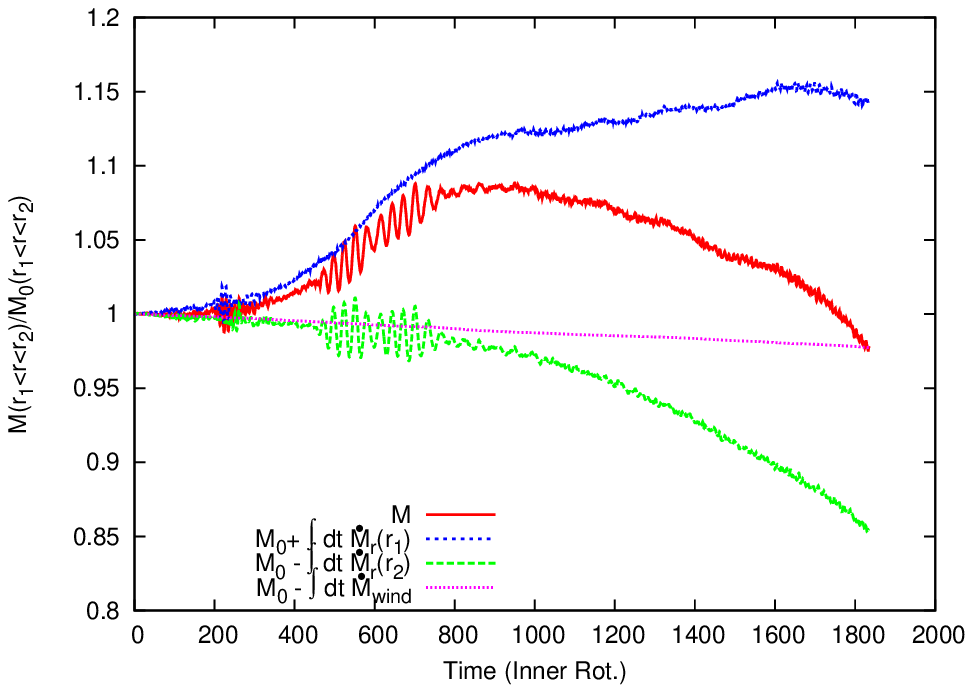}
\caption{Evolution of the mass in $r_1(=5r_{\rm in})<r<r_2(=10r_{\rm in})$ 
of Case I-high ({\it upper panel}) and Case II-high ({\it lower panel}). 
The mass normalized by the initial value, $M_0$, (red solid) is compared 
with the case that only takes into account the mass supply from the inner 
radius ($r_1$), 
$M_0 + \int dt \dot{M}_r(r_1)$ (blue short-dashed), the case with the mass 
supply from the outer radius ($r_2$), $M_0 - \int dt \dot{M}_r(r_2)$ 
(greed long-dashed), and the case with the mass loss by the disk winds, 
$M_0 - \int dt \dot{M}_{\rm wind}$ (magenta dotted). }
\label{fig:tevmass}
\end{figure}

The top panel of Figure \ref{fig:tevvint} shows that the mass of Cases I-high 
and I-low decreases quite rapidly, while the mass of Cases II-high and II-low 
is rather constant. 
The rapid decreases seen in Cases I-high and I-low are mainly due to 
the mass loss caused by the disk winds from the surfaces. In these cases, 
the simulation box covers a 
smaller vertical scale height ($\approx \pm 1.8 H$ at $r_1=5 r_{\rm in}$ and 
$\approx \pm 1.3 H$ at $r_2=10 r_{\rm in}$) than the simulation box 
of Cases II-high and II-low. 
As discussed in \citet{suz10} and \citet{fro13} by using MHD simulations in 
local shearing boxes \citep{hgb95}, the mass flux of the disk winds driven 
by MRI-triggered turbulence depends on the vertical box size; a smaller 
vertical box gives larger mass flux.  In the present global simulations, 
a larger amount of the gas streams out of the $\theta$ surfaces of the 
simulation box of Cases I-high and I-low because of the insufficient 
vertical box size. 

\begin{figure*}[h]
\begin{center}
\includegraphics[height=0.3\textheight,width=0.41\textwidth]{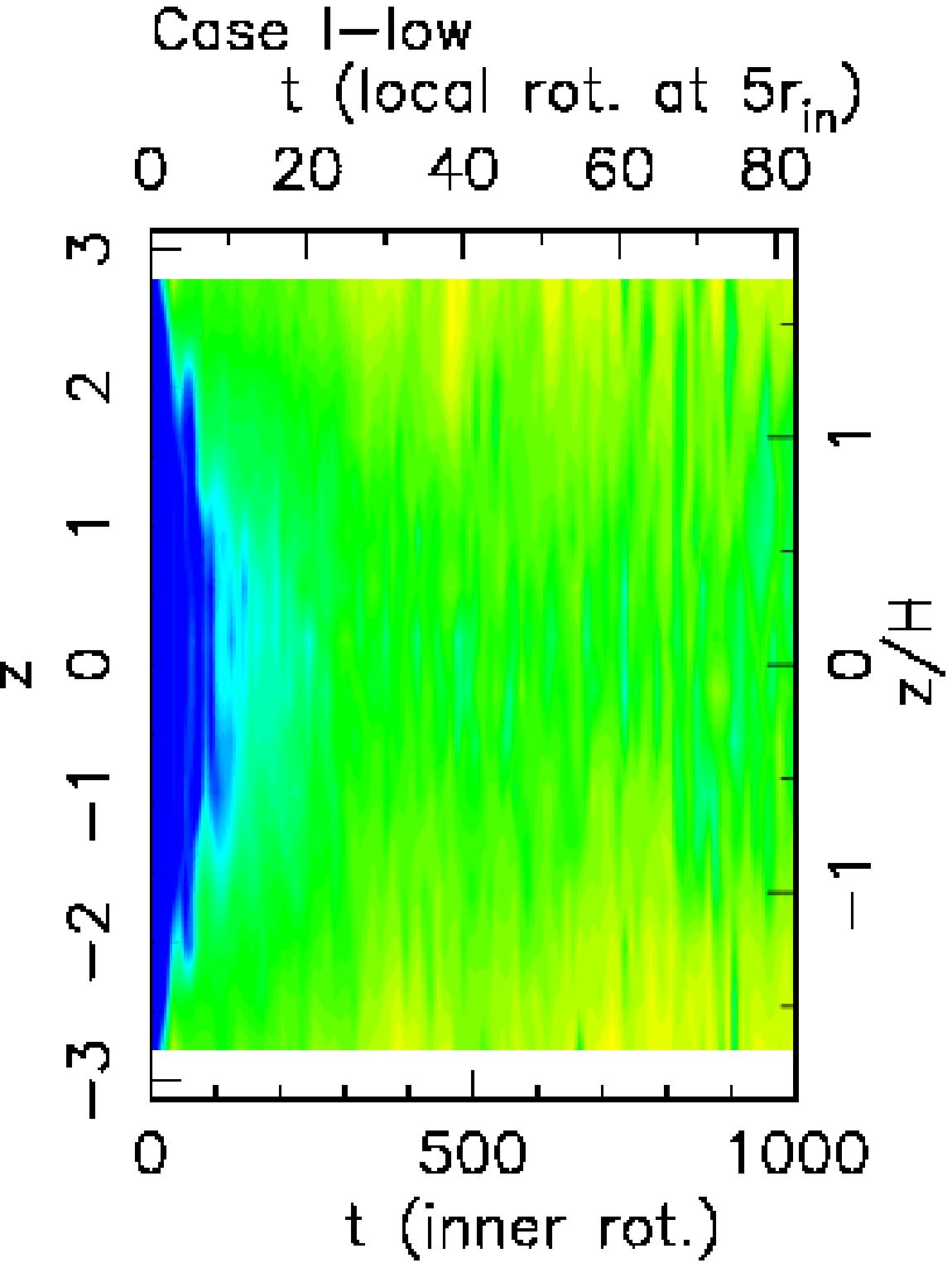}
\includegraphics[height=0.3\textheight,width=0.34\textwidth]{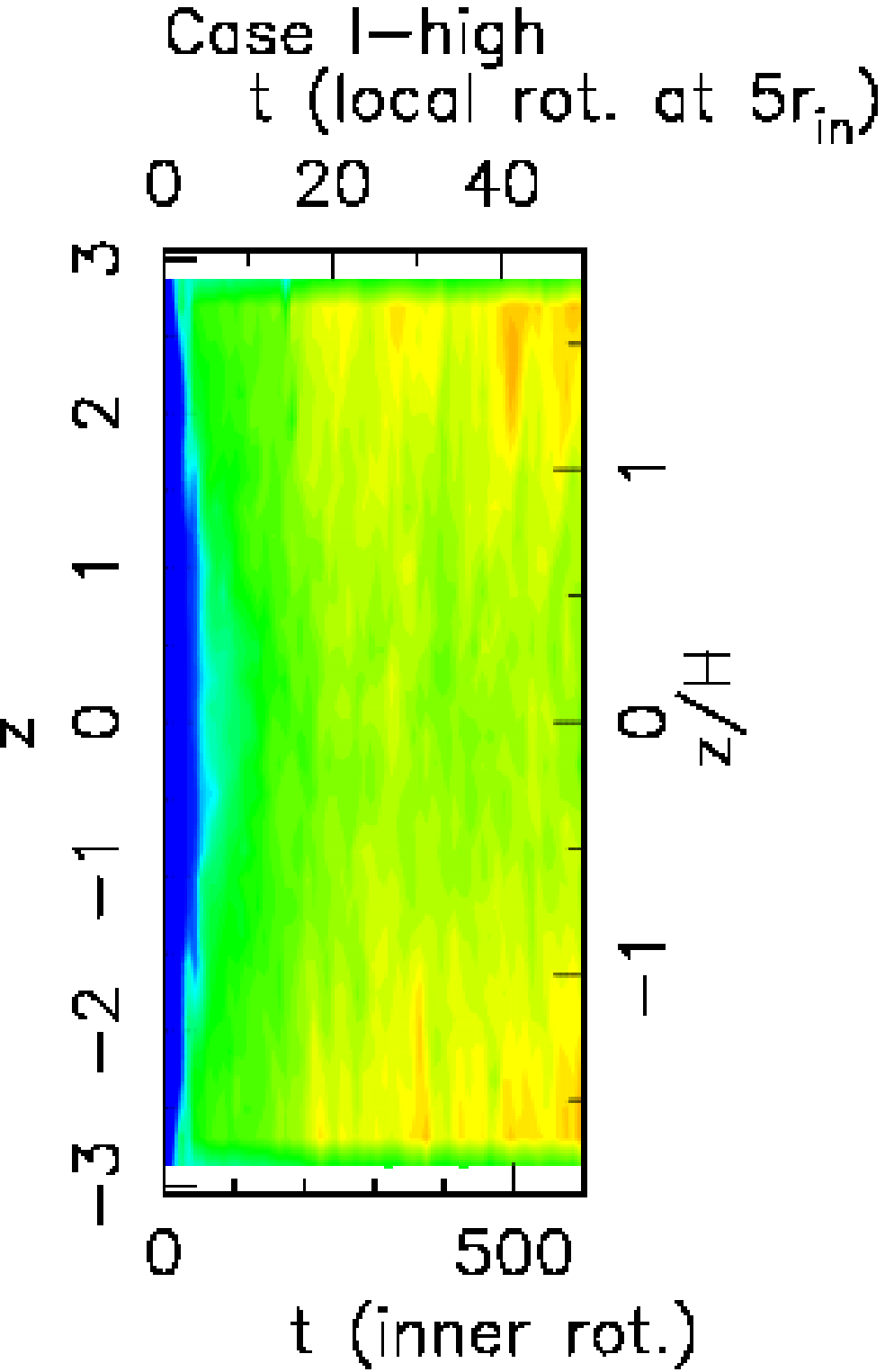}\\
\includegraphics[height=0.35\textheight,width=0.69\textwidth]{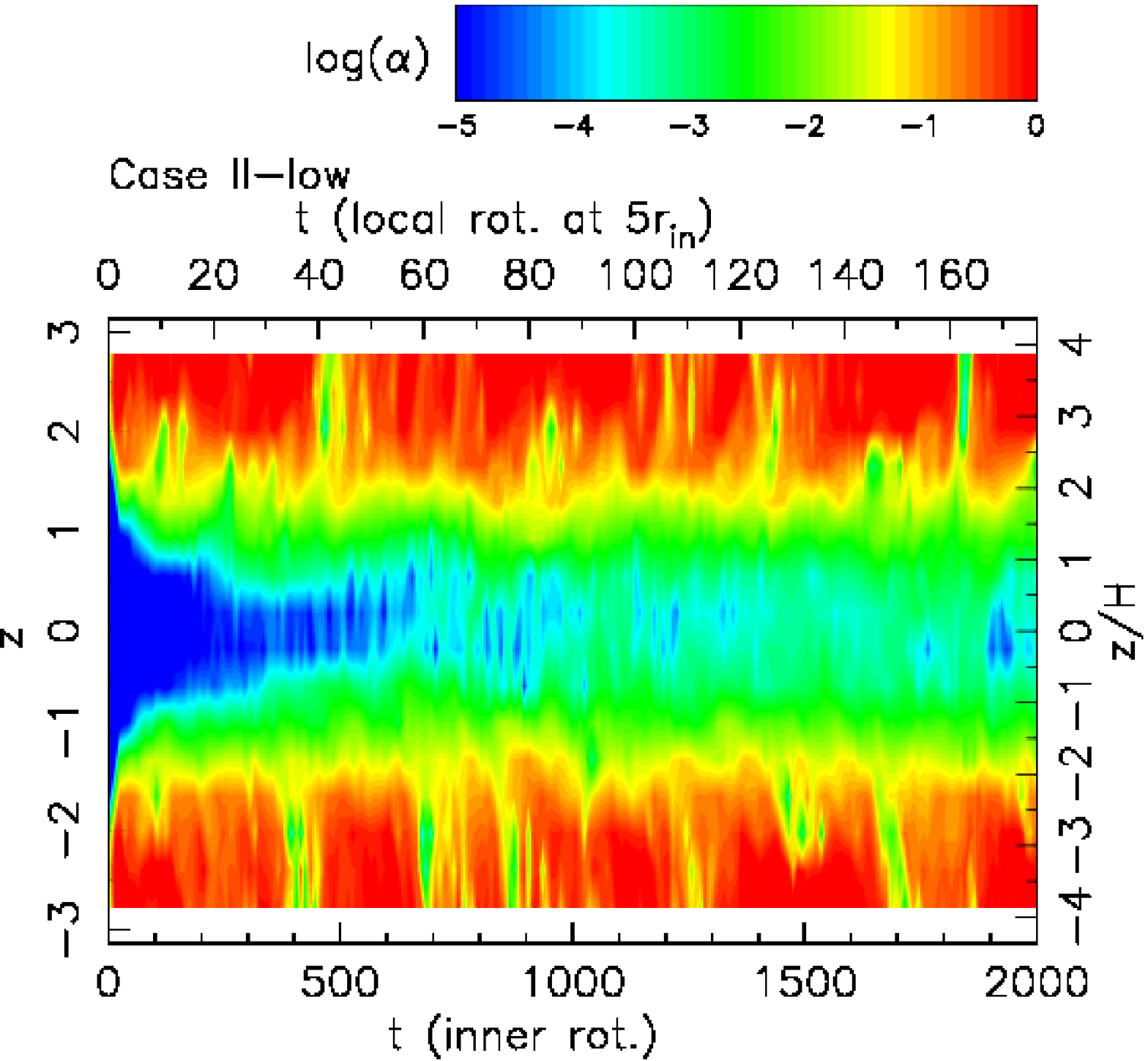}
\includegraphics[height=0.3\textheight,width=0.64\textwidth]{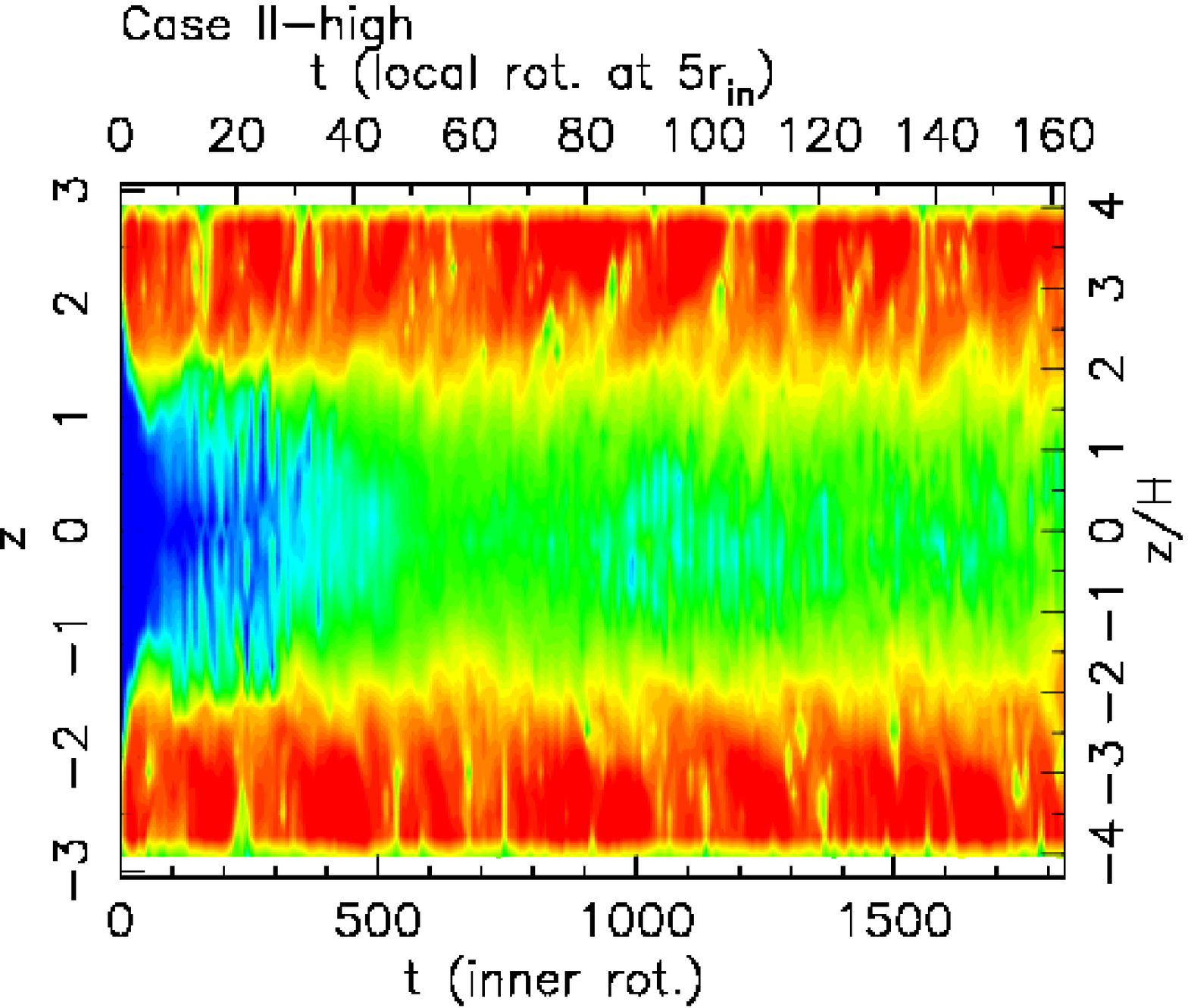}
\end{center}
\caption{$t$--$z$ diagrams of $\langle\alpha\rangle_{\phi}(r_1,z)$ 
at $R=r_1(=5r_{\rm in})$ for Case I-low ({\it top-left}), Case I-high 
({\it top-right}), Case II-low ({\it middle}), and Case II-high ({\it bottom}). 
On the horizontal axis are shown time in the inner rotation ({\it top} of 
panels) and time in the local rotation at $r=r_1$ ({\it bottom} of panels). 
On the vertical axis are shown $z$ ({\it left}) and $z/H$ 
({\it right}). }
\label{fig:tzalpha}
\end{figure*}


In Figure \ref{fig:tevmass} we examine the actual supply/loss of the mass 
to/from the region of $r_1<r<r_2$ for the high-resolution runs. 
The mechanisms are separated into two types, disk winds and radial flows. 
The mass loss caused by the disk winds can be measured via 
$$
\hspace{-3cm}
\dot{M}_{\rm wind}(r_1<r<r_2) = \int_{\phi_{\rm min}}^{\phi_{\rm max}} d\phi
$$
\begin{equation}
\int_{r_1}^{r_2} rdr
[-(\rho v_{\theta}\sin \theta)_{\theta_{\rm min}} 
+(\rho v_{\theta} \sin \theta)_{\theta_{\rm max}}],
\end{equation}
where we take into account the disk winds from both the upper surface at
$\theta=\theta_{\rm min} = \pi/2-0.5$ and the lower surface at 
$\theta=\theta_{\rm max} = \pi/2+0.5$. Note that $\dot{M}_{\rm wind}$ is 
defined in such a way that $\dot{M}_{\rm wind}<0$ when 
the mass is lost by the disk winds ($v_{\theta}(\theta_{\rm min})<0$ and 
$v_{\theta}(\theta_{\rm max})>0$). 
As for radial flows, we measure the mass flux across a $\theta-\phi$ surface at 
$r$($=r_1$ or $r_2$): 
\begin{equation}
\dot{M}_r(r) = \int_{\phi_{\rm min}}^{\phi_{\rm max}} d\phi
\int_{\theta_{\rm min}}^{\theta_{\rm max}} d\theta r^2 \sin \theta \rho v_r 
\label{eq:dotM_sp}
\end{equation}
$\dot{M}_r<0$ for accretion ($v_r<0$). $\dot{M}_r(r_1)>0$ and 
$\dot{M}_r(r_2)<0$ contribute to an increase in the mass in $r_1<r<r_2$, 
and vice versa. 

Figure \ref{fig:tevmass} shows that the disk winds continuously remove 
the mass ($\dot{M}_{\rm wind}<0$) in both cases. $\dot{M}_{\rm wind}$ of 
Case I-high is considerably larger than $\dot{M}_{\rm wind}$ of Case II-high, 
mainly because the vertical extent of the simulation box per 
scale height is smaller in Case I as discussed previously. 
As a result, after 600 inner rotations, more than 80 \% of the initial mass 
in $r_1<r<r_2$ is lost in Case I-high. On the other hand, the effect of 
the mass loss caused by the disk winds is not as large in Case II-high.

The radial flows show more complicated behaviors. In Case I-high, the mass is 
mostly supplied from the outer radius ($\dot{M}_r(r_2)<0$; green solid line) by 
accretion ($v_r<0$). From the inner radius ($r=r_1$; blue short-dashed line), 
the mass is initially supplied to the simulation box by outward flows 
($v_r>0$; $\dot{M}_r(r_1)>0$) but eventually the mass is lost by accretion 
($v_r<0$; $\dot{M}_r(r_2)<0$). 

On the other hand in Case II-high, the direction of the mass flow is initially 
outward ($v_r>0$) at both the inner and outer surfaces, namely the mass 
supply from $r=r_1$ (blue short-dashed line) and the mass loss from 
$r=r_2$ (green solid line). The radial flow at $r=r_2$ (green 
line) shows an oscillatory feature arising from epicycle motion during 
400 -- 800 inner rotations.
At later times ($t>1600$ inner rotations), the mass starts to accrete at 
$r=r_1$, while at $r=r_2$ the direction of the mass flow is kept outward. 
This tendency seems to follow the time evolution of a standard accretion 
disk \citep{lp74}; the mass diffuses inward in the inner region 
and outward in the outer region from the diffusion center that gradually moves 
outward with time.  
As a result, at early time the mass supply from the inner radius, $r=r_1$, 
dominates the other components and the net mass increases up to 
$t\approx 800$ inner 
rotations, but later decreases by the disk winds and the radial outflow 
from the outer radius, $r=r_2$. 
We discuss radial mass flows in more detail in \S \ref{sec:rf}.

Turning back to Figure \ref{fig:tevvint}, the middle and bottom panels 
show the time evolution of properties of the magnetic fields, 
Equations (\ref{eq:volbeta}) \& (\ref{eq:volmaxw}). 
One of the characteristic features of the present simulations with net vertical 
fields is that the magnetic energy and Maxwell stress monotonically 
increase and seem to saturate but never systematically decrease 
because the strength of the net vertical field is kept more or less globally 
constant (see the discussion on the net $\beta_z$ in \S \ref{sec:magf}). 
This is in contrast to global simulations without a net vertical magnetic 
flux, which exhibit a decrease of the magnetic energy caused by escaping net 
toroidal fields with vertical outflows after the initial amplification 
\citep[e.g.,][]{flo11,pb13b}. 

In Cases I-low and I-high the magnetic fields are amplified more rapidly 
than in Cases II-low and II-high and are saturated after $t\gtrsim 200-300$ 
inner rotations or 20-30 local rotations at $r=5r_{\rm in}$.   
In all the cases our simulations cannot initially resolve the 
wavelengths, 
\begin{equation}
\lambda_{\rm max} \approx 2\pi 
\frac{B/\sqrt{4\pi\rho}}{\Omega}, 
\label{eq:mrimax}
\end{equation}
of the most unstable mode of MRI with respect 
to the initial magnetic field strength \citep{bh91}. 
However, if Cases I and II are compared, one scale height can be resolved 
by a larger number of grid points in the outer regions of Cases I-high and 
I-low because of the dependence of the scale height, 
$H/R\propto R^{1/2}$ (Equation \ref{eq:cIhr}) as shown in Table \ref{tab:res}. 
Thus, MRI in smaller scales, which correspond to faster growing 
modes, can be captured from the beginning, which leads to faster 
amplification of the magnetic fields in Cases I-high and I-low.

On the other hand, the magnetic fields in Cases II-high and II-low grow 
quite slowly, and are finally saturated after $t\gtrsim 1200$ inner 
rotations or $\gtrsim 110$ local rotations at $r=5r_{\rm in}$, because 
the simulations cannot initially resolve small-scale modes of MRI. 
With the increase of the magnetic field, 
$\lambda_{\rm max}(\propto B)$ of the MRI increases, and can be marginally 
resolved at the midplane in Case II-high at later times, while it is 
underresolved for the $R$ and $z$ components at the midplane of Case II-low 
(see \S \ref{sec:vst}).

Comparing the high- and low-resolution runs, Cases I and II show 
a different trend. In Case I the high-resolution run shows the faster growth 
of the magnetic field and, consequently, a higher saturation level, which 
is expected from the amplification of the magnetic field by MRI; the 
higher-resolution run can resolve smaller scales with faster growth. 
On the other hand, in Case II the low-resolution run gives faster growth 
of the magnetic field, while the high-resolution run gives the higher 
saturation level. 
This indicates that a process other than from MRI operates in the amplification 
of the magnetic fields in Cases II-high and II-low. We suppose that the 
vertical differential rotation plays a key role, which will be discussed 
in \S \ref{sec:magf}.

In Figure \ref{fig:tzalpha}, we present the $\alpha$ values, 
\begin{equation}
\langle \alpha\rangle_{\phi}(t,r_1,z) = \frac{\langle \rho v_R \delta v_{\phi,0} 
- B_R B_{\phi}/4\pi \rangle_{\phi}(t,r_1,z)}{\langle p \rangle_{\phi}(t,r_1,z)},
\end{equation}
defined as the sum of Maxwell ($- B_R B_{\phi}/4\pi$) and Reynolds 
($\rho v_R \delta v_{\phi,0}$) stresses, at $R = r_1 (=5r_{\rm in})$ in time 
(horizontal axis) -- $z$ (vertical axis) diagrams, 
where 
\begin{equation}
\delta v_{\phi,0} = v_{\phi} - v_{\phi,0}
\label{eq:vpshft0}
\end{equation}
is the difference of the rotation speed from the initial equilibrium rotation 
speed, $v_{\phi,0}$. Note that using $\delta v_{\phi,0}$ to estimate 
the Reynolds stress might not be the best way because at later times 
the background rotation profile is modified by the change of the pressure 
gradient force due to the evolution of radial mass distribution. 
In the next section, we use the velocity shifts from the time-averaged 
$v_{\phi}$ rather than the initial $v_{\phi,0}$. However, the choice of 
the background $v_{\phi}$ does not affect the overall evolution of $\alpha$ 
since $\alpha$ is dominated by the Maxwell stress, and thus we simply use 
Equation (\ref{eq:vpshft0}) for Figure \ref{fig:tzalpha}. 

In all the cases, $\alpha$ starts to grow in the regions near the surfaces, 
$|z|\gtrsim 1.5$, which is similar to what is observed in our simulations 
with local shearing boxes \citep[e.g.][]{si09}. 
This is because the simulations can 
resolve $\lambda_{\rm max}$ there because of the smaller density and 
correspondingly larger $\lambda_{\rm max}$ than at the midplane. 
Around the midplane, $\alpha$ eventually increases and becomes saturated 
at later times, as discussed in Figure \ref{fig:tevvint}.

\section{Slice Images}
\label{sec:snp}
We investigate properties of the simulated accretion disks after the magnetic 
fields are amplified and saturated. 
In this section, we show typical slice images of the simulated 
accretion disks. 

\subsection{Face-on Views}
\label{sec:fov}
\begin{figure*}[h]
\begin{center}
\includegraphics[height=0.42\textheight]{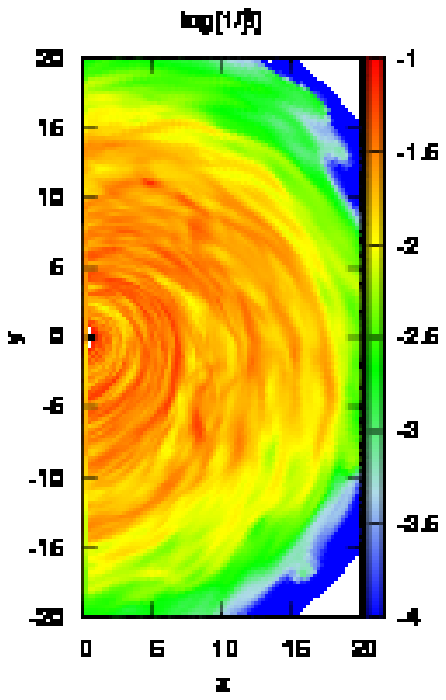}
\includegraphics[height=0.42\textheight]{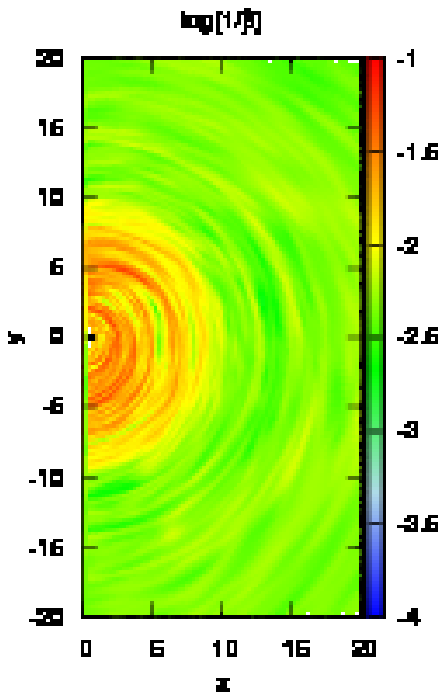}\\
\includegraphics[height=0.3\textheight]{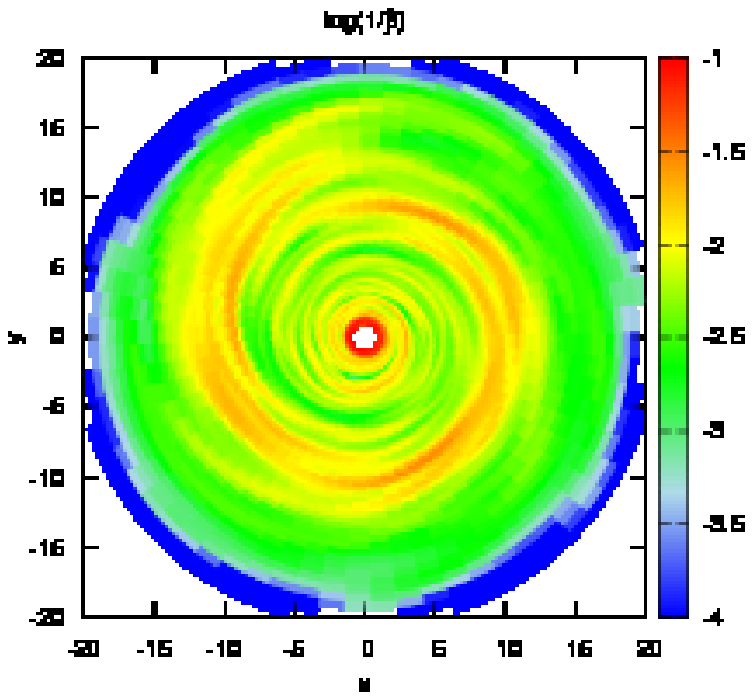}
\includegraphics[height=0.3\textheight]{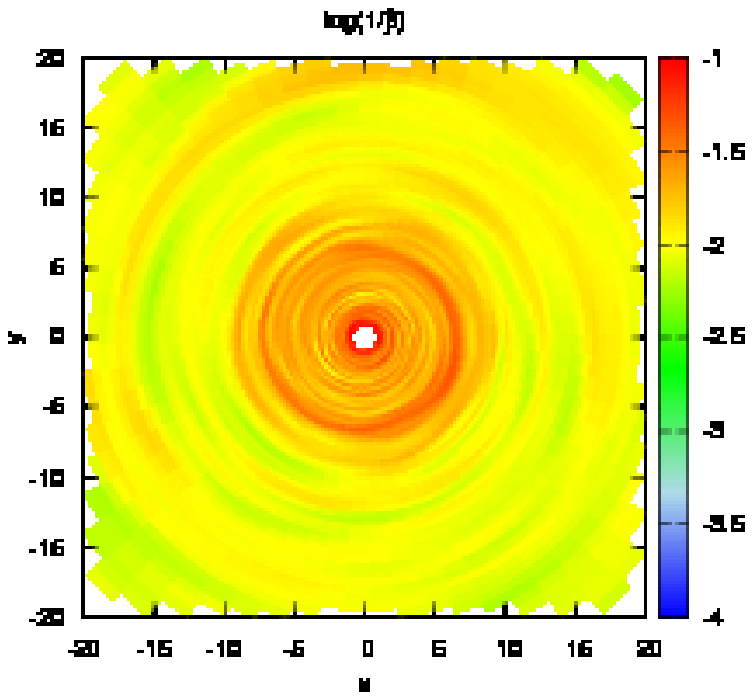}
\end{center}
\caption{Face-on views of the simulated accretion disks of Case I-high at 
$t=250$ inner rotations ({\it upper-left}), Case II-high at $t=1250$ inner 
rotations ({\it upper-right}), Case I-low at $t=250$ inner rotations 
({\it lower-left}), and Case II-low at $t=1250$ inner rotations 
({\it lower-right}). Colors show 
$1/\langle \beta \rangle_{z_{\rm tot}}$ in the logarithmic scale. 
Movies are available as {\it online materials} (also at 
www.ta.phys.nagoya-u.ac.jp/stakeru/research/glbdsk).}
\label{fig:faceon}
\end{figure*}

Figure \ref{fig:faceon} illustrates snapshots of face-on views 
of the disks when the MHD turbulence is almost in the saturated state. 
For Cases I-low and I-high the snapshots at $t=250$ inner rotations are 
shown, and for Cases II-low and II-high we show the snapshots at 
$t=1250$ inner rotations. 
The colors indicate the inverse of the plasma $\beta$ integrated with $\Delta 
z_{\rm tot}$ (Equation \ref{eq:dztot}) in the entire $z$ extent, 
\begin{equation}
\frac{1}{\langle \beta \rangle_{z_{\rm tot}}(t,R,\phi)} 
= \frac{\langle B^2\rangle_{z_{\rm tot}}(t,R,\phi)}{8\pi\langle p 
\rangle_{z_{\rm mid}}(t,R,\phi)}, 
\label{eq:sfbeta}
\end{equation}
and brighter colors correspond to regions with relatively larger magnetic 
pressure. 
Winding structures dominate in Cases II-high and II-low, while both winding 
and turbulent structures are distributed in Cases I-high and I-low. 
Although these winding structures are not so long-lived with 
typical lifetimes of the order of rotation time, they are ubiquitously 
created somewhere in the disks (see Movies for Figure \ref{fig:faceon} 
available as {\it online materials} and at 
www.ta.phys.nagoya-u.ac.jp/stakeru/research/glbdsk).

\subsection{Edge-on Views}
\begin{figure*}[t]
\begin{center}
\includegraphics[height=0.3\textheight]{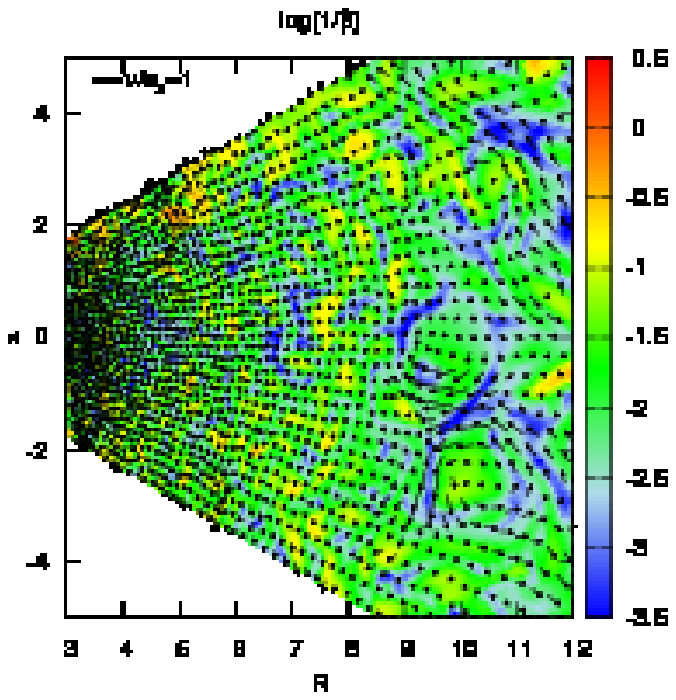}
\includegraphics[height=0.3\textheight]{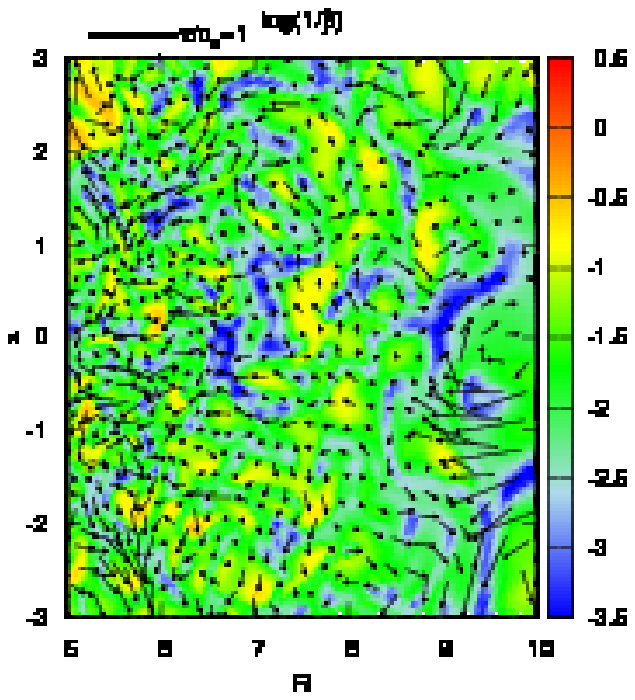}\\
\vspace{-1cm}
\includegraphics[height=0.3\textheight]{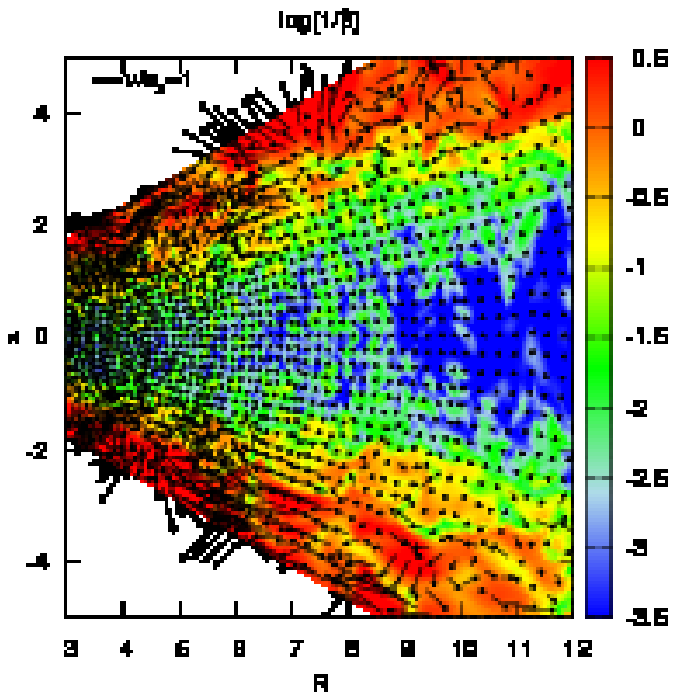}
\includegraphics[height=0.3\textheight]{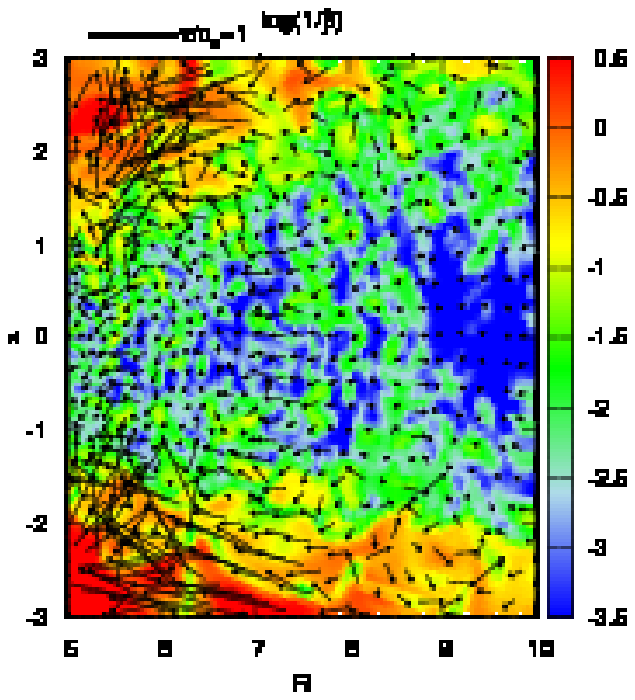}
\end{center}
\caption{Edge-on views of the simulated accretion disks of Case I-high at 
$t=250$ inner rotations ({\it upper}) and Case II-high at $t=1250$ inner 
rotations ({\it lower}) at $\phi = \pi/4$. The right panels are zoomed-in 
views of the left panels to inspect the region near the midplane. 
Colors show $1/\beta$ in the logarithmic scale. 
The arrows indicate velocities normalized by the local sound speed, 
of which the scale is shown at the upper left corner of each panel. 
Time-evolution movies are available as 
{\it online materials} (also at 
www.ta.phys.nagoya-u.ac.jp/stakeru/research/glbdsk).}
\label{fig:edgeon}
\end{figure*}

Figure \ref{fig:edgeon} presents edge-on views of the simulated disks 
in the saturated state. 
Here we present the results of only the high-resolution runs but 
zoomed-in views around the midplane (right panels) are displayed 
together with the views including the disk surfaces (left panels). 
Velocities normalized by the sound speed are shown by arrows, with 
the inverse of the plasma $\beta$ in color. 
The velocity fields show a quite complicated structure with both radially
inward ($v_{R}<0$) and outward ($v_{R}>0$) motions as well as vertical flows. 
Case II-high, which covers the larger vertical extent in scale height,  
captures detailed properties of the disk winds well. 
The zoomed-out panel for Case II-high (bottom left) shows that the 
velocities of the structured disk winds are about twice as fast as 
the local sound speed in some regions near the surfaces.  
The animation for the time-evolution of this figure 
\footnote{Time-evolution animations of Figure \ref{fig:edgeon} 
are available at www.ta.phys.nagoya-u.ac.jp/stakeru/research/glbdsk} 
further shows these vertical outflows are intermittent with time.

\section{Properties of Turbulent Disks}
\label{sec:trb}

\begin{figure}[h]
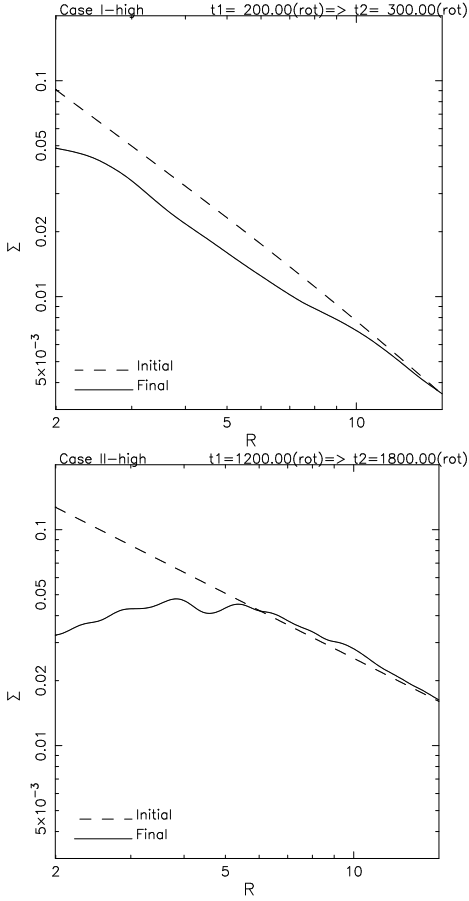

\begin{center}
\includegraphics[height=0.25\textheight,width=0.34\textwidth]{oneDdiag_tav_fig10_9_4.ps}
\includegraphics[height=0.25\textheight,width=0.34\textwidth]{oneDdiag_tav_fig11_9_6.ps}
\end{center}
\caption{Surface density (solid lines) of Case I-high averaged during 
$\Delta t_{\rm ave}=200-300$ inner rotations ({\it upper}) and of Case II-high 
averaged during $\Delta t=1200-1800$ inner rotations ({\it lower}) in 
comparison with the initial value (dashed line).}
\label{fig:sfdns_tav}
\end{figure}

For more quantitative studies of the magnetic fields and the turbulence in 
the disks, we examine several time-averaged quantities. 
In order to study the saturated state, 
we consider the time average during $\Delta t_{\rm ave}$
summarized in Table \ref{tab:models}.
Although we perform the simulations of Cases I-high and I-low to 
$t_{\rm end}=600$ and 1000 inner rotations respectively, 
a significant fraction of the mass is lost by the disk winds 
at later times (top panel of Figure \ref{fig:tevvint}), and 
the role of the magnetic field becomes relatively important (plasma $\beta$ 
decreases), because the magnetic field is not as dissipated, as will 
be discussed in \S \ref{sec:rf}. Thus, we take the 
averages well before $t_{\rm end}$. 
As for Cases II-high and II-low, the mass is kept almost constant during the 
simulations as shown in Figure \ref{fig:tevvint}. Although the middle 
and bottom panels of Figure \ref{fig:tevvint} show that in Case II-high 
the magnetic fields are still gradually growing during $\Delta t_{\rm ave}$, we 
take averages over $\Delta t_{\rm ave}=1200-1800$ inner rotations because 
of the limitation of the computational time.

Figure \ref{fig:sfdns_tav} displays the surface density of the high-resolution 
runs averaged over $\Delta t_{\rm ave}$ and $\phi_{\rm min}-\phi_{\rm max}$:
\begin{equation}
\Sigma(R) = \frac{\int_{\Delta t_{\rm ave}}dt\int_{\phi_{\rm min}}^{\phi_{\rm max}}d\phi 
\int_{z_{\rm bot}}^{z_{\rm top}}dz \rho}{\Delta t_{\rm ave} 
(\phi_{\rm max}-\phi_{\rm min})}
\end{equation} 
Note that the initial profiles, $\Sigma_0 \propto \rho_{\rm mid,0} H$, are
\begin{equation}
\Sigma_0 = \left\{
\begin{array}{l}
R^{-\frac{3}{2}} : {\rm Case \; I}\\
R^{-1} : {\rm Case \; II}
\end{array}
\right.
\end{equation} 
In both cases, the mass in the 
inner region $r\lesssim 3 r_{\rm in}$ is considerably lost by the accretion 
and the disk winds. On the other hand, the region in $r\gtrsim r_1 
(=5r_{\rm in})$ is not so severely affected. 

We now examine $z$ and $R$ dependences of various quantities. 
For the $z$ dependence, we consider the average of Equation 
(\ref{eq:qtavtp})
at $R=r_1 (=5 r_{\rm in})$. 
For the radial dependence, we take the average of Equation (\ref{eq:qtavtpz}),
whereas for the integration of $\Delta z$, we consider (i) the midplane 
region, $\Delta z_{\rm mid}$ (Equation \ref{eq:dzmid}), 
and (ii) the entire region, $\Delta z_{\rm tot}$ (Equation \ref{eq:dztot}). 
In Case II $\Delta z_{\rm tot}$ corresponds to $-4H \Rightarrow +4H$, while
in Case I, $\Delta z_{\rm tot}$ measured in $H$ varies with $R$ since 
$H/R\propto R^{1/2}$ (Equation \ref{eq:cIhr}). 

\subsection{Magnetic Fields}
\label{sec:magf}
In this subsection, we inspect various properties of the magnetic fields 
in the saturated state.

\subsubsection{Vertical Structure at $5r_{\rm in}$}
\label{sec:vst}

\begin{figure*}
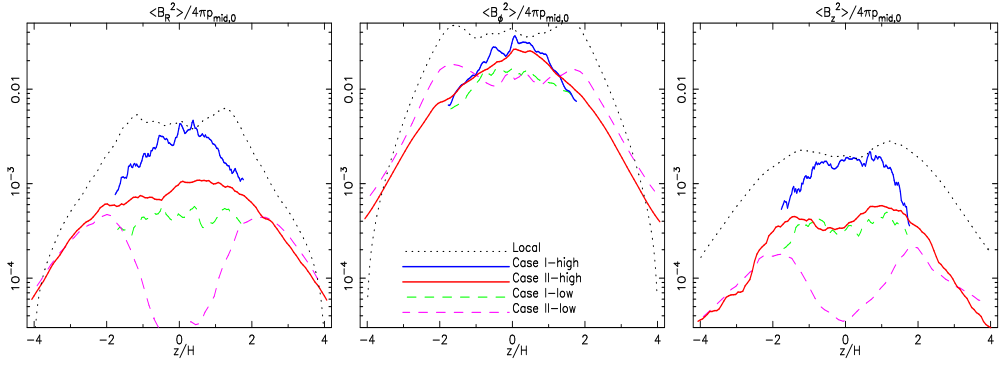

  \begin{center}
  \includegraphics[height=0.2\textheight,width=0.24\textwidth]{rdp-zdp_tav_cmp61_3.ps}
  \includegraphics[height=0.2\textheight,width=0.24\textwidth]{rdp-zdp_tav_cmp63_3.ps}
  \includegraphics[height=0.2\textheight,width=0.24\textwidth]{rdp-zdp_tav_cmp62_3.ps}
  \end{center}
\caption{Vertical structures of $\langle B_i^2\rangle_{t,\phi}(r_1,z)$ at 
$R=r_1(=5r_{\rm in})$ of Case I-high ({\it blue solid}), Case II-high ({\it 
red solid}), Case I-low ({\it green dashed}), and Case II-low ({\it 
magenta dashed}), in comparison with the result of the shearing box 
simulation ({\it black dotted}). From left to right, 
$i=R$, $\phi$, and $z$ components are displayed.}
\label{fig:zstBsq}
\end{figure*}

\begin{figure*}
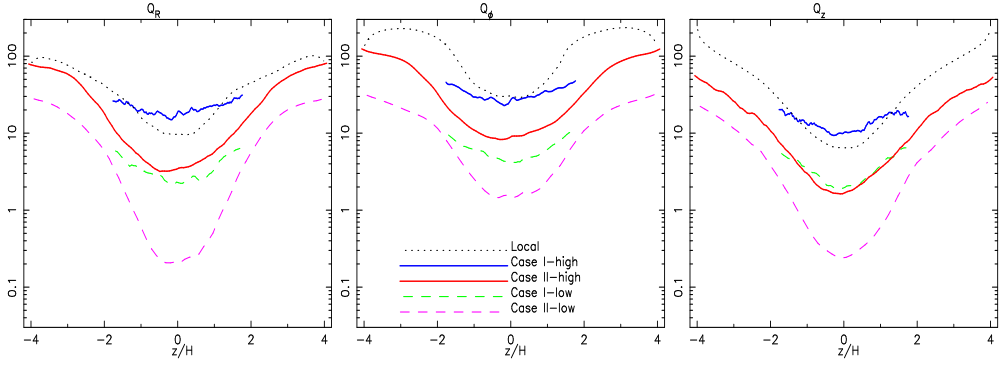

  \begin{center}
  \includegraphics[height=0.2\textheight,width=0.24\textwidth]{rdp-zdp_tav_cmp70_3.ps}
  \includegraphics[height=0.2\textheight,width=0.24\textwidth]{rdp-zdp_tav_cmp68_3.ps}
  \includegraphics[height=0.2\textheight,width=0.24\textwidth]{rdp-zdp_tav_cmp69_3.ps}
  \end{center}
\caption{Vertical structure of the quality factors, $Q_i$, of the MRI, 
Equation (\ref{eq:qfdfn}), 
of Case I-high ({\it blue solid}), Case II-high 
({\it red solid}), Case I-low ({\it green dashed}), 
and Case II-low ({\it magenta dashed}), in comparison with the result of 
the shearing box simulation ({\it black dotted}). 
From left to right, $i=R$, $\phi$, and $z$ components are displayed.}
\label{fig:zstqf}
\end{figure*}

\begin{figure*}
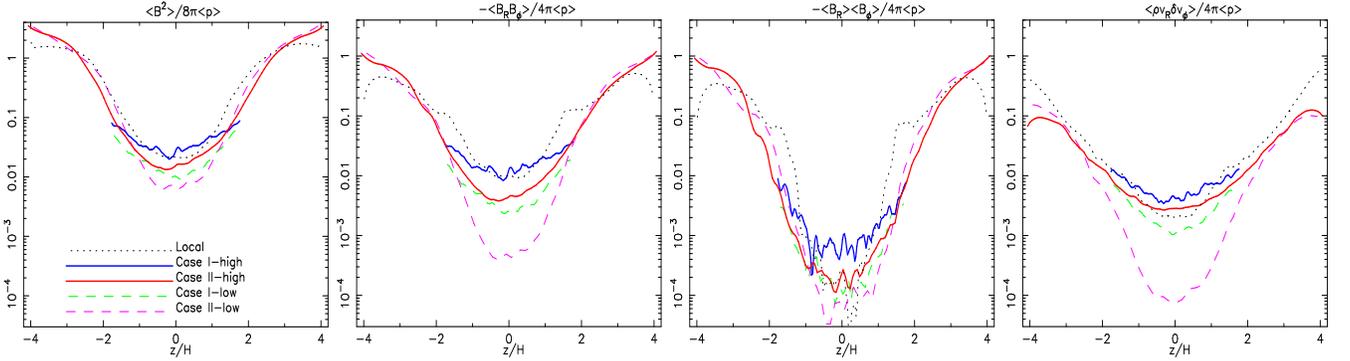

  \includegraphics[height=0.2\textheight,width=0.24\textwidth]{rdp-zdp_tav_cmp64_3.ps}
  \includegraphics[height=0.2\textheight,width=0.24\textwidth]{rdp-zdp_tav_cmp65_3.ps}
  \includegraphics[height=0.2\textheight,width=0.24\textwidth]{rdp-zdp_tav_cmp66_3.ps}
  \includegraphics[height=0.2\textheight,width=0.24\textwidth]{rdp-zdp_tav_cmp71_3.ps}
\caption{From left to right are shown vertical structures of 
$\frac{\langle B^2\rangle_{t,\phi}(r_1,z)}{8\pi\langle p 
\rangle_{t,\phi}(r_1,z)}$, $-\frac{\langle B_R B_{\phi}\rangle_{t,\phi}(r_1,z)}
{4\pi\langle p \rangle_{t,\phi}(r_1,z)}$ (total Maxwell stress), 
$-\frac{\langle\langle B_R\rangle_{\phi} \langle B_{\phi}\rangle_{\phi}
\rangle_{t}(r_1,z)}{4\pi\langle p \rangle_{t,\phi}(r_1,z)}$ (coherent Maxwell 
stress), and $\frac{\langle \rho v_R \delta v_{\phi}\rangle_{t,\phi}(r_1,z)}
{4\pi\langle p \rangle_{t,\phi}(r_1,z)}$ (Reynolds stress) of 
Case I-high ({\it blue solid}), Case II-high ({\it red solid}), 
Case I-low ({\it green dashed}), and Case II-low ({\it magenta dashed}), 
in comparison with the result of the shearing box simulation 
({\it black dotted}). See Equations (\ref{eq:2bttp}) 
-- (\ref{eq:Rytp}) for details. }
\label{fig:zstBBp}
\end{figure*}


\begin{table}
\begin{center}
\begin{tabular}{|c|c|c|}
\hline
Model & $H/\Delta z_{\rm mesh} (r_1)$ & $H/\Delta z_{\rm mesh} (r_2)$\\
\hline
\hline
Case I-high & 35 & 51\\
\hline
Case I-low & 18 & 25\\
\hline
Case II-high & 16 & 16\\
\hline
Case II-low & 8 & 8\\
\hline
Local & \multicolumn{2}{|c|}{32}\\
\hline
\end{tabular}
\end{center}
\caption{Resolution of each run at $R=r_1(=5r_{\rm in})$ ({\it 2nd column}) 
and $R=r_2(=10r_{\rm in})$ ({\it 3rd column}).\label{tab:res}} 
\end{table}

We examine the vertical structures of the magnetic fields at 
$R=r_1$, in comparison with results of a local shearing box simulation. 
As for the local simulation, we perform a 3D MHD simulation with the same 
strength for the net vertical field of $\beta=10^5$ at the midplane as in the 
global simulations in a simulation box with the size, 
$(x,y,z)=(2H,4H,8H)$, resolved by the uniform grid points of
$(N_x,N_y,N_z)=(64,128,256)$ \citep[see][for the details]{suz10}; namely, 
one scale height, $H$, is resolved by 32 grid points.
In Table \ref{tab:res}, we summarize the resolution, 
$H/\Delta z_{\rm mesh}$, with respect to the vertical direction 
of each case, where $\Delta z_{\rm mesh}$ indicates the size of a vertical 
mesh.
For the global simulations, we approximately use $\Delta z_{\rm mesh} 
\approx r \Delta \theta$.
Although at the inner radius, $R=r_{\rm in}$, the same resolution is set 
for Cases I and II, Case I gives a higher resolution at $R=r_1$ 
because of the different scalings of $H/R$ (Equations \ref{eq:cIhr} \& 
\ref{eq:cIIhr}).

Figure \ref{fig:zstBsq} compares the magnetic energies of each component of 
the four cases. We present here 
\begin{equation}
\frac{\langle B_{i}^2\rangle_{t,\phi}(r_1,z)}{4\pi p_{\rm mid,0}},
\end{equation}
of $i$-th components ($i=R$, $\phi$, and $z$ from left to right),
where the normalization in the denominator is the initial gas pressure, 
$p_{\rm mid,0}$, at the midplane.  
The saturated $B_{R}^2$ (left panel) and $B_{z}^2$ (right panel)
show roughly positive correlations with the 
resolution, $H/\Delta z_{\rm mesh}$ (Table \ref{tab:res}), whereas 
Case II-high gives the higher saturations than Case I-low though 
the $H/\Delta z_{\rm mesh}$ values are similar. 
In Case II-low (black dashed lines), which is the case with the lowest 
resolution at $R=r_1$, the strength of the magnetic field is too weak 
and shows a dip structure because of the insufficient resolution 
($H/\Delta z_{\rm mesh}=8$) at the midplane. 
On the other hand, in Case I-high, which has a resolution similar to 
the local simulation, the saturated magnetic field strength at the midplane 
is comparable to that of the local simulation. 

The dependence of $B_{\phi}^2$ (middle panel of 
Figure \ref{fig:zstBsq}) on the resolution is quite weak. 
Although Case I gives a positive dependence on the resolution, which is expected
from the amplification by MRI, Case II shows complicated behavior;  
while in the midplane region, the high-resolution run gives larger 
$B_{\phi}^2$, in the surface regions, $|z|\gtrsim 1.5H$, the low-resolution 
run gives larger $B_{\phi}^2$. This indicates that the magnetic field strength 
in the surface regions of Case II is not regulated by the MRI but mainly 
by the vertical differential rotation, which does not have a positive 
dependence on the numerical resolution.   
This effect cannot be handled in local shearing box simulations, indicating 
the importance of studies using global simulations.  

In order to further study the amplification of the magnetic fields, in 
Figure \ref{fig:zstqf} we inspect a quality factor, $Q_i$, of $i-$th 
components ($i=R,\phi,z$) for MRI \citep[][]{nob10,haw11},
which is defined as the ratio of the $\lambda_{\rm max}$ for MRI to a 
mesh size, 
\begin{equation}
\langle Q_{i}\rangle_{t,\phi}(r_1,z) = 2\pi 
\frac{\sqrt{\langle v_{{\rm A},i}^2\rangle_{t,\phi}(r_1,z)} }{\Omega \Delta l_i}, 
\label{eq:qfdfn}
\end{equation}
where $\langle v_{{\rm A},i}^2\rangle=\langle B^2_{i}\rangle/4\pi
\langle\rho\rangle$ and we approximately 
use $\Delta l_i=\Delta r, r\sin \theta \Delta \phi$, 
and $r \Delta \theta$ for $i=R$, $\phi$, and $z$ components, respectively, 
to convert the spherical coordinates used in the simulations to the 
cylindrical coordinates for the data analyses. 
According to \citet{san04}, $Q_z\gtrsim 6$ is a necessary condition for a
vertical magnetic field to get a linear growth rate close to the analytic 
prediction from MRI. 

The saturations of $B_R^2$ and $B_z^2$ shown in Figure \ref{fig:zstBsq} 
are well-explained by the profiles of $Q_R$ (left panel) and 
$Q_z$ (right panel) in Figure \ref{fig:zstqf}. In the midplane region of 
Case II-low $Q_R, Q_z<1$, which leads to the low levels of $B_R^2$ and 
$B_z^2$. Case I-high gives $Q_R,Q_z>10$ in the entire region, and then, 
the obtained saturation levels are supposed to be reasonable. 
In the other two cases, $Q_R,Q_z\approx 2-3$ at the midplane, which is
probably marginally insufficient to resolve 
the MRI. Therefore, the saturation levels of these cases are lower than 
those of Case I-high and the local simulation.   
Since the toroidal component is amplified by the winding involving the 
differential rotation in addition to the MRI, $Q_{\phi}\gtrsim 5$ 
in the entire regions except in the midplane region of Case II-low. 
 
Figure \ref{fig:zstBBp} presents the following four nondimensional quantities: 
\begin{equation}
\frac{\langle B^2\rangle_{t,\phi}(r_1,z)}{8\pi\langle p \rangle_{t,\phi}(r_1,z)} 
(=\frac{1}{\langle \beta \rangle_{t,\phi,}(r_1,z)}), 
\label{eq:2bttp}
\end{equation}
\begin{equation}
-\frac{\langle B_R B_{\phi}\rangle_{t,\phi}(r_1,z)}
{4\pi\langle p \rangle_{t,\phi}(r_1,z)}: \; {\rm total\; Maxwell\; stress},
\label{eq:tMxtp}
\end{equation}
\begin{equation}
-\frac{\langle\langle B_R\rangle_{\phi} \langle B_{\phi}\rangle_{\phi}
\rangle_{t}(r_1,z)}{4\pi\langle p \rangle_{t,\phi}(r_1,z)}: \; 
{\rm coherent\; Maxwell\; stress}, 
\label{eq:tMBtp}
\end{equation}
\begin{equation}
\frac{\langle \rho v_R \delta v_{\phi}\rangle_{t,\phi}(r_1,z)}
{4\pi\langle p \rangle_{t,\phi}(r_1,z)}: \; {\rm Reynolds\; stress}, 
\label{eq:Rytp}
\end{equation}
where the time average is taken over $\Delta t_{\rm ave}$ in Table 
\ref{tab:models}. 
In Equation (\ref{eq:tMBtp}) we pick up the coherent part of the Maxwell 
stress in Equation (\ref{eq:tMxtp}), by 
taking the $\phi$ average of $B_r$ and $B_{\phi}$ separately before 
multiplying them. This term is supposed to roughly correspond to the 
transport of angular momentum by magnetic braking \citep{wd67}.  
The total Maxwell stress in 
Equation (\ref{eq:tMxtp}) contains both coherent and turbulent components. 
When estimating Reynolds stress, we use the difference of $v_{\phi}$ from 
the time-averaged value, 
\begin{equation}
\delta v_{\phi} \equiv v_{\phi} - \langle v_{\phi}\rangle_{t,\phi} \equiv v_{\phi} 
- \frac{\langle \rho v_{\phi }\rangle_{t,\phi}}{\langle \rho \rangle_{t,\phi}}, 
\label{eq:vshftave}
\end{equation}
instead of the initial value 
(Equation \ref{eq:vpshft0}) because the background rotation profile is slightly 
modified because of the change of the pressure gradient force through the 
evolution of radial density distribution, 
As shown in Equation (\ref{eq:vshftave}), the background velocity, 
$\langle v_{\phi}\rangle_{t,\phi}$, is derived from the density-weighted 
average, $\langle \rho v_{\phi }\rangle_{t,\phi}/\langle \rho \rangle_{t,\phi}$.

The left panel of Figure \ref{fig:zstBBp} shows that the four cases mostly 
follow the trend of the local simulation; the plasma $\beta$ 
values are $\approx 10-100$ at the midplane and decrease to $<1$ 
in the surface regions owing to the decrease of the density by the gravity 
of the central object. 
The overall trends are similar for all these cases and only weakly depend 
on the resolution, because the total magnetic fields are dominated by 
the $\phi$ component, which weakly depends on the resolution.
The magnetic field strengths in the surface regions 
of Cases II-high and II-low are larger than the value obtained in the 
local simulation, 
because the coherent magnetic fields are amplified by the vertical differential 
rotation. 

The total Maxwell stress (the second panel from the left in 
Figure \ref{fig:zstBBp}) 
around the midplane exhibits the positive dependence on the resolution, 
which is quite similar to $B_{R}^2$ and $B_z^2$ in Figure \ref{fig:zstBsq}. 
The strength of the Maxwell stress near the midplane is determined by the 
MRI. On the other hand, the Maxwell stress in the surface regions of Cases 
II-high and II-low is considerably larger than that of the local simulation. 
Comparing the middle two panels, the Maxwell stress in the surface regions 
is mostly by the coherent component. The larger coherent Maxwell stresses in 
Cases II-high and II-low in the surface regions are a consequence of the 
wound-up magnetic field lines by the vertical differential rotation.

The Reynolds stress (the right panel of Figure \ref{fig:zstBBp}) is 
systematically smaller than the Maxwell stress (second panel from the left) of 
each case by a factor of 2-3. The difference in the Reynolds stresses 
among the four models is similar to the tendency obtained for the total 
Maxwell stresses. 

\subsubsection{Radial Profile}
\label{sec:rpB}
\begin{figure*}[t]
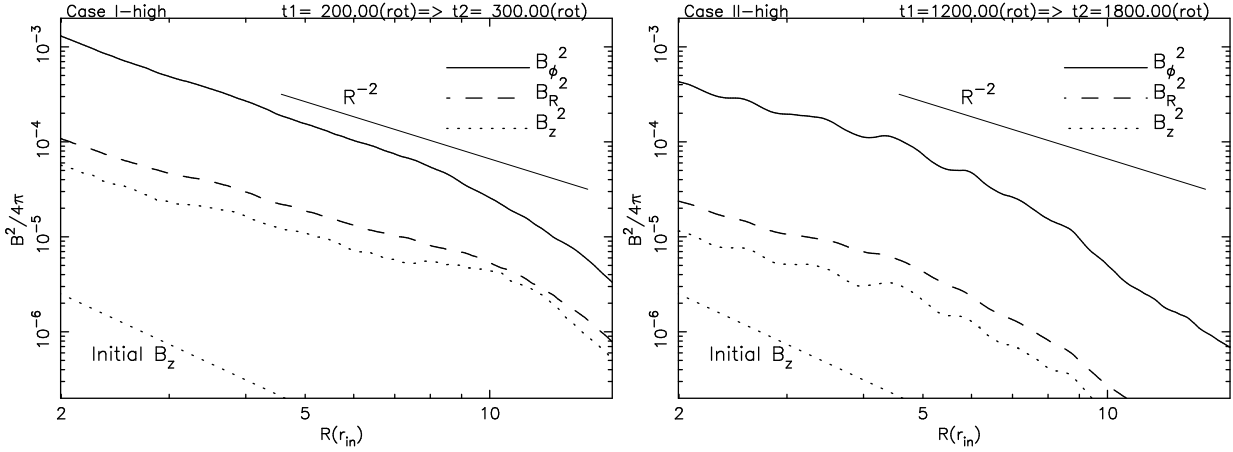

\includegraphics[height=0.25\textheight,width=0.45\textwidth]{oneDdiag_tav_fig10_21_1.ps}
\includegraphics[height=0.25\textheight,width=0.45\textwidth]{oneDdiag_tav_fig11_21_1.ps}
\caption{$\langle \frac{B_R^2}{4\pi}\rangle_{R,\phi,z_{\rm tot}}$ ({\it solid}), 
$\langle \frac{B_z^2}{4\pi}\rangle_{R,\phi,z_{\rm tot}}$ ({\it dashed}), 
and $\langle \frac{B_{\phi}^2}{4\pi}\rangle_{R,\phi,z_{\rm tot}}$ 
({\it dash-dotted}) of Case I-high 
({\it left}) and Case II-high ({\it right}). 
A slope, $R^{-2}$, ({\it thin solid}) and the initial 
condition, $\frac{B_{z}^2}{4\pi} = 2\times 10^{-5} 
\left(\frac{R}{r_{\rm in}}\right)^{-3}$, ({\it dotted}) are overplotted. 
The vertical averages are taken over the entire region, 
$\Delta z_{\rm tot}$.}
\label{fig:Bsq}
\end{figure*}


Figure \ref{fig:Bsq} compares the radial dependence of the $i=R$, $z$, and 
$\phi$ components of the $t,\phi,z$ averaged $\langle B_i^2\rangle_{t,\phi,z}
/4\pi$ of Case I-high (left panel) and Case II-high 
(right panel), where the $z$ average is taken over the entire region, 
$\Delta z_{\rm tot}$. 
In both cases, the toroidal component dominates by the winding owing 
to the radial differential rotation, which is consistent with 
the result of previous local simulations \citep[e.g.,][]{suz10}. 
Examining quantitative ratios of 
different components, Case II-high gives larger $B_{\phi}^2 / (B_z^2 + B_R^2)$ 
than Case I-high, because of the contribution from the vertical differential 
rotation. 

Focusing on the radial dependences, the two 
cases give different trends. Here we again concentrate on the region between 
$R=r_1(=5r_{\rm in})$ and $R=r_2(=10r_{\rm in})$ to avoid effects of the inner 
boundary. Starting from the initial vertical fields with 
$B_{z,0}^2\propto R^{-3}$, each component of Case I-high is amplified as it
approaches $R^{-2}$, which is expected from the force balance between 
magnetic hoop stress and magnetic pressure \citep{flo11}, 
\begin{equation}
F = \frac{1}{\rho R^2}\frac{\partial}{\partial R}(R^2 \frac{B_{\phi}^2}{8\pi}) 
= 0.
\end{equation}
$B_R^2$ and $B_z^2$, which are not explained 
by the force balance, are supposed to be subject simply to the largest 
$\phi$ component.
On the other hand, each component of Case II-high is amplified while keeping 
the initial profile $\propto R^{-3}$,
whereas the magnetic field quickly decreases with $R$ in $R>10 r_{\rm in}$ 
because it is still in the growth phase.  
The $\propto R^{-3}$ profile corresponds to the constant plasma $\beta$ with 
$R$, as will be discussed below. 

\begin{figure}[h]
\begin{center}
\includegraphics[height=0.4\textheight,width=0.45\textwidth]{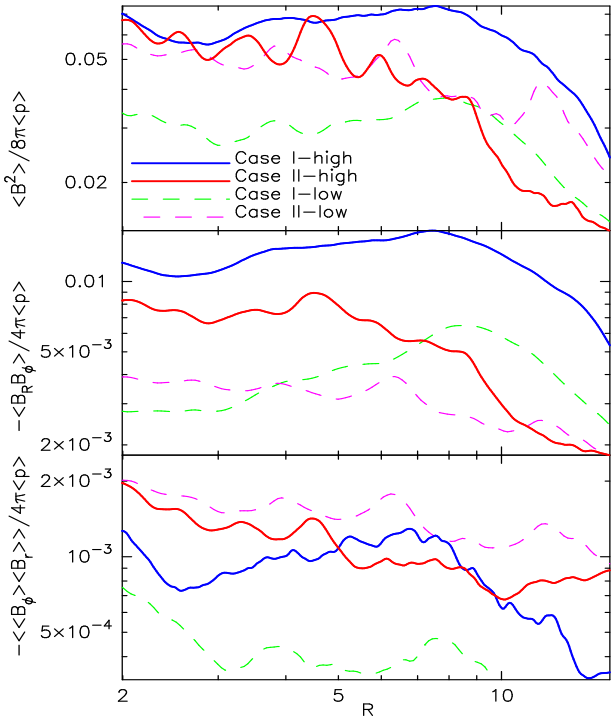}
\end{center}
\caption{$\langle B^2\rangle_{t,\phi,z_{\rm tot}}
/4\pi\langle p \rangle_{t,\phi,z_{\rm tot}}$ 
({\it top}) and total ($-\langle B_R B_{\phi}\rangle_{t,\phi,z_{\rm tot}}
/4\pi\langle p \rangle_{t,\phi,z_{\rm tot}}$; {\it middle}) and coherent ($-\langle 
\langle B_R\rangle_{\phi} \langle B_{\phi}\rangle_{\phi}\rangle_{t,z_{\rm tot}}
/4\pi\langle p \rangle_{t,\phi,z_{\rm tot}}$; {\it bottom}) components 
of Maxwell stress of Case I-high ({\it blue solid}), Case II-high 
({\it red solid}), Case I-low ({\it green dashed}), and Case II-low 
({\it magenta dashed}). 
In these plots, the integrations with $z$ are performed in 
the entire vertical extent, $\Delta z_{\rm tot}$. }
\label{fig:BBpcomp}
\end{figure}

Figure \ref{fig:BBpcomp} compares, from top to bottom 
$\frac{\langle B^2\rangle_{t,\phi,z_{\rm tot}}}{8\pi\langle p \rangle_{t,\phi,z_{\rm tot}}} (=\frac{1}
{\langle \beta \rangle_{t,\phi,z_{\rm tot}}})$, 
$-\frac{\langle B_R B_{\phi}\rangle_{t,\phi,z_{\rm tot}}}
{4\pi\langle p \rangle_{t,\phi,z_{\rm tot}}},$ 
and
$-\frac{\langle\langle B_R\rangle_{\phi} \langle B_{\phi}\rangle_{\phi}
\rangle_{t,z_{\rm tot}}}{4\pi\langle p \rangle_{t,\phi,z_{\rm tot}}}$,
which are the vertically integrated quantities in $\Delta z_{\rm tot}$ in 
Equations (\ref{eq:2bttp}) -- (\ref{eq:tMBtp}) 
for the vertical structures (Figure \ref{fig:zstBBp}).  

${\langle B^2\rangle}/{8\pi\langle p\rangle}$ of Cases I-high 
({\it blue solid}) and I-low ({\it green dashed}) show increasing trends 
with $R$ in $R<8r_{\rm in}$, while those of Case II-high ({\it red solid}) 
and II-low ({\it magenta dashed}) show slightly decreasing trends 
with $R$.   
The scalings of ${\langle B^2\rangle}/4\pi$ (Figure \ref{fig:Bsq} for 
the high resolution runs)
and ${\langle B^2\rangle}/4\pi\langle p \rangle$ in Figure \ref{fig:BBpcomp} 
are different by the scaling of $\langle p \rangle$, which is proportional to 
$(\Sigma/H)c_{\rm s}^2$.  In all the cases the final profiles of 
$\Sigma$ (Figure \ref{fig:sfdns_tav}) and correspondingly the profiles of 
$\langle p\rangle\propto R^{-\xi_p}$ become slightly shallower with $2.5<\xi_p<3$ 
in $r_1<R<r_2$ than the initial profile, $\langle p \rangle\propto R^{-3}$. 
Therefore, the obtained trend, $\langle B^2\rangle/8\pi\propto R^{-\xi_B}$, 
with a shallower index $\xi_B\approx 2$ in Case I-high (Figure \ref{fig:Bsq}) 
as well as I-low results in the trend of ${\langle B^2\rangle}
/8\pi\langle p \rangle$ increasing with $R$ (Figure \ref{fig:BBpcomp}). 
On the other hand, in Case II-high as well as II-low the initial profile 
of $\langle B^2\rangle/8\pi$ with $\xi_B\approx 3$ is almost maintained 
(Figure \ref{fig:Bsq}), and the slow decreasing trend of ${\langle B^2\rangle}
/8\pi\langle p \rangle$ results (Figure \ref{fig:BBpcomp}). 
 
The Maxwell stresses (middle panel of Figure \ref{fig:BBpcomp}) follow the 
trends of ${\langle B^2\rangle}/8\pi\langle p \rangle$ (top panel): 
increasing with $R$ in Cases I-high and I-low, and flat or 
slightly decreasing with $R$ in Cases II-high and II-low. 
The coherent component of the Maxwell stresses (bottom panel of  
Figure \ref{fig:BBpcomp}) show different behaviors. 
In Cases I-high and I-low, the contributions from the coherent component 
are quite small. On the other hand, in Cases II-high and II-low, 
the roles of the coherent component are not negligible because of the 
vertical differential rotation of the equilibrium profile, and their 
radial dependences almost follow those of the total Maxwell stresses.  


Comparing the high- and low-resolution runs, Cases I and II give totally 
different results. These three quantities of Case I-high (blue solid lines 
in Figure \ref{fig:BBpcomp}) 
are larger than those of Case I-low (dashed green lines),  
which is understandable from the amplification of the magnetic field 
by the MRI. The low-resolution run cannot resolve smaller-scale 
turbulence by the MRI, which leads to the smaller saturation level (Figure 
\ref{fig:zstqf}). 
On the other hand, the comparison between the high- (red solid lines 
in Figure \ref{fig:BBpcomp}) and low-resolution (magenta dashed lines) runs 
of Case II shows a different trend. Although the higher 
saturation level of the Maxwell stress is obtained in the high-resolution 
run ({\it middle}), 
the coherent component of the Maxwell stress ({\it bottom}) shows the 
opposite behavior. The saturation of the total $B^2$ ({\it top}), 
which is dominated by $B_{\phi}^2$, looks almost 
independent of the resolution. These tendencies imply that, in addition 
to the MRI, the vertical differential rotation plays a role in the 
amplification of the magnetic fields in Cases II-high and II-low, which is 
consistent with the tendency obtained from the vertical structures (Figure 
\ref{fig:zstBBp}).

\begin{figure}[h]
\begin{center}
\includegraphics[height=0.44\textheight,width=0.45\textwidth]{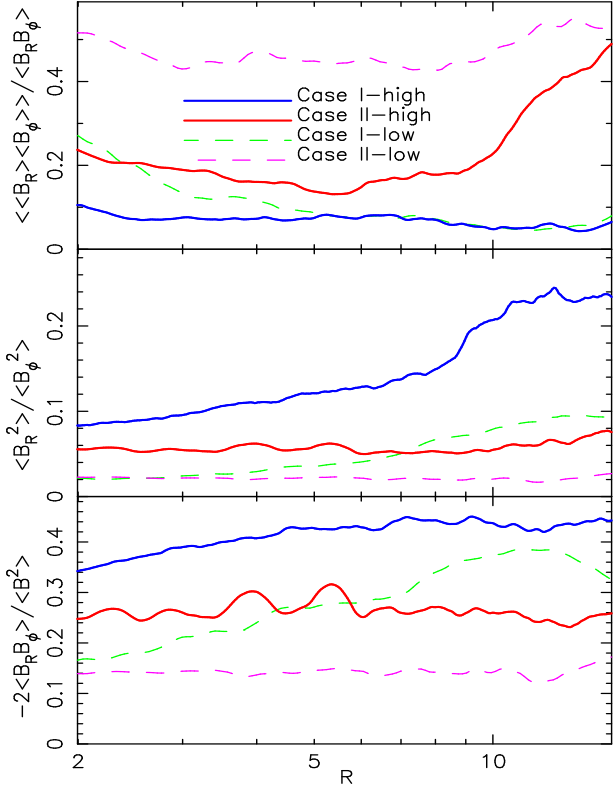}
\end{center}
\caption{Ratios of various quantities concerning the magnetic fields of 
Case I-high ({\it blue solid}), Case II-high ({\it red solid}), 
Case I-low ({\it green dashed}), and Case II-low ({\it magenta dashed}). 
In these plots, the integrations with $z$ are performed in 
the entire vertical extent, $\Delta z_{\rm tot}$. 
{\it top}: Relative contribution of the coherent component to the 
total Maxwell stress. 
{\it middle}: $\langle B_R^2\rangle_{t,\phi,z_{\rm tot}} 
/\langle B_{\phi}^2\rangle_{t,\phi,z_{\rm tot}}$. 
{\it bottom}: Maxwell stress, $-\langle B_{R}B_{\phi}\rangle_{t,\phi,z_{\rm tot}}
/4\pi$, to magnetic pressure $\langle B^2\rangle_{t,\phi,z_{\rm tot}}/8\pi$. }
\label{fig:BBrel}
\end{figure}

In Figure \ref{fig:BBrel}, we compare nondimensional quantities 
concerning magnetic fields. 
The top panel of Figure \ref{fig:BBrel} compares the ratio of 
the coherent to total Maxwell stresses, $\langle \langle B_R
\rangle_{\phi} \langle B_{\phi}\rangle_{\phi}\rangle_{t,z_{\rm tot}}
/\langle B_R  B_{\phi} \rangle_{t,\phi,z_{\rm tot}}$, which indicates the relative 
importance of the winding of magnetic field lines in the Maxwell stresses. 
The figure shows that this quantity has a negative correlation with 
the quality factor $Q_i$ (Equation \ref{eq:qfdfn}).
For instance, Case II-low, which has the smallest $Q$ (Figure 
\ref{fig:zstqf}), gives the largest $\langle \langle B_R
\rangle_{\phi} \langle B_{\phi}\rangle_{\phi}\rangle_{t,z_{\rm tot}}
/\langle B_R  B_{\phi} \rangle_{\phi,t,z_{\rm tot}}$ among the four cases. 
In this case the total Maxwell stress 
is the smallest because of the insufficient resolution for the MRI, which 
leads to the relatively large contribution of the coherent component 
by the winding. In addition to the dependence on the resolution, 
the vertical differential rotation originating from the temperature profile 
of Case II gives a larger fraction of the coherent component.  
Cases II-high and I-low give similar initial resolution, 
$H/\Delta_{z_{\rm mesh}}$, at $R=r_1$ (Table \ref{tab:res}). 
Although Case II-high gives slightly larger $Q$ there 
(see Figure \ref{fig:zstqf}), 
the fraction of the coherent component is larger there, which is the 
opposite of the tendency expected from the dependence on the resolution. 
This is also indirect evidence that in Case II-high 
the vertical differential rotation plays a role in the amplification of 
the magnetic field.

\begin{figure*}[t]
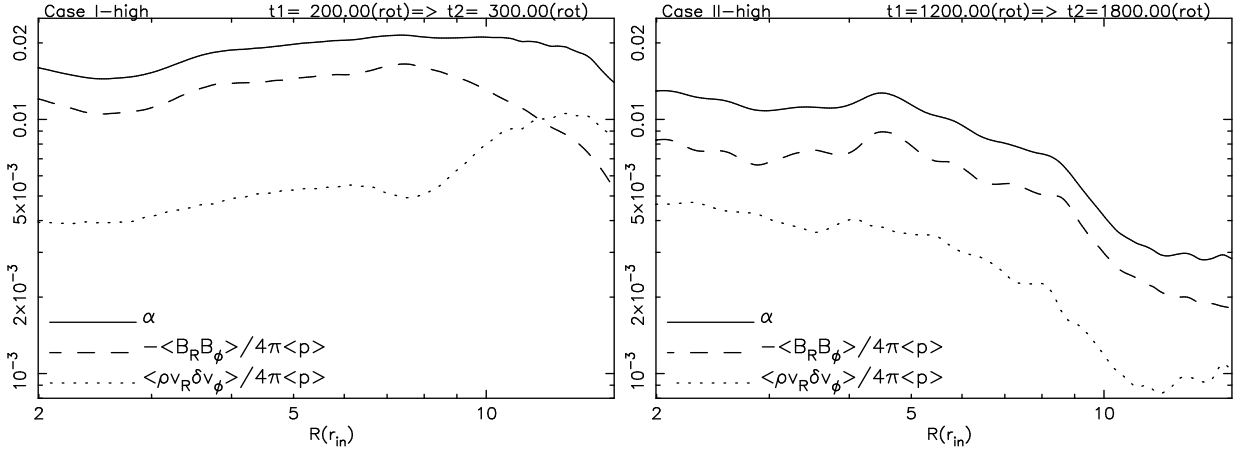

\includegraphics[height=0.25\textheight,width=0.45\textwidth]{oneDdiag_tav_fig10_23_1.ps}
\includegraphics[height=0.25\textheight,width=0.45\textwidth]{oneDdiag_tav_fig11_23_1.ps}
\caption{Maxwell stress ($-\langle B_R B_{\phi}\rangle_{t,\phi,z_{\rm tot}}
/4\pi\langle p \rangle_{t,\phi,z_{\rm tot}}$; {\it dashed}), Reynolds stress 
($\langle \rho v_R \delta v_{\phi}\rangle_{t,\phi,z_{\rm tot}}
/4\pi\langle p \rangle_{t,\phi,z_{\rm tot}}$; {\it dotted}), 
and their sum, $\alpha$, of Case I-high ({\it left}) and Case II-high 
({\it right}). The vertical averages are taken over the entire region, 
$\Delta z_{\rm tot}$. }
\label{fig:MRstr}
\end{figure*}

\citet{haw11} introduced several diagnostics that are related to 
properties of MRI-driven turbulence in numerical simulations. 
In the middle and bottom panels of Figure \ref{fig:BBrel} we show 
two such indices; the middle panel plots the ratio of the $R$ component 
of the magnetic energies to the $\phi$ component$R$,  $\langle B_R^2
\rangle_{t,z_{\rm tot},\phi}/\langle B_{\phi}^2\rangle_{t,z_{\rm tot},\phi}$,    
and the bottom panel displays the ratios of the Maxwell stress to the magnetic 
pressure, which is defined as $\alpha_{\rm mag}=-2\langle B_R B_{\phi}\rangle 
/ \langle B^2\rangle$ in \citet{haw11}. 
These quantities, which exhibit similar trends, have positive 
correlations with the quality factor, which is the 
opposite of the trend obtained for $\langle \langle B_R\rangle_{\phi} 
\langle B_{\phi}\rangle_{\phi}\rangle_{t,z_{\rm tot}}/\langle B_R  B_{\phi} 
\rangle_{\phi,t,z_{\rm tot}}$ in the top panel.  Cases I-high (blue solid lines) 
and I-low (green dashed lines)
show increasing trends with $R$, which reflects the adopted increasing trend 
of the numerical resolution, $H/\Delta z_{\rm mesh}\propto R^{1/2}$. 
On the other hand, Cases II-high (red solid lines) and II-low (magenta dashed
lines), which have resolutions that remain constant with $R$, show rather 
flat trends with $R$. 

On the basis of the shearing box simulations by \citet{sim11}, 
$\langle B_R^2\rangle/\langle B_{\phi}^2\rangle$ approaches 0.2 with 
sufficient resolution. 
In the outer region of Case I-high, the value of $\langle B_R^2
\rangle_{t,z_{\rm tot},\phi}/\langle B_{\phi}^2\rangle_{t,z_{\rm tot},\phi}$ 
is this saturation value, while the other cases 
give smaller $\langle B_R^2\rangle_{t,z_{\rm tot},\phi}
/\langle B_{\phi}^2\rangle_{t,z_{\rm tot},\phi}$ probably because 
the resolution is not sufficient. However we need to carefully consider 
global effects such as radial flows, meridional flows and vertical 
differential rotation, which are not taken into account in local simulations. 
The vertical differential rotation in Cases II-high and II-low is supposed 
to give smaller $\langle B_R^2\rangle_{t,z_{\rm tot},\phi}
/\langle B_{\phi}^2\rangle_{t,z_{\rm tot},\phi}$ 
because $B_{\phi}$ is systematically amplified. 

The ratio of the Maxwell stress to the magnetic pressure, 
$-2\frac{\langle B_R B_{\phi}\rangle_{t,\phi,z_{\rm tot}}}
{\langle B^2 \rangle_{t,\phi,z_{\rm tot}}}$, also shows a positive correlation with 
the quality factor. 
Local simulations \citep{shi10,dav10,sim11,gg11} with sufficient resolution
give $-2\frac{\langle B_R B_{\phi}\rangle}{\langle B^2 \rangle}\approx 
0.3-0.4$ \citep{haw11}. Case I-high gives a quite large value, 
$-2\frac{\langle B_R B_{\phi}\rangle_{t,\phi,z_{\rm tot}}}
{\langle B^2 \rangle_{t,\phi,z_{\rm tot}}}\approx 0.4$, while smaller values 
are obtained in the other cases. 
As we did for $\langle B_R^2\rangle/\langle B_{\phi}^2\rangle$, we also 
carefully take into account the global effects. 
Cases II-high and II-low are expected to give systematically lower 
$-2\frac{\langle B_R B_{\phi}\rangle_{t,\phi,z_{\rm tot}}}
{\langle B^2 \rangle_{t,\phi,z_{\rm tot}}}$,
because the vertical differential rotation tends to increase $B^2$ through the 
amplification of $B_{\phi}^2$ rather than $\langle B_R B_{\phi}\rangle$ 
through the coherent component $\langle B_R\rangle\langle B_{\phi}\rangle$ 
in a quantitative sense (Figure \ref{fig:zstBBp}).

Figure \ref{fig:MRstr} compares the radial profiles of 
$\alpha$ values (solid lines), 
\begin{equation}
\langle \alpha \rangle_{t,\phi,z_{\rm tot}}(R) 
\equiv \frac{\langle \rho v_R \delta v_{\phi} \rangle_{t,\phi,z_{\rm tot}}(R)}
{\langle p \rangle_{t,\phi,z_{\rm tot}}(R)}-\frac{\langle 
B_R B_{\phi}\rangle_{t,\phi,z_{\rm tot}}(R)}{4\pi\langle p \rangle_{t,\phi,z_{\rm tot}}(R)}, 
\label{eq:atpz}
\end{equation}
of Case I-high (left panel) and Case II-high (right panel), and their 
breakdowns to the Reynolds (dotted lines) and Maxwell (dashed lines) 
stresses, which are the first and second terms of Equation (\ref{eq:atpz}), 
respectively. 
Here $\delta v_{\phi}$ is the difference of the rotation velocity from the 
time-averaged value, Equation (\ref{eq:vshftave}).
Although the Maxwell stresses are larger than the Reynolds stresses in both 
the cases, the ratios are slightly different. 
Case I-high gives a ratio of the Maxwell stress to Reynolds stress of 
$\approx 3:1$, while Case II-high gives $\approx 2:1$; both these values are 
smaller than the typical value of $\approx 5:1$ obtained in local shearing 
box simulations \citep{san04}. 
However, we cautiously note that there are ambiguities particularly in the 
choice of $\delta v_{\phi}$ used to estimate the Reynolds stresses in the global 
simulations. 
Here we derive $\delta v_{\phi}$ by subtracting $v_{\phi}$ from the $\phi$ and 
$\Delta t_{\rm ave}$-averaged $v_{\phi}$ (Equation \ref{eq:vshftave}). 
If we used a shorter duration for 
the time average, $\delta v_{\phi}$ would be smaller because $\langle v_{\phi}
\rangle$ gradually changes with time owing to the change in the radial density 
profile. In this case the derived Reynolds stress would be smaller.  
Then, the above estimated ratios of the Maxwell stress to Reynolds stress are 
lower limits in a sense, and we need to be careful when comparing Reynolds  
stresses of global simulations with those of local simulations. 
 
 
\begin{figure}[h]
\includegraphics[height=0.3\textheight,width=0.45\textwidth]{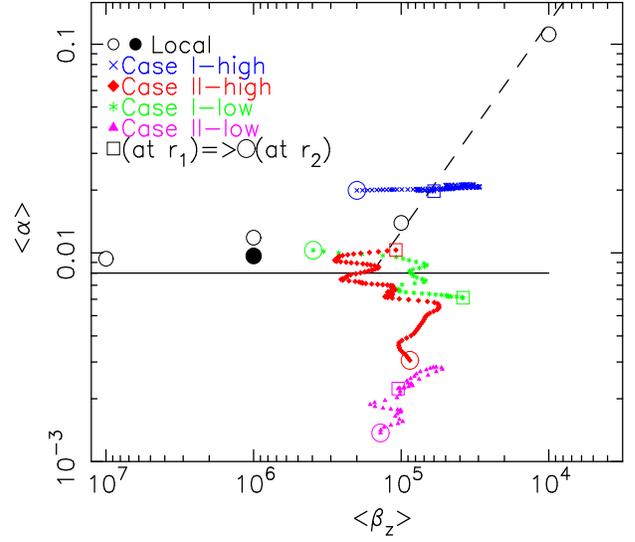}
\caption{Dependence of time and $\phi$-$z$ plane averaged 
$\langle \alpha \rangle_{t,\phi,z_{\rm tot}}$ on plasma 
$\langle \beta_{z} \rangle_{t,\phi,z_{\rm tot}}$ for net vertical magnetic fields 
in $r_1<R<r_2$. 
Multiple data points for each case of the global simulations correspond
to different radial locations. Colored squares and circles indicate the data 
at $R=r_1$ and $R=r_2$. Case I-high ({\it blue crosses}), 
Case II-high ({\it red diamonds}), Case I-low ({\it green asterisks}), 
and Case II-low ({\it magenta triangles}) are compared with the local 
simulations by \citet[][;{\it open circles} for low-resolution runs and 
{\it filled circle} for high-resolution run]{suz10}. The solid line is 
a `floor' value based on the local simulations and the dashed line 
is a fit for the increasing trend of $\langle\alpha\rangle$ \citep{suz10}.}
\label{fig:bzalp}
\end{figure}

The dependence of the saturations of MRI-triggered turbulence on net vertical 
magnetic field strengths has been extensively studied by using local 
shearing box simulations to date, and it has been discussed that the net 
$B_z$ plays an important role in determining the $\alpha$ 
parameter \citep{hgb95,mt95,bra95,san04,hir06,pes07,si09,oh11}. 
Figure \ref{fig:bzalp} shows the relation between the net vertical 
field strengths (horizontal axis) and the $\alpha$ values (vertical axis) 
derived from our global simulations. 
The net vertical magnetic flux at each $R$ is not conserved in the global 
simulations because of radial motions of the field lines, in contrast to 
the treatment by local shearing boxes in which the net vertical flux is 
strictly conserved within round-off errors. 
Thus, we derive the time-averaged net vertical field strength 
in the form of plasma $\beta$ in the following way:
\begin{equation}
\langle \beta_{z} \rangle_{t,\phi,z_{\rm tot}}(R) \equiv \frac{8\pi \langle 
p_{\rm mid}\rangle_{t,\phi}(R) }{\langle \langle B_z 
\rangle^2_{t,\phi}\rangle_{z_{\rm tot}}(R)}, 
\label{eq:bttpz}
\end{equation}
where we take the entire vertical box, $\Delta z_{\rm tot}$, for the $z$ 
average of the net $\phi$- and $\Delta t_{\rm ave}$-averaged 
$\langle B_z\rangle_{t,\phi}$.
The meaning of Equation (\ref{eq:bttpz}) is probably 
straightforward. Net vertical field strength is first estimated by the 
$\phi$ and time average, which is squared before being integrated over $\Delta 
z_{\rm tot}$. 
The normalization is taken with the $\phi$- and time-averaged gas pressure 
at the midplane. Note that $\langle \beta \rangle_{t,\phi,z_{\rm tot}}(R)$ is also 
affected by the change of local density in addition to radial motions of 
vertical field lines. $\langle \alpha \rangle_{t,\phi,z_{\rm tot}}(R)$ 
(Equation \ref{eq:atpz}) 
are plotted with the derived $\langle \beta \rangle_{t,\phi,z_{\rm tot}}(R)$ 
values in Figure \ref{fig:bzalp}, in comparison with the results of 
the local shearing box simulations by \citet{suz10}. 
A number of the data points for each case of the global simulations correspond
to different radial locations, and here we pick up the data in 
$r_1\le R \le r_2$. The data points at $R=r_1$ and $R=r_2$ are indicated by 
squares and circles.  
The simulation box size for the local simulations adopted in \citet{suz10}
is $(x,y,z) = (H,4H,8H)$ (smaller than that used in \S \ref{sec:vst}), and 
the grid numbers are (32,64,256) for the low-resolution runs (open circles) 
and (64,128,512) for the high-resolution run (filled circle).

The saturation levels of $\langle \alpha \rangle_{t,\phi,z_{\rm tot}}$ are 
roughly correlated with the resolution (Table \ref{tab:res}). 
Case I-high (blue crosses; 
$H/\Delta z_{\rm mesh} =35$ at $R=r_1$ and 51 at $R=r_2$) seems to capture 
MRI turbulence well and gives $\langle \alpha \rangle\approx 0.02$, which is 
comparable to the level obtained in the local simulations. On the other hand, 
Case II-low (magenta triangles; $H/\Delta z_{\rm mesh} =8$) does not resolve 
small-scale turbulence and $\langle \alpha \rangle$ is far below the floor 
value ($=8\times 10^{-3}$; solid line) based on the local simulations. 
In the outer region of Case II-high (red diamonds; $H/\Delta z_{\rm mesh} = 16$), 
which is close to the red open circle at $R=r_2$, the turbulence is still 
supposed to be in the developing phase. 

It seems that the global simulations do not show correlations of 
the saturated $\langle \alpha \rangle$ with the net vertical field strength. 
However, when examining the results of Cases I-high and I-low, we 
should carefully take into account the radial change of the resolution. 
For instance, in Case I-high (blue crosses), $\langle \alpha \rangle 
(r_2)\approx \langle \alpha \rangle (r_1)$ ({\it squares} for $R=r_1$ and 
{\it circles} for $R=r_2$), although the net vertical 
field strength at $R=r_2$ is weaker than at $R=r_1$ ($\langle \beta_z 
\rangle(r_2) > \langle \beta_z \rangle(r_1)$). This is partly because 
the scale height is resolved by the larger number of grid points at $R=r_2$ 
($H/\Delta_{\rm mesh} z=51$) than at $R=r_1$ ($H/\Delta z_{\rm mesh}=35$); a positive 
correlation of $\langle \alpha \rangle$ might be smeared out by the dependence 
on the resolution.

\subsubsection{Azimuthal Power Spectra}
\label{sec:apsb}

\begin{figure*}[t]
\begin{center}
\includegraphics[height=0.37\textheight,width=0.8\textwidth]{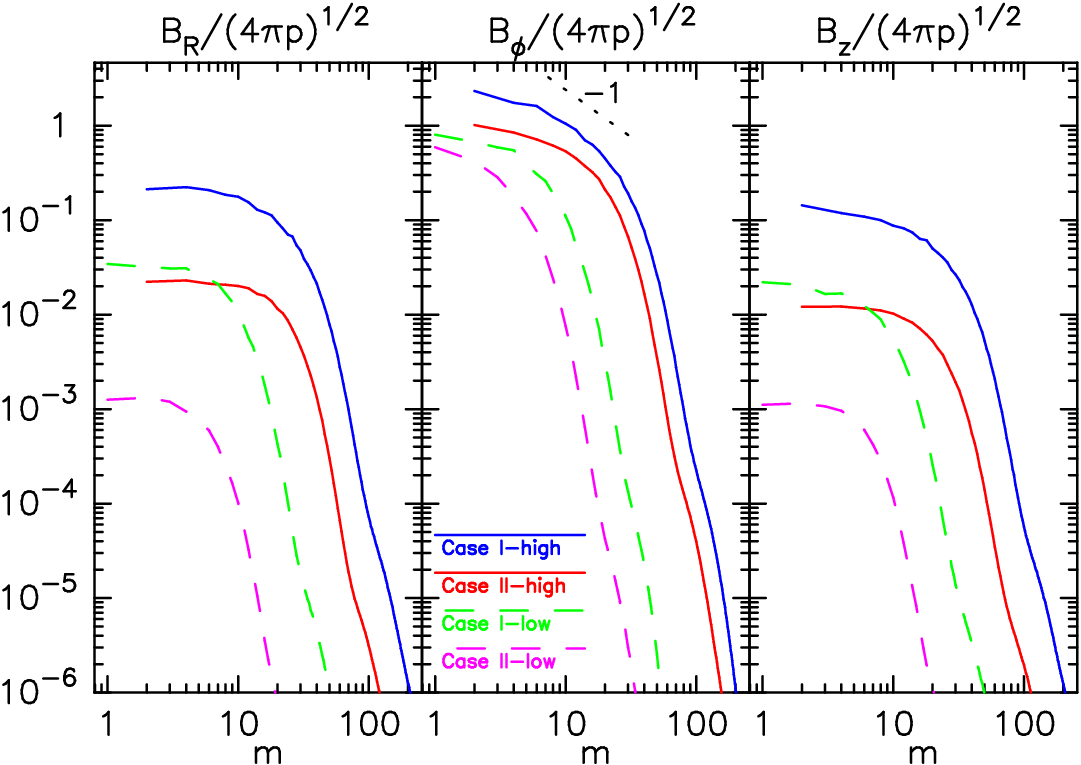}
\end{center}
\caption{Azimuthal power spectra of nondimensional magnetic fields,
$B/\sqrt{4\pi p}$, around the midplane of Case I-high ({\it blue solid}), 
Case II-high ({\it red solid}), Case I-low ({\it green dashed}), and 
Case II-low ({\it magenta dashed}). From left to right, the $R$, $\phi$, and 
$z$ components are displayed. The data are averaged over 
$\Delta z_{\rm mid}$, $\Delta R = r_1\Rightarrow r_2$, and $\Delta t_{\rm ave}$.
The low resolution runs (Cases I-low \& 
II-low) cover $m=1-64$, and the high resolution runs (Cases I-high \& II-high) 
cover $m=2-256$.}
\label{fig:apsb}
\end{figure*}

In Figure \ref{fig:apsb} we present azimuthal power spectra of each component 
of the magnetic fields. We first take the Fourier 
transformation of $B_i/\sqrt{4\pi p}$ ($i=R,z,\phi$) by using  
Equation (\ref{eq:aps}). Then we derive the power spectrum
from Equation (\ref{eq:apsave}) by taking the average over 
$\Delta z_{\rm mid}$ ($\pm H$ around the midplane; Equation \ref{eq:dzmid}), 
$\Delta R = r_1\Rightarrow r_2$, and $\Delta t_{\rm ave}$. 
The azimuthal mode is covered from $m_{\rm l}=1$ to $m_{\rm h}=64$ 
in the low-resolution runs that cover the full $2\pi$ disk by 128 grid points.  
On the other hand, in the high-resolution runs that treat the half ($\pi$) disk 
by 256 grid points, the azimuthal mode is covered from $m_{\rm l}=2$ to 
$m_{\rm h}=256$. Since regions close to $m_{\rm h}$ are strongly affected 
by numerical dissipation, we focus on the regions in 
$m\lesssim m_{\rm h}/10$.  

If MRI dominantly contributes to the generation of turbulent magnetic 
field, the injection scale is supposed to be comparable to 
the wavelength, $\lambda_{\rm max}$, of the most unstable mode 
(Equations \ref{eq:mrimax}). The corresponding injection scale in terms of 
mode $m_{\rm inj}$ can be estimated as 
\begin{eqnarray}
m_{\rm inj} &=& Rk_{\phi,{\rm inj}} \approx R\frac{2\pi}{\lambda_{\rm max}}
\approx \frac{2\pi/\Delta \phi }{Q_{\phi}} \nonumber \\
&\approx& 50 \left(\frac{2\pi/\Delta \phi}{512}\right) 
\left(\frac{Q_{\phi}}{10}\right)^{-1},
\label{eq:minj}
\end{eqnarray}
where we use the relation, $\lambda_{\rm max}\approx Q_{\phi}
\Delta l_{\phi}=Q_{\phi} R \Delta \phi$ (Equations \ref{eq:mrimax} and 
\ref{eq:qfdfn}), and the normalization in the second line is done for 
typical values of Case II-high.
This estimate shows that the energy injection is from high $m$ modes
(small scales).

The $R$ (left panel of Figure \ref{fig:apsb}) and $z$ (right panel) 
components show flat spectra in $m\lesssim m_{\rm h}/10$.  
The $\phi$ component (middle panel) shows slightly steeper spectra with 
$\propto m^{-1}$, 
probably because large-scale (small $m$) fields are amplified by the winding 
due to the differential rotation. 
These obtained power spectra are shallower than theoretical 
predictions based on incompressible MHD turbulence consisting of \Alfvenic 
wave packets \citep[e.g.,][]{gs95,cl03}. For example, \citet{gs95} show that 
well-developed \Alfvenic turbulence gives anisotropic power-law indices with 
respect to a background magnetic field with power $\propto k_{\perp}^{-5/3}$ and 
$\propto k_{\parallel}^{-2}$, where $k_{\perp}$ and $k_{\parallel}$ are 
wave numbers perpendicular and parallel to the background field. 
In the present global disk simulations, the magnetic fields are dominated 
by the $\phi$ component. Thus, we expect $m=Rk_{\phi}\approx Rk_{\parallel}$, 
and power $\propto m^{-2}$. A unique character of MRI in accretion disks 
is that the turbulent energy is injected from small scales 
(Equation \ref{eq:minj}). This is in contrast to the above picture for 
\Alfvenic turbulence, in which the energy is injected from a large scale 
and cascades to smaller scales. 
This difference can explain the obtained shallow power spectra of 
the magnetic fields in the accretion disks.

\subsection{Velocity \& Density Fluctuations}
We examine fluctuations of velocity and density here. 
Compared to magnetic fields, extracting the fluctuation components of 
velocity and density in global simulations is not straightforward, because 
of ambiguities in measuring the ``average'' quantities. 
In this paper, we use the $\phi$- and time-integrated quantities 
as the average values. 
We begin with vertical structures at $R=r_1(=5r_{\rm in})$ and later 
inspect radial profiles in the same manner as in the previous subsection 
for the magnetic fields. 

\subsubsection{Vertical Structure at $R=5r_{\rm in}$}
\label{sec:vsvrho}
\begin{figure*}[t]
\includegraphics[height=0.45\textheight,width=0.9\textwidth]{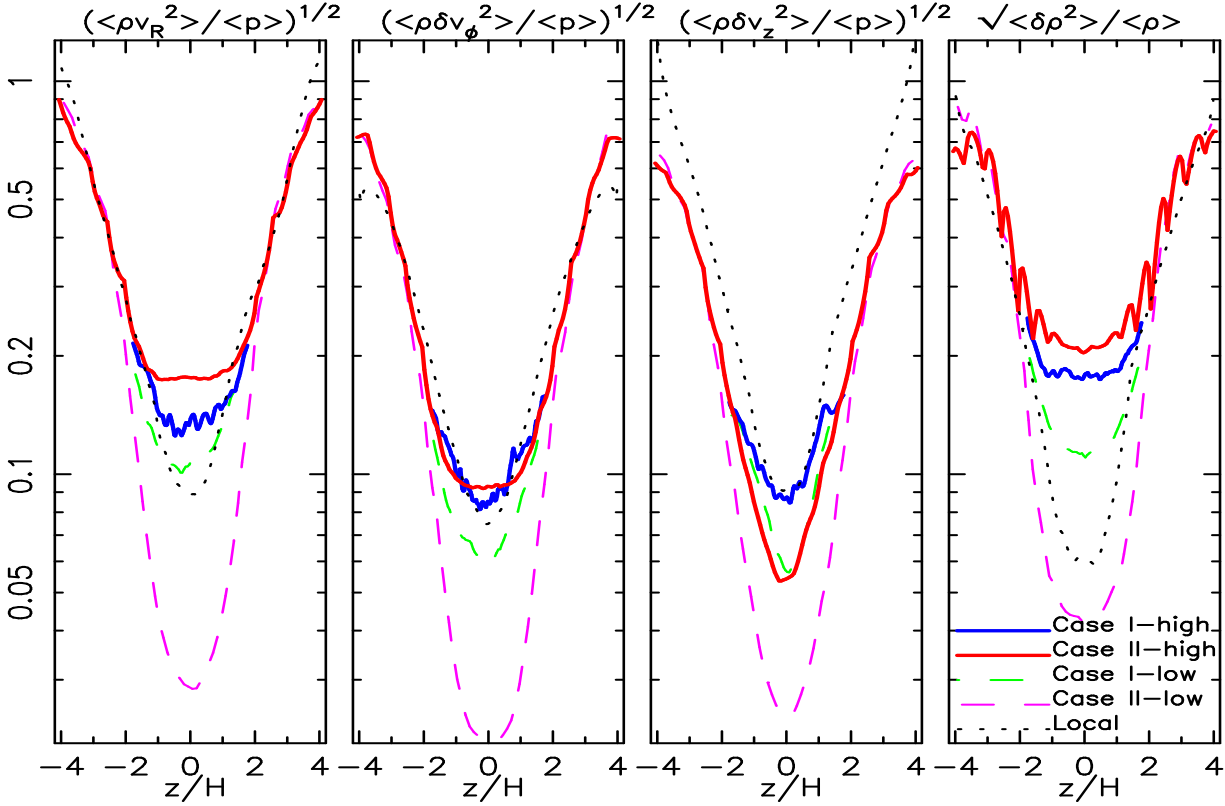}
\caption{Vertical structures of velocity and density fluctuations of 
Case I-high (blue solid), Case II-high (red solid), Case I-low (green 
dashed), and Case II-high (magenta dashed) at $R=r_1(=5 r_{\rm in})$. 
For comparison, the result of the local shearing box simulation is also 
shown ({\it black dotted}). 
From left to right are shown $\sqrt{\langle v_R^2\rangle_{t,\phi}(r_1,z)}
/c_{\rm s}$, $\sqrt{\langle \delta v_\phi^2\rangle_{t,\phi}(r_1,z)}/c_{\rm s}$, 
$\sqrt{\langle \delta v_z^2\rangle_{t,\phi}(r_1,z)}/c_{\rm s}$, and 
$\langle \delta \rho/\rho \rangle_{t,\phi}(r_1,z)$ 
(See Equations \ref{eq:dv} \& \ref{eq:drho}). }
\label{fig:zstdvdrh}
\end{figure*}
We evaluate velocity fluctuations of the $i$--th component normalized by 
the local sound speed from the simulations by taking density weighted 
averages,  
\begin{equation}
\frac{\sqrt{\langle \delta v_i^2\rangle_{t,\phi}(r_1,z)}}{c_{\rm s}} 
\equiv \sqrt{\frac{\langle \rho \delta v_i^2\rangle_{t,\phi}(r_1,z)}
{\langle p \rangle_{t,\phi}(r_1,z)}}, 
\label{eq:dv}
\end{equation}
where for the $i=R$ component we simply use $\delta v_i = v_R$, and for 
the $i={\phi}$ component $\delta v_{\phi} = v_{\phi}-\langle \rho v_{\phi} 
\rangle_{t,\phi} / \langle \rho \rangle_{t,\phi}$ 
(Equation \ref{eq:vshftave}).  
As for the $i=z$ component we use the same subtraction from the averaged 
value, 
\begin{equation}
\delta v_z = v_z - \langle v_z\rangle_{t,\phi} = v_z - \frac{\langle \rho v_z
\rangle_{t,\phi}}{\langle \rho \rangle_{t,\phi}}
\label{eq:dvzave}
\end{equation}
to remove the effect of the disk winds, which is not negligible 
in the regions near the surfaces (\S \ref{sec:dwd}). 
We evaluate density fluctuations as the root mean squared difference from 
the time- and $\phi$-averaged density:
\begin{equation}
\langle \frac{\delta \rho}{\rho}\rangle_{t,\phi}(r_1,z) \equiv
\sqrt{\frac{\langle (\rho(t,r_1,\phi,z)-\langle\rho\rangle_{t,\phi}(r_1,z))^2
\rangle_{t,\phi}(r_1,z) }{(\langle\rho\rangle_{t,\phi}(r_1,z))^2}}. 
\label{eq:drho}
\end{equation}
The time averages are again taken over $\Delta t_{\rm ave}$ 
(Table \ref{tab:models}). 

The four panels of Figure \ref{fig:zstdvdrh} display the vertical structures of 
the three components of the velocity fluctuations and the density 
fluctuations. The four cases of the global simulations are compared with the 
local shearing box simulation (black dotted lines).

Three cases of the global simulations (all except Case II-low) 
give similar trends of the velocity fluctuations; the $R$ component dominates 
the other components and the total fluctuations $\sqrt{\langle \delta v^2
\rangle} /c_{\rm s} \approx 0.1-0.2$ at the midplane increase towards 
the surfaces, where $\delta v^2 = v_R^{2} + \delta v_{\phi}^{2} + \delta v_z^{2}$. 
The fluctuation component is larger than the mean component of the mass 
flows, 
$\langle \rho v\rangle/\langle \rho \rangle$ (\S \ref{sec:dwd} 
\& \ref{sec:rf}), by more than 
an order of magnitude near the midplane.    
Case II-low shows quite small velocity fluctuations at the midplane because 
MRI-triggered turbulence is not well developed there because of the 
insufficient resolution.

The $R$ component of the velocity fluctuations in the midplane region of all 
the global cases except for Case II-low is systematically larger than the value 
obtained in the local simulation, while the $z$ component is smaller. 
The global simulations can handle net radial flows, which cannot 
be taken into account in the local shearing box approximation. Such radial 
flows contribute to the obtained $\sqrt{\langle v_R^2\rangle}/c_s$, in addition 
to the pure fluctuating component. 

The velocity fluctuations of Case II-high give a level similar to 
that of Case I-high, whereas the detailed profile is slightly different in 
each component.
This is in contrast to the results where Case II-high gives a lower 
saturation of the magnetic field (Figures \ref{fig:zstBsq} 
\& \ref{fig:zstBBp}). This implies that global mass flows involving 
the vertical differential rotation
contribute to the velocity fluctuations. 


\begin{figure*}[t]
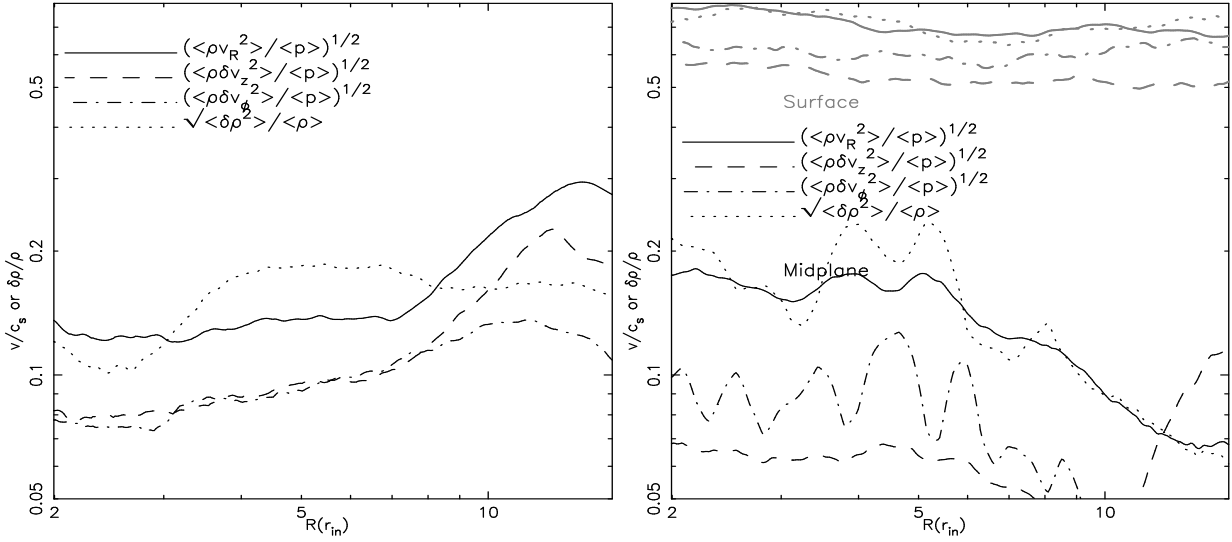

\includegraphics[height=0.3\textheight,width=0.45\textwidth]{rdc-snp-prj2tsum2zsum10_3_0.3s0.7r.ps}
\includegraphics[height=0.3\textheight,width=0.45\textwidth]{rdc-snp-prj2tsum2zsum11_5_0.3s0.7r.ps}
\caption{Radial structures of the velocity and density perturbations of 
Case I-high ({\it left}) and Case II-high ({\it right}). 
$\sqrt{\langle v_R^2\rangle_{t,\phi,z}(r_1)}/c_{\rm s}$ ({\it solid}), 
$\sqrt{\langle \delta v_\phi^2\rangle_{t,\phi,z}(r_1)}/c_{\rm s}$ ({\it dashed}), 
$\sqrt{\langle \delta v_z^2\rangle_{t,\phi,z}(r_1)}/c_{\rm s}$ ({\it dot-dashed}), 
and $\langle \delta \rho/\rho \rangle_{t,\phi,z}(r_1)$ (dotted) are plotted 
together in each panel. For Case II-high ({\it right}) the fluctuations 
integrated in the surface regions, $\Delta z_{\rm sfc}$ (Equation 
\ref{eq:dzsfc}), are shown ({\it thick gray}), in addition to those 
integrated around the midplane, $\Delta z_{\rm mid}$ ({\it black}). 
In Case I-high the only data of the 
midplane integration are shown. }
\label{fig:dvdrh}
\end{figure*}

The density fluctuations (right-most panel of Figure \ref{fig:zstdvdrh})
of all the global cases except Case II-low are considerably larger than 
that of the local simulation around the midplane. In particular, the 
high-resolution runs give quite large 
$\langle \frac{\delta \rho}{\rho}\rangle \approx 0.2$ around the midplane.
Case I-high and Case II-high give similar $\langle \frac{\delta \rho}
{\rho}\rangle$, although Case II-high gives smaller $\langle B^2\rangle$ 
around the midplane (Figures \ref{fig:zstBsq} \& \ref{fig:zstBBp}). 
This implies that in Case II-high global effects such as the 
vertical differential rotation contribute to the density fluctuations 
in addition to the MRI. As will be examined by using a power 
spectrum in \S \ref{sec:apsv}, the large value of 
$\langle \frac{\delta \rho}{\rho}\rangle$ in Case II-high comes
from a large-scale structure. We infer that there is a connection between 
the vertical differential rotation and the large-scale density structure, which 
will be the subject of our future work. 

In the context of the evolution of protoplanetary 
disks, such large density fluctuations greatly affect the dynamics of solid 
particles and subsequent planet formation \citep{np04,oo13a,oo13b}. 
Note, however, that non-isothermal local calculations show that the ratio of 
specific heats also affects $\frac{\delta \rho}{\rho}$ \citep{san04,is13}; 
realistic thermal physics is important in determining actual values of 
$\frac{\delta \rho}{\rho}$.

\subsubsection{Radial Profile}
\label{sec:rpvrho}

We examine the radial dependences of the velocity and density fluctuations by 
using Equation (\ref{eq:qtavtpz}). 
The averages are taken with density weighted for the velocity fluctuations, 
\begin{equation}
\frac{\sqrt{\langle \delta v_i^2\rangle_{t,\phi,z}(R)}}{c_{\rm s}} \equiv 
\sqrt{\frac{\langle \rho \delta v_i^2\rangle_{t,\phi,z}(R)}
{\langle p \rangle_{t,\phi,z}(R)}}, 
\label{eq:dvrd}
\end{equation}
and with the subtraction from the time- and $\phi$-averaged quantity for the 
density fluctuations, 
\begin{equation}
\langle \frac{\delta \rho}{\rho}\rangle_{t,\phi,z}(R) \equiv
\sqrt{\frac{\langle (\rho(t,R,\phi,z)-\langle\rho\rangle_{t,\phi}(R,z))^2
\rangle_{t,\phi,z}(R) }{(\langle\rho\rangle_{t,\phi,z}(R))^2}}, 
\label{eq:drhord}
\end{equation}
in a manner similar to that for the vertical profiles (Equations \ref{eq:dv} \& 
\ref{eq:drho}).
Figure \ref{fig:dvdrh} compares the fluctuations of velocity and density of 
Case I-high (left panel) and Case II-high (right panel). 
In Case I-high, we only 
show the values averaged in the midplane region, $\Delta z_{\rm mid}$ 
(Equation \ref{eq:dzmid}). In Case II-high, we also plot 
the values averaged in the surface regions over $\Delta z_{\rm sfc}$, 
Equation (\ref{eq:dzsfc}).  
Since Case II covers the vertical box from $z_{\rm bot}=-4H$ to $z_{\rm top} 
= 4H$, the integration over $\Delta z_{\rm sfc}$ corresponds to the sum of 
the integrations in the top and bottom surface regions between $\pm 3H$ 
and $\pm 4H$. 

\begin{figure*}[t]
\begin{center}
\includegraphics[height=0.37\textheight,width=0.8\textwidth]{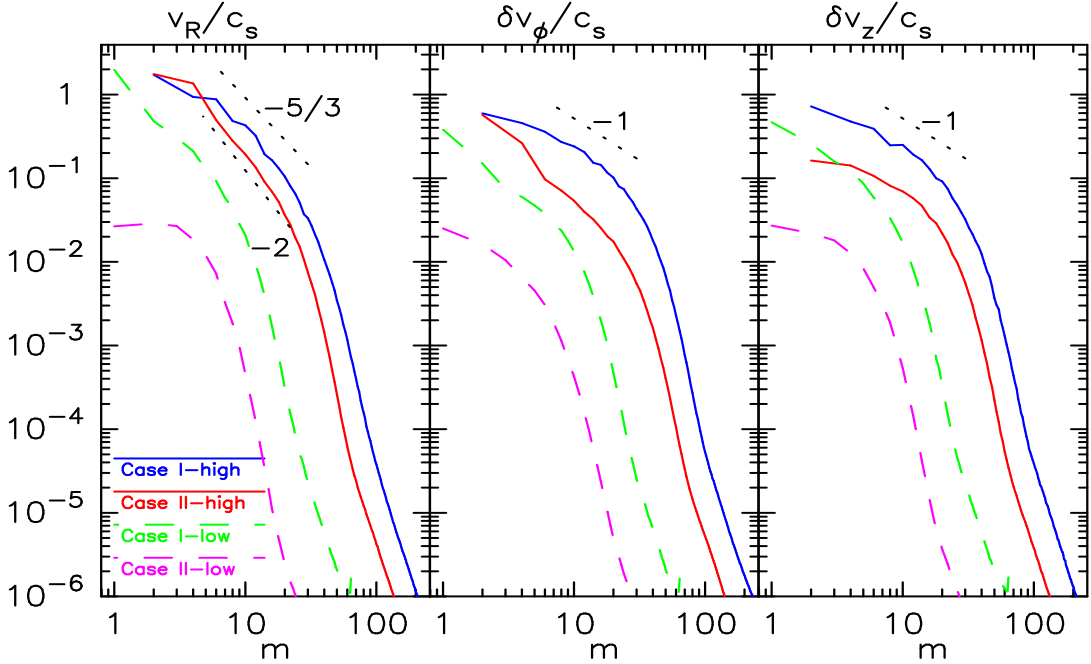}
\end{center}
\caption{Azimuthal power spectra of nondimensional velocity fields, 
$\delta v_i/c_{\rm s}$, 
around the midplane of Case I-high ({\it blue solid}), 
Case II-high ({\it red solid}), Case I-low ({\it green dashed}), and 
Case II-low ({\it magenta dashed}). From left to right, the $R$, $\phi$, and 
$z$ components are displayed. The data are averaged over 
$\Delta z_{\rm mid}$, $\Delta R = r_1\Rightarrow r_2$, and $\Delta t_{\rm ave}$.
The low resolution runs (Cases I-low \& 
II-low) cover $m=1-64$, and the high resolution runs (Cases II-low \& II-high) 
cover $m=2-256$.}
\label{fig:apsv}
\end{figure*}

Both cases show similar values of the velocity fluctuations in 
$R\lesssim 8 r_{\rm in}$; $\sqrt{\langle \delta v^2 \rangle }/c_{\rm s}\approx 
0.1-0.2$, with values mostly dominated by the $R$ component. 
These values are slightly larger than a typical value, $\approx 0.1$, 
obtained in global simulations without a net vertical flux \citep{flo11}. 
The difference might imply the importance 
of the net vertical field in the velocity fluctuations. 
In $R\gtrsim 10r_{\rm in}$ in Case II-high, the velocity fluctuations  
decline to $<0.1$ because the turbulence is still developing 
there. On the other hand, in $R\gtrsim 8r_{\rm in}$ in Case I-high, the 
velocity fluctuations increase. This trend is quite similar to that of 
$\frac{\langle B_{R}^2\rangle_{t,\phi,z_{\rm tot}}}{\langle B_{\phi}^2
\rangle_{t,\phi,z_{\rm tot}}}$ in Figure \ref{fig:BBrel}, which is a good 
indicator for measuring the role of MRI in turbulence \citep{haw11}. 

Both Cases I-high and II-high give quite large density fluctuations, 
$\langle \delta \rho / \rho \rangle_{t,\phi,z_{\rm mid}}(R) \approx 0.2$ 
in the midplane region (see also Figure \ref{fig:zstdvdrh} for the vertical 
structure). 
In particular, Case II-high shows wavy structure of $\langle \delta \rho 
/\rho \rangle_{t,\phi,z_{\rm mid}}(R)$, which is anticorrelated with 
$\sqrt{\langle \delta v_{\phi}^2\rangle_{t,\phi,z_{\rm mid}}(R)}
/c_{\rm s}$. Compared with the bottom panel of 
Figure \ref{fig:BBrel}, the wavy pattern is well correlated with 
$-2\langle B_R B_{\phi}\rangle_{t,\phi,z_{\rm tot}} 
/ \langle B^2 \rangle_{t,\phi,z_{\rm tot}}$ (Maxwell stress to 
magnetic pressure), which is a good indicator for MRI turbulence 
\citep{haw11}. In other words, the density perturbations are more strongly 
excited in the regions with higher activities of MRI turbulence. 
An interesting thing is that these regions remain for a rather long time
during $\Delta t_{\rm ave}$=600 inner rotations or 55 local rotations at 
$R=5r_{\rm in}$ in Case II-high, which might be related to zonal flows 
observed in local simulations \citep{joh09}.

\subsubsection{Azimuthal Power Spectra}
\label{sec:apsv}

\begin{figure}[h]
\includegraphics[height=0.3\textheight,width=0.4\textwidth]{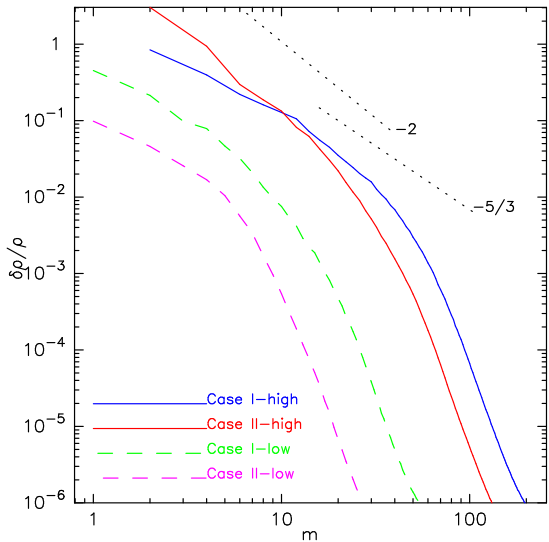}
\caption{Azimuthal power spectra of density perturbations, 
$\delta \rho/\rho$, 
around the midplane of Case I-high ({\it blue solid}), 
Case II-high ({\it red solid}), Case I-low ({\it green dashed}), and 
Case II-low ({\it magenta dashed}). The data are averaged over 
$\Delta z_{\rm mid}$, $\Delta R = r_1\Rightarrow r_2$, and $\Delta t_{\rm ave}$.
The low resolution runs (Cases I-low \& 
II-low) cover $m=1-64$, and the high resolution runs (Cases II-low \& II-high) 
cover $m=2-256$.}
\label{fig:apsrho}
\end{figure}

We inspect azimuthal power spectra of velocity and density perturbations in a 
manner similar to inspection of the magnetic fields in \S \ref{sec:apsb}. 
Figure \ref{fig:apsv} presents each component of the velocity power spectra. 
After taking the Fourier transformation of $\delta v_i/c_{\rm s}$ 
($i=R,z,\phi$) by Equation (\ref{eq:aps}), the power spectra 
are derived from Equation (\ref{eq:apsave}) by averaging over 
$\Delta z_{\rm mid}$ ($\pm H$ around the midplane; Equation \ref{eq:dzmid}), 
$\Delta R = r_1\Rightarrow r_2$, and $\Delta t_{\rm ave}$. Here the fluctuation
component is derived in the same way as in the analyses in real space 
(\S \ref{sec:vsvrho} \& \ref{sec:rpvrho}), 
$\delta v_i=(v_R,\delta v_{\phi},\delta v_z)$, by using Equations 
(\ref{eq:vshftave}) \& (\ref{eq:dvzave}). 
The azimuthal mode is covered from $m_{\rm l}=1$ to $m_{\rm h}=64$ 
in the low-resolution runs, and from $m_{\rm l}=2$ to 
$m_{\rm h}=256$ in the high-resolution runs.

Figure \ref{fig:apsv} presents the derived power spectra of the velocity 
perturbations. Compared to the power spectra of the magnetic fields 
(Figure \ref{fig:apsb}), all the cases except for Case II-low exhibit 
steeper slopes. The $R$ component of the high-resolution runs gives the 
spectra, $\propto m^{-5/3} - m^{-2}$. Here, $m^{-5/3}$ is the Kolmogorov 
scaling derived from incompressible hydrodynamical turbulence \citep{k41}, 
and $m^{-2}$ corresponds to the \citet{bur39}' type shock 
dominated spectrum. 

Figure \ref{fig:apsrho} displays the power spectra of the density 
perturbations. The obtained slopes are roughly similar to those for 
$v_R/c_{\rm s}$. An interesting feature is that in Case II-high the lowest 
$m=2$ mode largely dominates higher $m$ modes, compared to Case I-high. 
This implies the existence of a non-axisymmetric but small-$m$ (large-scale) 
zonal structure with a density bump or dip for a rather long time, 
$\sim \Delta t_{\rm ave}$ in Case II-high.

\subsection{Vertical Outflows \& Wave Phenomena}
\label{sec:dwd}

\begin{figure}
\includegraphics[height=0.5\textheight,width=0.4\textwidth]{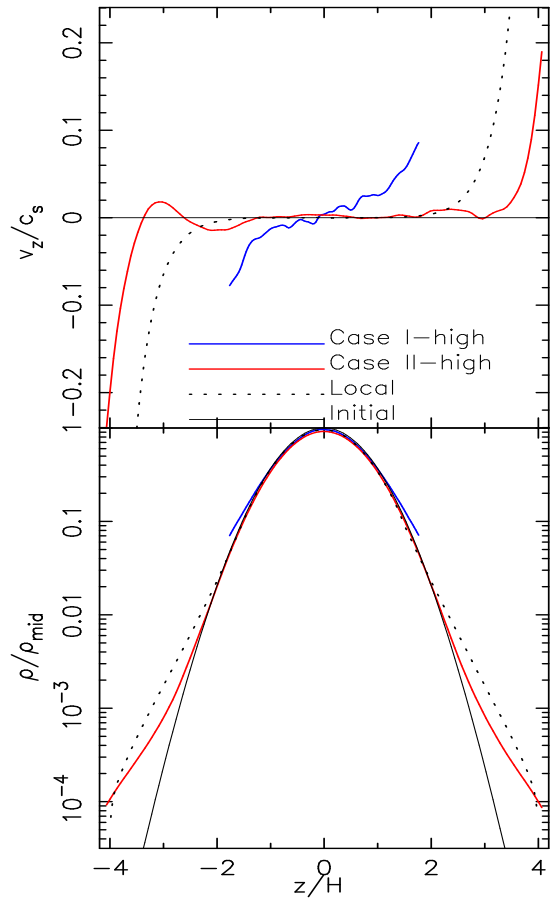}
\caption{Comparison of vertical velocities ({\it upper}) and densities 
({\it lower}) of Case I-high ({\it blue solid}) at $R=r_1(=5R)$, Case II-high 
({\it red solid}) at $R=r_1$, the local simulation ({\it black dotted}), 
and the initial condition ({\it black solid}) on $z/H$ . 
$v_z$ in the upper panel is normalized by the local sound speed, and 
$\rho$ in the lower panel is normalized by the time and $\phi$ averaged 
density at the midplane. }
\label{fig:dskwnd}
\end{figure}

\citet{si09} \& \citet{suz10} pointed out MRI turbulence in accretion disks 
could play a role in driving disk winds particularly in mass loading 
to the surface regions. Such vertical outflows were also reported 
by 2D axisymmetric global simulations \citep{sp01,pb03,mp09}. 
Recently, various aspects of relations between MRI turbulence and disk winds 
have been studied by both local simulations 
\citep{bs13a,bs13b,fro13,les13} and global simulations \citep{flo11}. 
In addition to MRI, Parker instability is also studied as a reliable mechanism 
in driving vertical outflows \citep{nis06,mac13}.

The upper panel of Figure \ref{fig:dskwnd} shows that disk winds are 
also observed in our global simulations. In Case I-high (solid) the disk winds 
are driven from the near-midplane regions, because the vertical extent 
of the simulation box is insufficient and only $\pm 1.8H$ at 
$R=r_1(=5r_{\rm in})$. 
As discussed in \citet{suz10} and \citet{fro13} by using local 
shearing boxes, the mass flux of the 
disk winds depends on the vertical box size; a smaller vertical size gives 
a larger mass flux. The result of Case I-high is consistent with this trend.

\begin{figure*}[t]
\begin{center}
\includegraphics[height=0.35\textheight,width=0.64\textwidth]{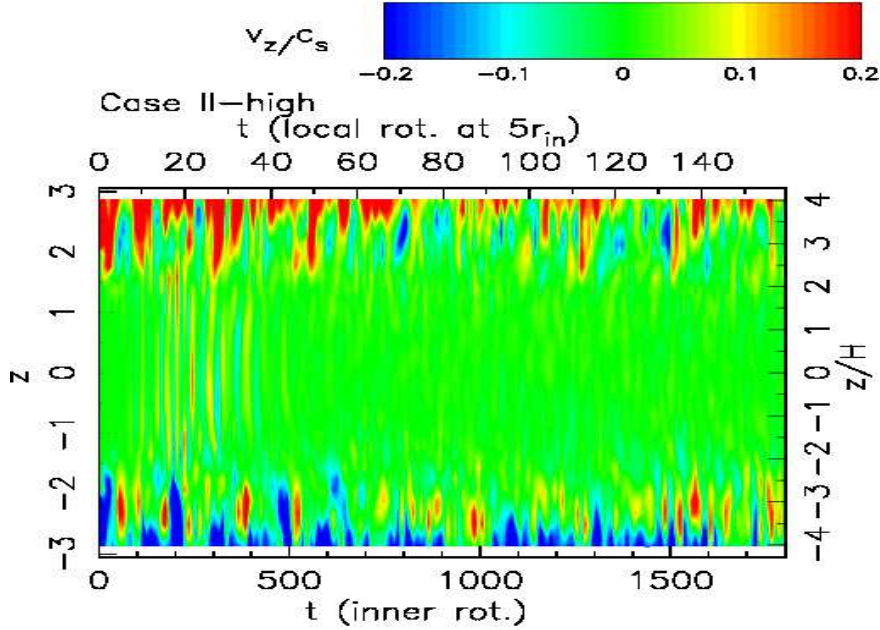}
\end{center}
\caption{$t$--$z$ diagrams of $\langle v_z\rangle_{\phi}(r_1,z)$ normalized 
by the sound speed at $R=r_1(=5r_{\rm in})$ for Case II-high. 
On the horizontal axis are shown time in units of the inner rotation 
({\it top} of panel) and time in units of the local rotation at 
$r=r_1$ ({\it bottom} of panel). 
On the vertical axis are shown $z$ ({\it left} to panel) and 
$z/H$ ({\it right} to panel).}
\label{fig:tzvz}
\end{figure*}

Case II-high (dashed lines in Figure \ref{fig:dskwnd}) has the same vertical 
box size $=\pm 4H$ 
as the local simulation. The onset positions of the disk winds in Case II-high 
are located at slightly higher altitudes than those of the local simulation. 
A reason for the difference is related to the intermittent natures of 
MRI-driven disk winds. In the local simulations \citep{si09,suz10}, we observed 
quasi-periodicity of the driving disk winds with 5-10 rotation times, 
caused by the breakups of channel-mode flows. In contrast, in Case 
II-high the intermittency is more random with time as in 
Figure \ref{fig:tzvz},  mainly because quasi-periodic channel 
flows seen in the local simulation are distorted by the vertical differential 
rotation.
During the time integration, $\Delta t_{\rm ave} =1200-1800$ inner rotations, 
the disk wind from the upper surface ceases for a while, which causes 
the slower onset of the disk wind from the upper surface in the time-averaged 
structure (upper panel of Figure \ref{fig:dskwnd}). 
However, during strong wind phases, the wind speed at the surfaces far 
exceeds the sound speed (Figure \ref{fig:edgeon}) and the onset 
heights are comparable to that observed in the local simulation.

The bottom panel of Figure \ref{fig:dskwnd} shows that the gas is lifted up 
near the surface regions from the initial density distribution; the simulated 
density structures deviate from the initial profile around the onset 
locations of the disk winds.   

\begin{figure}
\includegraphics[height=0.27\textheight,width=0.5\textwidth]{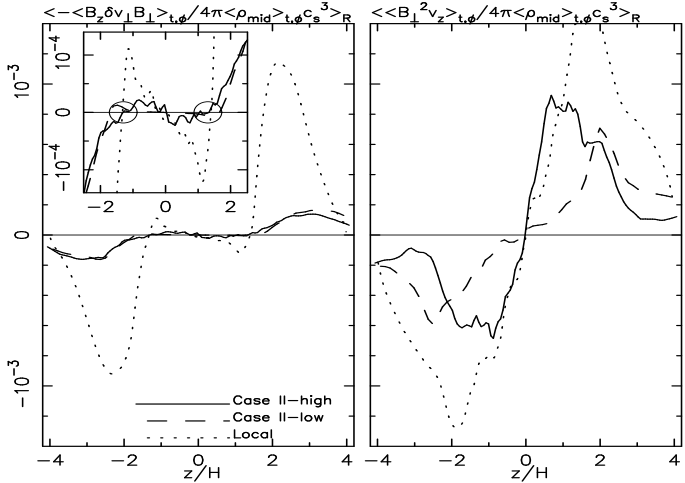}
\caption{Comparison of the vertical structures of the Poynting flux associated 
with magnetic tension ({\it left}) and Poynting flux associated with 
the magnetic pressure and the advected magnetic energy ({\it right}) 
of Case II-high ({\it solid}), Case II-low ({\it dashed}), and the local 
simulation ({\it dotted}). 
Here the values of the global simulations 
are averaged from $R=r_1$ to $r_2$ in addition to the time and $\phi$ averaging.
In the inset in the left panel a region 
around the midplane is zoomed in. The two circles at $z=\pm 1.3H$ indicate 
the injection regions of the tension-associated Poynting flux. }
\label{fig:pyflx}
\end{figure}

In \citet{si09}, we claimed that the disk winds are driven by 
the Poynting flux associated with the MHD turbulence from the results of 
the local shearing box simulations. 
We also found that the Poynting flux 
of the magnetic tension is injected from $z\approx \pm 1.5H$, which we call 
``injection regions'', toward the surfaces and the midplane because of the 
intermittent breakups of channel-mode flows. 
We inspect whether these characteristic features of the disk winds 
are also observed in the global simulations.  
 
We can write the $z$ component of energy flux as follows: 
\begin{equation}
\rho v_z\left(\frac{1}{2}v^2 + h + \Phi \right) + v_z \frac{B_{\perp}^2}{4\pi} 
-\frac{B_z}{4\pi}(v_{\perp}B_{\perp}),
\end{equation}
where $B_{\perp}^2 = B_R^2+B_{\phi}^2$ and $v_{\perp}B_{\perp} = v_R B_R 
+ v_{\phi}B_{\phi}$ \citep{si09}. 
The last two terms originate from the Poynting 
flux; the first term is related to the sum of magnetic pressure and advected 
magnetic energy \footnote{Here, the term $\frac{B_{\perp}^2}{4\pi}$ consists 
of work per time ({\it i.e.} power) done by magnetic pressure, 
$\frac{B_{\perp}^2}{8\pi}$ and magnetic energy, $\frac{B_{\perp}^2}{8\pi}$, 
advected with velocity, $v_z$}, and the second term is associated with magnetic 
tension.  
Here we pick up the fluctuation component of perpendicular velocity, 
$\delta \mbf{v_{\perp}}
=(v_R,\delta v_{\phi})$, as in Equations (\ref{eq:vshftave}). 
Then the fluctuation component of the Poynting flux with magnetic tension 
can be rewritten as 
\begin{equation}
-\frac{1}{4\pi} B_z \delta v_{\perp}B_{\perp}
= \rho v_{{\rm A},z}(\delta v_{\perp,+}^2-\delta v_{\perp,-}^2),
\end{equation}
where $\delta v_{\perp}B_{\perp}=v_R B_R + \delta v_{\phi}B_{\phi}$, 
$v_{{\rm A},z} = B_z/\sqrt{4\pi\rho}$, and $\delta v_{\perp,\pm} 
=\frac{1}{2}(\delta v_{\perp}\mp B_{\perp}/\sqrt{4\pi\rho})$ are 
Els\"{a}sser variables, which correspond to the amplitudes of \Alfven waves 
propagating in the $\pm B_z$ directions. 

Figure \ref{fig:pyflx} compares the vertical structures of the Poynting flux. 
Since we would like to study the structures at high altitudes, we present 
the results of Cases II-high (solid lines) and II-low (dashed lines) in 
comparison with the result of the local simulation (dotted line), and 
we do not show the results of Cases I-high and I-low, of which the vertical 
box sizes are $|z|<2H$. What are shown here are the Poynting flux 
with magnetic tension, $\langle-\langle B_z \delta v_{\perp} B_{\perp}
\rangle_{t,\phi}(z) / 4\pi\langle \rho_{\rm mid}\rangle_{t,\phi} c_{\rm s}^3\rangle_R$ ({\it left}),  
and the Poynting flux of the sum of magnetic pressure 
and advected energy, $\langle \langle B_{\perp}^2 v_z \rangle_{t,\phi}(z)/4\pi
\langle \rho_{\rm mid}\rangle_{t,\phi} c_{\rm s}^3\rangle_R$ 
(right panel), where we take the averages from $R=r_1$ to $r_2$ 
after normalizing by $\langle \rho_{\rm mid}\rangle_{t,\phi} c_{\rm s}^3$ 
at the midplane, in addition to the usual averages over $\Delta t_{\rm ave}$ 
and $\phi$ (Equation \ref{eq:qtavtp}). 

Figure \ref{fig:pyflx} shows that the global simulations (solid and dashed 
lines) give magnitudes for both components of the Poynting fluxes that are 
smaller than those of the local simulation (dotted line). 
Then, the mass fluxes of 
the disk winds in the global simulations are smaller than that of the 
local simulation. In the global simulations (solid and dashed lines) 
the Poynting fluxes with the magnetic tension ($-\langle B_z \delta v_{\perp} 
B_{\perp}\rangle / 4\pi$) give smaller contributions to the disk winds than 
the Poynting flux associated with the magnetic pressure and energy 
($\langle B_{\perp}^2 v_z\rangle/4\pi$), while in the local simulation 
(dotted lines) both components give comparable contributions.  

Although the magnitudes are smaller, the Poynting fluxes with the magnetic 
tension in the global simulations show qualitative vertical structures similar 
to that of the local simulation; at $z\approx \pm 1.3 H$ the solid and 
dashed lines cross $x=0$, indicated by the circles in the left panel 
of Figure \ref{fig:pyflx}. This indicates that from these regions the 
Poynting fluxes with the magnetic tension are injected in both directions.  

\begin{figure}
\begin{center}
\includegraphics[height=0.3\textheight,width=0.3\textwidth]{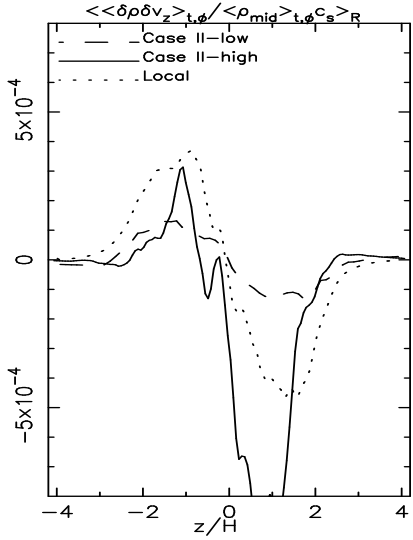}
\end{center}
\caption{Comparison of the vertical structures of the normalized energy flux 
of sound waves of Case II-high ({\it solid}), Case II-low ({\it dashed}), and 
the local simulation ({\it dotted}).}
\label{fig:sound}
\end{figure}

In \citet{si09} we also found sound waves traveling from the surface 
regions to the midplane, which is related to the injection regions of the 
Poynting flux with the magnetic tension. The energy flux of sound waves 
in static media is expressed as 
\begin{equation}
\delta \rho \delta v_z c_{\rm s}^2 = \rho c_{\rm s} (\delta v_{\parallel,+}^2 
- \delta v_{\parallel,-}^2), 
\end{equation}
where $\delta \rho = \rho - \langle \rho \rangle_{t,\phi}$, 
$\delta v_z = v_z - \langle \rho v_z\rangle_{t,\phi}/\langle \rho 
\rangle_{t,\phi}$ (Equation \ref{eq:dvzave}), and 
$\delta v_{\parallel,\pm} = \frac{1}{2}\left(\delta v_z \pm c_{\rm s} 
\frac{\delta \rho}{\rho}\right)$ are the amplitudes of sound waves propagating 
in the $\pm z$ directions. 
Figure \ref{fig:sound} compares the energy flux of sound waves,  
$\langle \langle \delta \rho \delta v_z\rangle_{t,\phi}(z)c_{\rm s}^2/\langle 
\rho_{\rm mid}\rangle_{t,\phi}c_{\rm s}^3\rangle_R=
\langle \langle \delta \rho \delta v_z
\rangle_{t,\phi}(z)/\langle \rho_{\rm mid}\rangle_{t,\phi}c_{\rm s}\rangle_R$, 
where the average and normalization are taken in the same manner as for 
the Poynting flux. The global simulations show that the sound waves are 
directed toward the midplane from the surface regions, which is consistent with 
the trend obtained in the local simulations.  The sound waves themselves 
carry mass flux ($\delta \rho \delta v_z$), and the direction (to the 
midplane) is opposite the direction (to the surfaces) of the mass flux 
carried by the disk winds. 
In protoplanetary disks, these sound waves possibly play a role in the dynamics 
of solid materials \citep{si09,oh11,oh12}.  

\begin{figure}[h]
\includegraphics[height=0.3\textheight,width=0.45\textwidth]{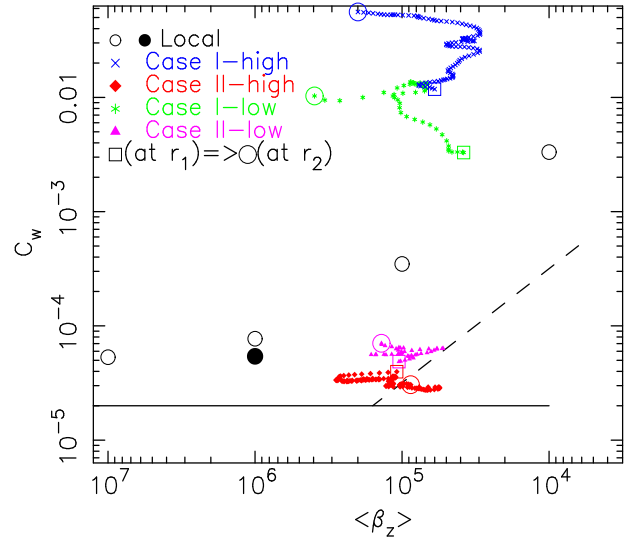}
\caption{Dependence of normalized mass flux, $C_{\rm w}$, of the disk winds 
on plasma $\langle \beta_{z} \rangle_{t,\phi,z_{\rm tot}}$ for net vertical magnetic 
fields in $r_1<R<r_2$. 
Multiple data points for each case of the global simulations correspond
to different radial locations. Colored squares and circles indicate the data 
at $R=r_1$ and $R=r_2$ of each case. Case I-high ({\it blue crosses}), 
Case II-high ({\it red diamonds}), Case I-low ({\it green asterisks}), 
and Case II-low ({\it magenta triangles}) are compared with the local 
simulations by \citet[][;{\it open circles} for low-resolution runs and 
{\it filled circle} for high-resolution run]{suz10}. The solid line is 
a 'floor' value based on the local simulations and the dashed line 
is a fit for the increasing trend of $\langle C_{\rm w}\rangle$ \citep{suz10}.}
\label{fig:bzcw}
\end{figure}

Before concluding this subsection, we examine the dependence of the mass flux 
of the disk winds on the net vertical magnetic field strength. 
Figure \ref{fig:bzcw} presents the sum of the nondimensional mass fluxes of 
the disk winds from the top (subscript ``top'') and bottom (subscript ``bot'') 
surfaces,
\begin{equation}
\langle C_{\rm w}\rangle_{t,\phi}(R) 
= \frac{\langle (\rho v_z)_{\rm top}-(\rho v_z)_{\rm bot}\rangle_{t,\phi}}
{\langle \rho_{\rm mid} \rangle_{t,\phi} c_{\rm s}}, 
\end{equation}
as a function of $\langle \beta_z \rangle_{t,\phi,z_{\rm tot}}(R)$ 
(Equation \ref{eq:bttpz}), 
corresponding to (the inverse of) the net vertical field strength. 
Note that the negative sign for $(\rho v_z)_{\rm bot}$ is because $v_z<0$ 
corresponds to an outflow from the bottom surface.

Cases I-high and I-low give systematically larger $\langle C_{\rm w}\rangle$ 
because the vertical box size in units of scale height in these cases is 
smaller ($\pm 1.8H$ at $R=r_1$ and $\pm 1.3H$ at $R=r_2$) and the gas streams 
out rapidly, as shown in Figure \ref{fig:dskwnd}. Since this is an unphysical 
reason, we mainly discuss the results of Cases II-high and II-low here.   
Contrary to the $\langle \alpha \rangle_{t,\phi,z_{\rm tot}}(R)$ values 
(Figure \ref{fig:bzalp}), the mass flux shows almost no dependence on the 
resolution;   
the lower-resolution case (Case II-low; magenta triangles) rather gives 
slightly higher $\langle C_{\rm w}\rangle$. This is because the disk winds 
are driven from the surface regions where 
both the low-resolution and high-resolution runs give the comparable 
magnetic field strengths (Figure \ref{fig:zstBsq}).
Compared to the local simulation, Cases II-high and II-low give smaller 
$\langle C_{\rm w}\rangle$ because of the more random time dependency 
discussed in Figure \ref{fig:dskwnd}.

\subsection{Radial Flows}
\label{sec:rf}

\begin{figure}
\includegraphics[height=0.28\textheight,width=0.5\textwidth]{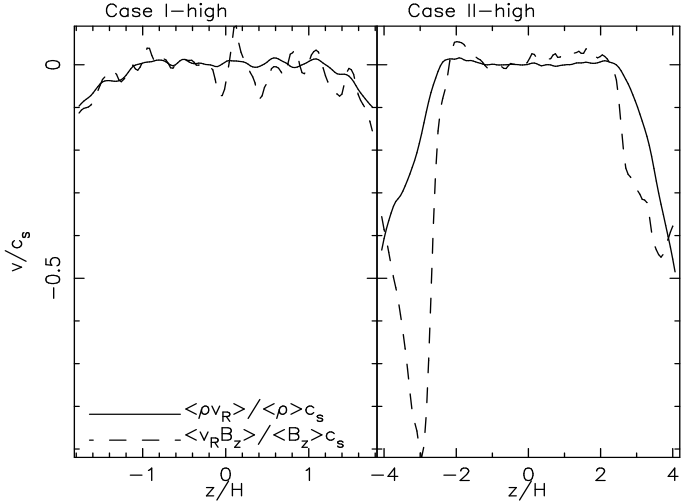}
\caption{Vertical structure of time and $\phi$ averaged radial 
velocity at $R=r_1(=5 r_{\rm in})$ normalized by the local 
sound speed of the gas ({\it solid}; Equation \ref{eq:rdfl_gas}) 
and the vertical magnetic field ({\it dashed}; Equation \ref{eq:rdfl_B}) 
of Case I-high ({\it left}) and Case II-high ({\it right}). }
\label{fig:zstacc}
\end{figure}

\begin{figure}
\includegraphics[height=0.28\textheight,width=0.5\textwidth]{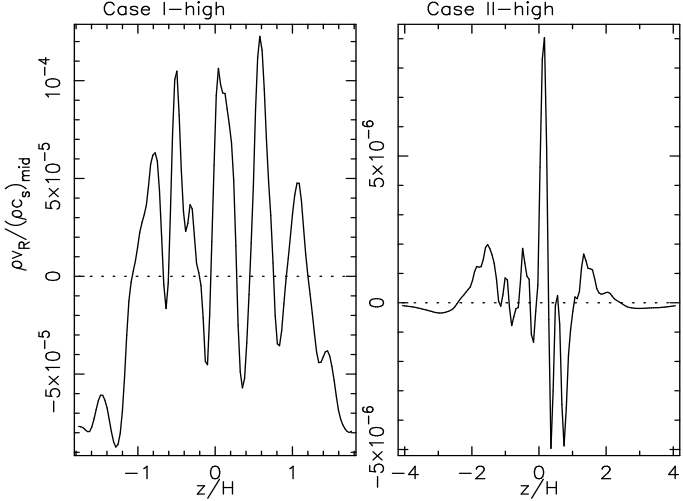}
\caption{Vertical structure of time and $\phi$ averaged radial 
mass flux, $\rho v_R$, at $R=r_1(=5 r_{\rm in})$ normalized 
by the local sound speed multiplied by the density at the midplane, 
$(\rho c_{\rm s})_{\rm mid}$, ({\it solid}) of Case I-high ({\it left}) 
and Case II-high ({\it right}). 
The zero level is shown by the dotted line.}
\label{fig:zstacc2}
\end{figure}

In this subsection, we inspect the radial motions of the gas and the 
vertical magnetic field of the simulated accretion disks. 
Figure \ref{fig:zstacc} presents the vertical structures of the radial 
velocities of Case I-high ({\it left}) 
and Case II-high ({\it right}). We derive the radial velocity of the gas by 
density-weighted average over time, $\Delta t_{\rm ave}$, and $\phi$ at 
$R=r_1(=5r_{\rm in})$, 
\begin{equation}
\langle \frac{v_{R,{\rm gas}}}{c_{\rm s}}\rangle_{t,\phi} (r_1,z)
= \frac{\langle \rho v_R \rangle_{t,\phi}
(r_1,z)}{\langle \rho \rangle_{t,\phi}(r_1,z)c_{\rm s}}. 
\label{eq:rdfl_gas}
\end{equation}
As for the movement of the vertical magnetic field, we adopt a similar 
averaging procedure: 
\begin{equation}
\langle \frac{v_{R,B_z}}{c_{\rm s}}\rangle_{t,\phi}(r_1,z) = \frac{\langle B_z v_R 
\rangle_{t,\phi}(r_1,z)}{\langle B_z \rangle_{t,\phi}(r_1,z)c_{\rm s}}. 
\label{eq:rdfl_B}
\end{equation}
Both Cases I-high and II-high show that the mass accretions (solid lines) 
take place near the surfaces, while in the midplane regions the radial 
velocities are quite small with the averages being slightly positive with 
fluctuations. This is because the Maxwell stresses ($-\langle B_R B_{\phi}
\rangle/4\pi\langle p\rangle$) are larger in the surface regions (Figure 
\ref{fig:zstBBp}), and the outward transport of the angular momentum is 
more effective there. 

To see the mass flows, we plot the vertical structure of $\langle 
\rho v_R\rangle_{t,\phi}(r_1,z)$ normalized by 
$[\langle \rho \rangle_{t,\phi}(r_1)c_{\rm s}]_{\rm mid}$ at the midplane 
in Figure \ref{fig:zstacc2}.  
As expected, one can observe accretion in the surface regions and fluctuating 
outward and inward mass flows near the midplane in both Cases I-high 
(left panel) and II-high (right panel). In Case I-high the accretion in 
the surface regions dominates and the vertically integrated mass flux, 
$\int dz \langle \rho v_R\rangle_{t,\phi}(r_1,z)$, is directed inward. 
On the other hand, in Case II-high the magnitude of the mass flux near 
the surfaces is small because of the low density and is dominated by the 
outward mass flows in $1<|z/H|<2$. Thus the direction of the vertically 
integrated mass flux is outward. 

The radial velocities of the vertical magnetic fields roughly follow the  
trend of the gas flows in both cases. 
However, $\langle \frac{v_{R,{\rm gas}}}{c_{\rm s}}
\rangle$ and $\langle \frac{v_{\small R,B_z}}{c_{\rm s}}\rangle$ do not exactly 
follow each other, which means that the gas and the vertical magnetic flux 
are not exactly frozen-in because of magnetic reconnections. Since our 
numerical code adopts the ideal MHD equations, this is purely a numerical 
effect; magnetic reconnections occurs because of small-scale turbulent fields 
in the sub-grid scales. 

\begin{figure}
\includegraphics[height=0.25\textheight,width=0.4\textwidth]{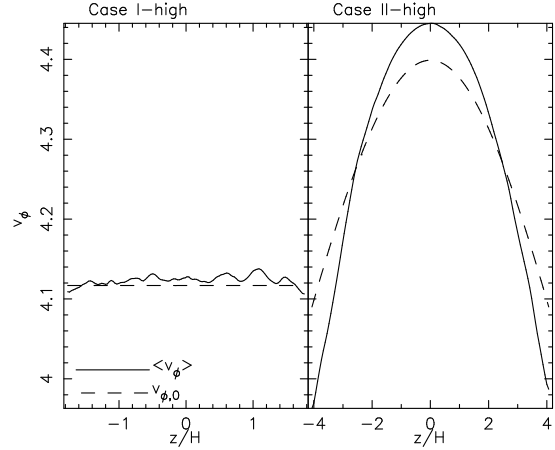}
\caption{Vertical structure of time and $\phi$ averaged azimuthal 
velocity at $R=r_1(5r_{\rm in})$ ({\it solid}; Equation \ref{eq:azvl}) 
of Case I-high ({\it left}) and Case II-high ({\it right}) in comparison 
with the initial value ({\it dashed}).}
\label{fig:zstvph}
\end{figure}

What we observe in the simulations is similar to the layered accretion 
which is proposed for the evolution of protoplanetary disks \citep{gam96}. 
In that scenario, the accretion is inhibited around the midplane because 
a so-called dead zone, which is an inactive region with respect to MRI 
due to insufficient ionization, forms there. On the other hand, the layered 
accretion obtained in our simulations occurs simply because of the vertical 
structure of the $\alpha$ stress. Figure \ref{fig:zstvph} 
compares the time- and $\phi$-averaged vertical structure of density-weighted 
azimuthal velocity (solid line), 
\begin{equation}
\langle v_{\phi} \rangle_{t,\phi}(r_1,z) = \frac{\langle \rho v_{\phi} 
\rangle_{t,\phi}(r_1,z)}{\langle \rho \rangle_{t,\phi}(r_1,z)} , 
\label{eq:azvl}
\end{equation}
with the initial value (dashed line). 
In Case II-high (right panel), the rotation speed becomes slower in the 
surface regions than the initial state because a larger angular momentum 
is extracted there as a result of the larger $\alpha$ stress. 
This further leads to the layered accretion discussed so far.

\citet{tl02} also derived a similar vertical profile for the 
layered radial velocity of the gas, mainly because of 
the vertical differential rotation, which is considered in our Case II. 
\citet{tl02} as well as \citet{kg04} and \citet{jac13} apply 
this radial velocity profile to 
the outward migration of dust particles, which is consistent with 
observed crystalline solid particles in the outer regions of protoplanetary 
disks \citep{bou08}. 
Our simulations also support the radial outward flows at the midplane. 

The vertical variation of radial flows also triggers instability 
even in the hydrodynamical gas without magnetic field \citep{gs67,fri68,nel13}. 
We need further studies to determine whether this type of instability is 
effective in the presence of vertical magnetic flux. 

The layered accretion we observe is totally opposite to a trend 
obtained in simulations without a net vertical magnetic flux 
\citep{fro11,flo11}; 
in their simulations, the gas moves outwardly in the surface regions. 
This might indicate the importance of vertical magnetic fields in 
detailed properties of mass accretion. 
However, we should be careful, because the layered accretion obtained 
in our simulations might be due to the boundary condition at the disk 
surfaces, $\theta = \theta_{\rm max}$ \& $\theta_{\rm min}$. 
Since our simulations handle the net vertical magnetic fields in 
the spherical coordinates, we need to take special care at the boundaries, 
which we discuss further later in this subsection.

\begin{figure*}
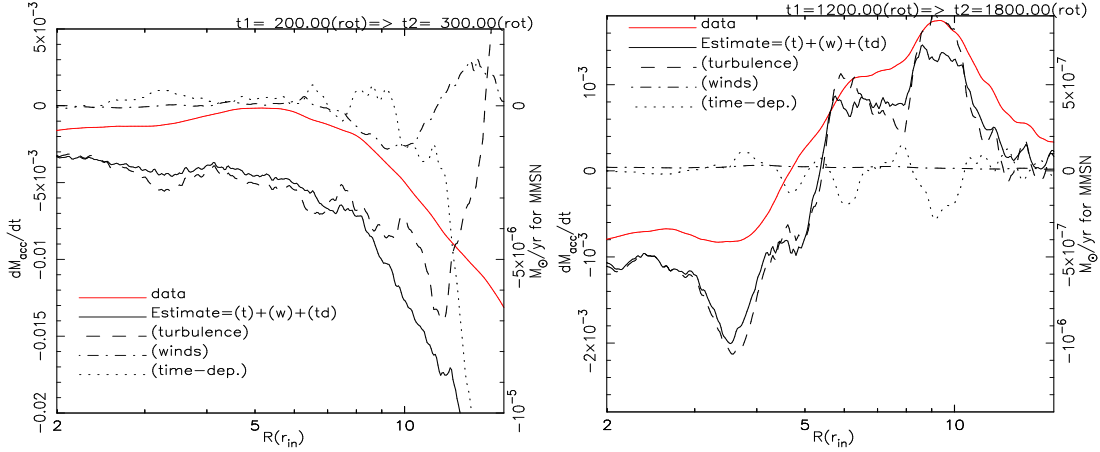

\includegraphics[height=0.25\textheight,width=0.4\textwidth]{accvel_fig10_13.ps}
\includegraphics[height=0.25\textheight,width=0.4\textwidth]{accvel_fig11_13_2.ps}
\caption{Comparison of the mass accretion rates of Case I-high ({\it left}) and 
Case II-high ({\it right}). In each panel, the measured accretion rate 
({\it red solid}; Equation \ref{eq:dotM_cl}) is compared with the accretion 
rate estimated from the transport of the angular momentum, Equation 
(\ref{eq:dotM_est}; {\it black solid}). The other three lines are the breakdown 
of the estimated accretion rate; the dashed line corresponds to the mass 
accretion driven by the turbulent stress, the dot-dashed line is that by the 
disk winds, and the dotted line is the contribution from the time-dependent 
term in Equation (\ref{eq:vrest}).}
\label{fig:accelm}
\end{figure*}

Transport of angular momentum is a key to control radial flows in 
accretions disks \citep[e.g.,][]{lp74}. In order to quantitatively study 
the mechanisms that drive the radial flows, we examine the transport of the 
angular momentum in the simulated disks. 
For our analyses, we begin with an equation for the conservation of momentum 
in the $\phi$ direction ({\it i.e.} conservation of angular momentum) 
in the cylindrical coordinates. 
Since we take the $\phi$ averages, we start from an 
axisymmetric equation: 
\begin{eqnarray}
\frac{\partial}{\partial t}(\rho R v_R) &+& \frac{1}{R}
\frac{\partial}{\partial R}\left[R^2\left(\rho v_R v_{\phi} 
- \frac{B_R B_{\phi} }{4\pi}\right)\right] \nonumber \\
&+& \frac{\partial}{\partial z}
\left[R\left(\rho v_R v_z - \frac{B_R B_z }{4\pi}\right)\right]= 0. 
\label{eq:angmom}
\end{eqnarray}
Taking the integration from $z=z_{\rm bot}$ to $z=z_{\rm top}$, we have
$$
R^2 \Omega \frac{\partial \Sigma}{\partial t} + R \frac{\partial}{\partial t}
(\Sigma \delta v_{\phi}) + \frac{1}{R}\frac{\partial}{\partial R}
(R^3\Sigma \Omega v_R + R^2 \Sigma w_{R\phi}) 
$$
\begin{equation}
+\left[\rho R^2 \Omega v_z 
+R\left(\rho\delta v_{\phi}v_z - \frac{B\phi B_z}{4\pi}\right)
\right]_{z_{\rm bot}}^{z_{\rm top}} =0,
\label{eq:angmom_zint}
\end{equation}
where $\Omega = v_{\phi}/R$ is rotation frequency and 
\begin{equation}
\Sigma w_{R\phi} \equiv \int_{z_{\rm bot}}^{z_{\rm top}} dz \left(\rho v_R v_{\phi} 
- \frac{B_R B_{\phi} }{4\pi}\right)
\end{equation}
is $\alpha$ multiplied by $p$ and integrated with $z$. 
The last term of Equation (\ref{eq:angmom_zint}) indicates the loss of 
angular momentum by the disk winds from the top ($z=z_{\rm top}$) and the 
bottom ($z=z_{\rm bot}$) of a simulation box.  

In order to rearrange Equation (\ref{eq:angmom_zint}) to a more useful 
form, we use an equation for the mass conservation,  
\begin{equation}
\frac{\partial \rho}{\partial t} + \frac{1}{R}\frac{\partial}{\partial R}
(\rho v_R R) + \frac{\partial}{\partial z}(\rho v_z) = 0.
\label{eq:masscns}
\end{equation}
As we did for Equation (\ref{eq:angmom}), we integrate Equation 
(\ref{eq:masscns}) from $z=z_{\rm bot}$ to $z=z_{\rm top}$ to have
\begin{equation}
R^2\Omega \frac{\partial \Sigma}{\partial t} + R \Omega 
\frac{\partial}{\partial R}(\Sigma v_R R) + R^2 \Omega 
[\rho v_z]_{z_{\rm bot}}^{z_{\rm top}} = 0
\label{eq:masscns_zint}
\end{equation}
Combining Equations (\ref{eq:angmom_zint}) and (\ref{eq:masscns_zint}), 
we can derive an equation that determines radial velocity as
$$
v_{R,{\rm ang}} = \left[-R\frac{\partial}{\partial t}(\Sigma \delta v_{\phi})
-\frac{1}{R}\frac{\partial}{\partial R}(R^2 \Sigma w_{R\phi}) \right.
$$
\begin{equation}
\left. 
-\left\{R\left(\rho \delta v_{\phi}v_z - \frac{B_{\phi}B_z}{4\pi}\right)
\right\}_{z_{\rm bot}}^{z_{\rm top}} \right]\left[\Sigma \frac{\partial}{\partial R}
(R^2\Omega)\right]^{-1}, 
\label{eq:vrest}
\end{equation}
where we use the subscript ``ang'' to explicitly show that this $v_{R}$ is 
estimated from the balance of angular momentum, and the physical meaning 
of each term in the numerator would be clear. The 
second and third terms denote the change of angular momentum by 
magneto-turbulent stresses and disk winds respectively. 
The first term arises from the change of mass distribution with time. 

Figure \ref{fig:accelm} presents the mass accretion rates, 
\begin{equation}
\dot{M}_R(R) = 2\pi R \int_{z_{\rm bot}}^{z_{\rm top}}dz \rho v_R
\label{eq:dotM_cl}
\end{equation}
of Case I-high (left panel) and II-high (right panel) and the contributions 
from each term in Equation (\ref{eq:vrest}), where we use $\dot{M}_R$ 
in the cylindrical coordinates instead of $\dot{M}_r$ 
(Equation \ref{eq:dotM_sp}). Note that $\dot{M}_R<0$ corresponds to accretion 
and $\dot{M}_R>0$ corresponds to radial outflows.  
The red lines are the measured mass accretion rates from the simulations and 
the black lines are the estimated accretion rates using $v_{R,{\rm ang}}$ 
of Equation (\ref{eq:vrest}), 
\begin{equation}
\dot{M}_{R,{\rm ang}} = 2\pi R \Sigma v_{R,{\rm ang}}. 
\label{eq:dotM_est}
\end{equation}
In the figure, the breakdown of the three contributions to the accretion rate 
is also shown. The contribution from the turbulent stress (dashed lines) 
is calculated by using the only 2nd term of Equation (\ref{eq:vrest}) when 
deriving $v_{R,{\rm ang}}$. The contributions from the disk winds (dot-dashed 
lines) and the time-dependent term (dotted lines) can be derived in the same 
manner.  

\begin{figure*}
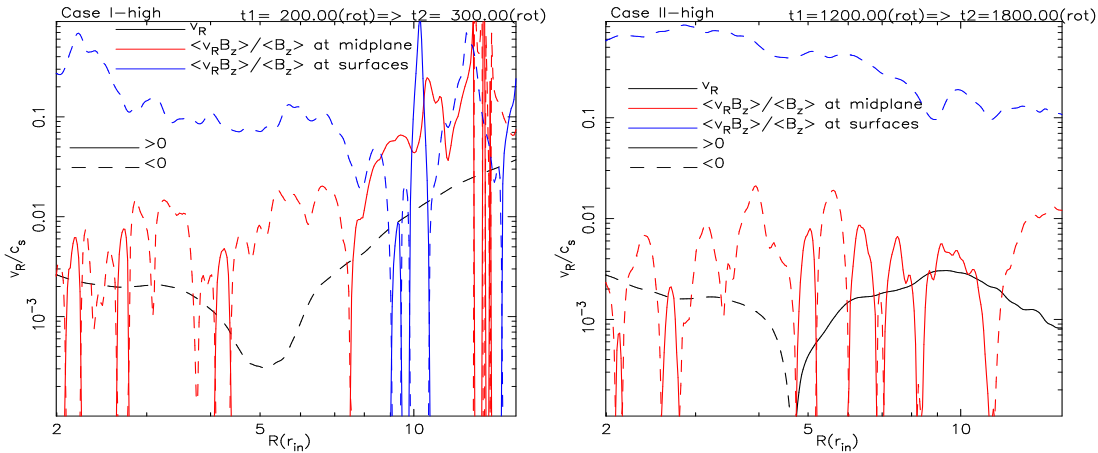

\includegraphics[height=0.25\textheight,width=0.4\textwidth]{accvel_fig10_5_3.ps}
\includegraphics[height=0.25\textheight,width=0.4\textwidth]{accvel_fig11_5_3.ps}
\caption{Comparison of radial velocities of the gas ({\it black}), and the 
net vertical fields in the midplane region ({\it red}) and in the surface 
regions ({\it blue}) of Case I-high ({\it left}) and Case II-high 
({\it right}). Dashed lines correspond to inward velocities, $v_R<0$, and solid 
lines indicate outward velocities, $v_R>0$.}
\label{fig:Bzflow}
\end{figure*}

Figure \ref{fig:accelm} shows that the estimated accretion rates 
($\dot{M}_{R,{\rm ang}}$; black solid lines) explain the overall trends of 
the measured accretion rates ($\dot{M}_R$; red solid lines), whereas 
the deviations of $\dot{M}_R$ from $\dot{M}_{R,{\rm ang}}$ are not small 
because of the truncation errors when converting from the spherical 
coordinates used in the simulations to the cylindrical coordinates, 
which are further accumulated during the long time integration.  
In both the cases, the radial mass flows are mainly determined by the 
turbulent stress (dashed lines). The time-dependent term also gives 
significant contributions in some regions, which indicates that the assumption 
of the steady state is not good for the simulated disks. 
The contribution from the disk winds (dotted lines) is much smaller than 
the other two components. Although in the outer region ($R>15r_{\rm in}$) of 
Case I-high the disk winds become significant, the effect of the disk winds 
is overestimated in this region because the simulation box can cover 
up to only $z\approx \pm H$ in Case I-high.

While in Case I-high the mass is accreting ($\dot{M}_R<0$) in the entire 
region, in Case II-high the mass is going outward ($\dot{M}_R>0$) in 
$R>5r_{\rm in}$. This radially outward flow is natural during 
the evolution of accretion disks \citep[e.g.,][]{lp74}. In this region 
of Case II-high, the angular momentum supplied from the inner region is larger
than the angular momentum lost to the outer region. The net angular momentum 
increases in a ring located in $R>5r_{\rm in}$, and then the gas moves outward. 

Figure \ref{fig:Bzflow} displays radial dependences of the motion of the 
vertical magnetic field,  
\begin{equation}
\langle \frac{v_{R,B_z}}{c_{\rm s}}\rangle_{t,\phi,z} = \frac{\langle B_z v_R 
\rangle_{t,\phi,z}(R)}{\langle B_z \rangle_{t,\phi,z}(R)c_{\rm s}}, 
\label{eq:vrBztpz}
\end{equation}
in comparison with the radial velocity of the gas, 
\begin{equation}
\langle \frac{v_{R,{\rm gas}}}{c_{\rm s}}\rangle_{t,\phi,z} 
= \frac{\langle \rho v_R \rangle_{t,\phi,z}(R)}
{\langle \rho \rangle_{t,\phi,z}(R)c_{\rm s}}. 
\label{eq:vrgastpz}
\end{equation}
For the radial velocity of $B_z$ in Equation (\ref{eq:vrBztpz}) we take 
the average in the midplane region, $\Delta z_{\rm mid}$ 
(Equation \ref{eq:dzmid}), and in the 
surface regions, $\Delta z_{\rm sfc}$ (Equation \ref{eq:dzsfc}), to compare 
the motions of the vertical magnetic 
flux at the midplane and in the surface regions. For the radial flow of 
gas in Equation (\ref{eq:vrgastpz}), we average $v_R$ over the 
entire surface $\Delta z_{\rm tot}$ (Equation \ref{eq:dztot}) to see the net
gas flow. As shown in Figures \ref{fig:zstacc} \& 
\ref{fig:accelm}, the radial velocities could be either positive or negative. 
In order to display both positive and negative values in the logarithmic scale, 
we take the absolute values and use dashed lines for radially inward flows 
($v_R<0$) and solid lines for radially outward flows ($v_R>0$). 

In both Cases I-high (left panel in Figure \ref{fig:Bzflow}) and 
II-high (right panel), the motions of 
the net $\langle B_z\rangle$ near the midplane and in the surface regions 
are very different. 
In the surface regions, the vertical magnetic flux mostly moves inward at a 
quite high speed, $\gtrsim 0.1 c_{\rm s}$. On the other hand, at the midplane, 
no clear tendencies are observed in either case; $\langle B_z\rangle$ 
moves outward in some regions and inward in other regions at 
slow speeds, $\lesssim 0.01 c_{\rm s}$. (Note that in the outer region of 
Case I-high, $v_R/c_{\rm s}$ becomes large because of the effect of the 
surface boundaries.) 
These different properties of the net $\langle B_z\rangle$ indicate that 
the vertical magnetic field lines are not connected from the midplane to 
the surface regions when considering the long time integration, 
$\Delta t_{\rm ave}$. 
The inward dragged magnetic field lines in the surface regions 
continuously reconnect with field lines in the midplane region because of 
the numerical resistivity; although our simulations assume the ideal MHD, 
magnetic reconnections could take place in the sub-grid scales as a result of 
the numerical diffusion. 
For the same reason, the motions of the vertical magnetic fields are also 
not strictly coupled to the motions of the gas. 

As shown so far, our simulations show the inward dragging of the vertical 
magnetic flux in the surface regions, which follows the trend of the layered 
accretion of the gas component. Interestingly enough, this is consistent 
with a recent result based on an analytic model \citep{rl08}, while different 
trends could be realized with different settings \citep{lub94}.
The inward advection of $B_z$ in the surface regions will cause 
a concentration of the magnetic flux around a central object, which is 
suitable for driving strong jets \citep{bec09}. 
However, we should carefully state that this
trend might be affected by the boundary condition at the disk surfaces.   
Our simulations adopt the outgoing condition based on a characteristic 
method \citep{si06} at the surface boundaries (see \S \ref{sec:bc}). 
Thus, no information comes into the simulation box. However, in realistic 
situations, a global magnetic field is probably anchored in somewhere 
above a disk, {\it e.g.} in coronal regions above the disk surfaces 
\citep[e.g.,][]{kat04,ohs09,bec09}. 
In this case, the inward advection of the vertical magnetic flux observed 
in our simulations would eventually be stopped as well as further wound up 
by the vertical differential rotation. Later on, the configuration of 
the global magnetic field would be suitable for magneto-centrifugal driven 
winds \citep{bp82,kud98}, which also contributes to the transport of the 
angular momentum of the disk. 
For simulations in more realistic situations, we need a larger simulation 
box particularly in the $\theta$ direction, although it is a trade-off for 
the numerical resolution in a disk region. 

\subsection{Toroidal $B$ field}


\begin{figure*}[h]
\begin{center}
\includegraphics[height=0.3\textheight,width=0.41\textwidth]{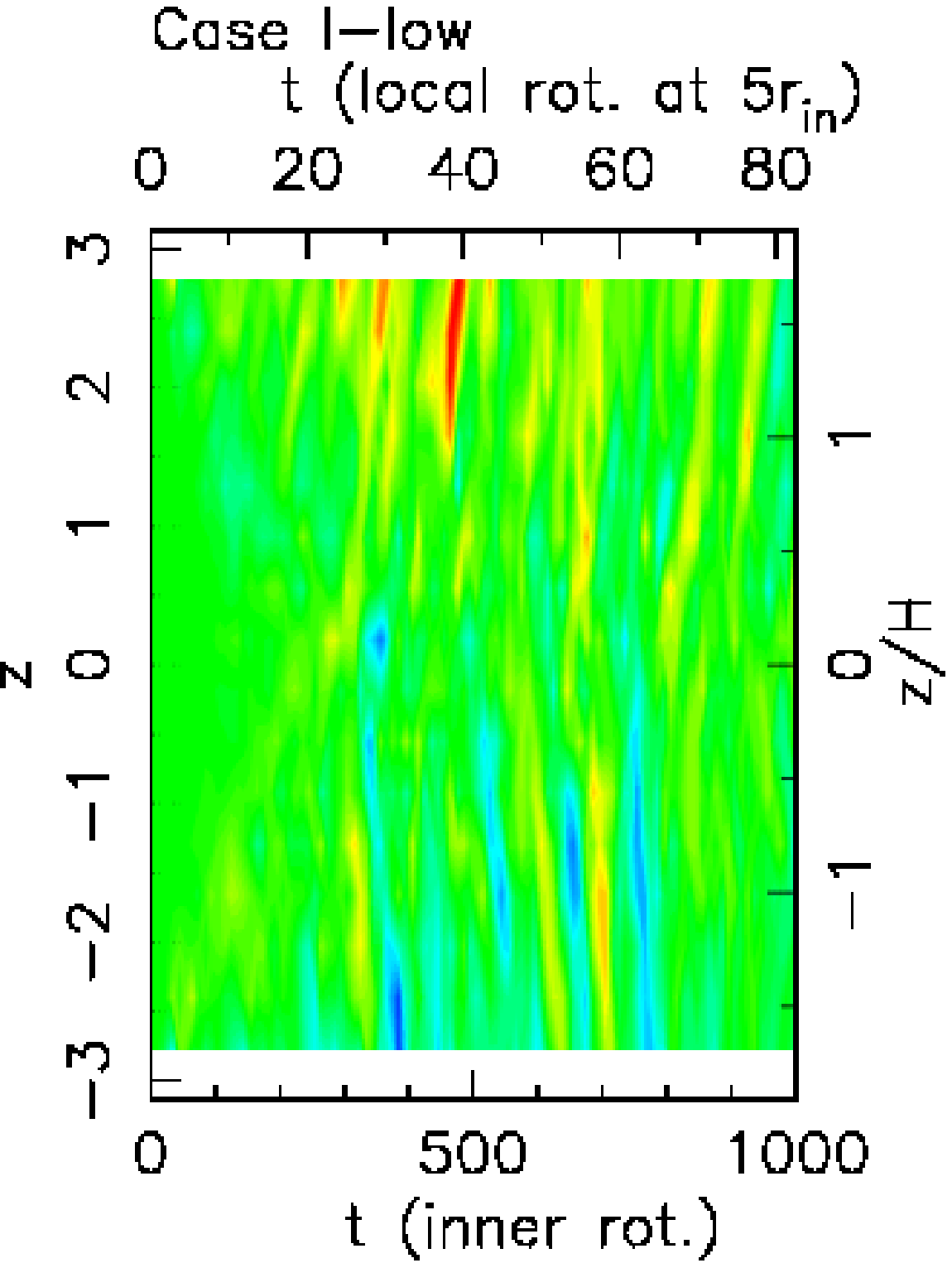}
\includegraphics[height=0.3\textheight,width=0.34\textwidth]{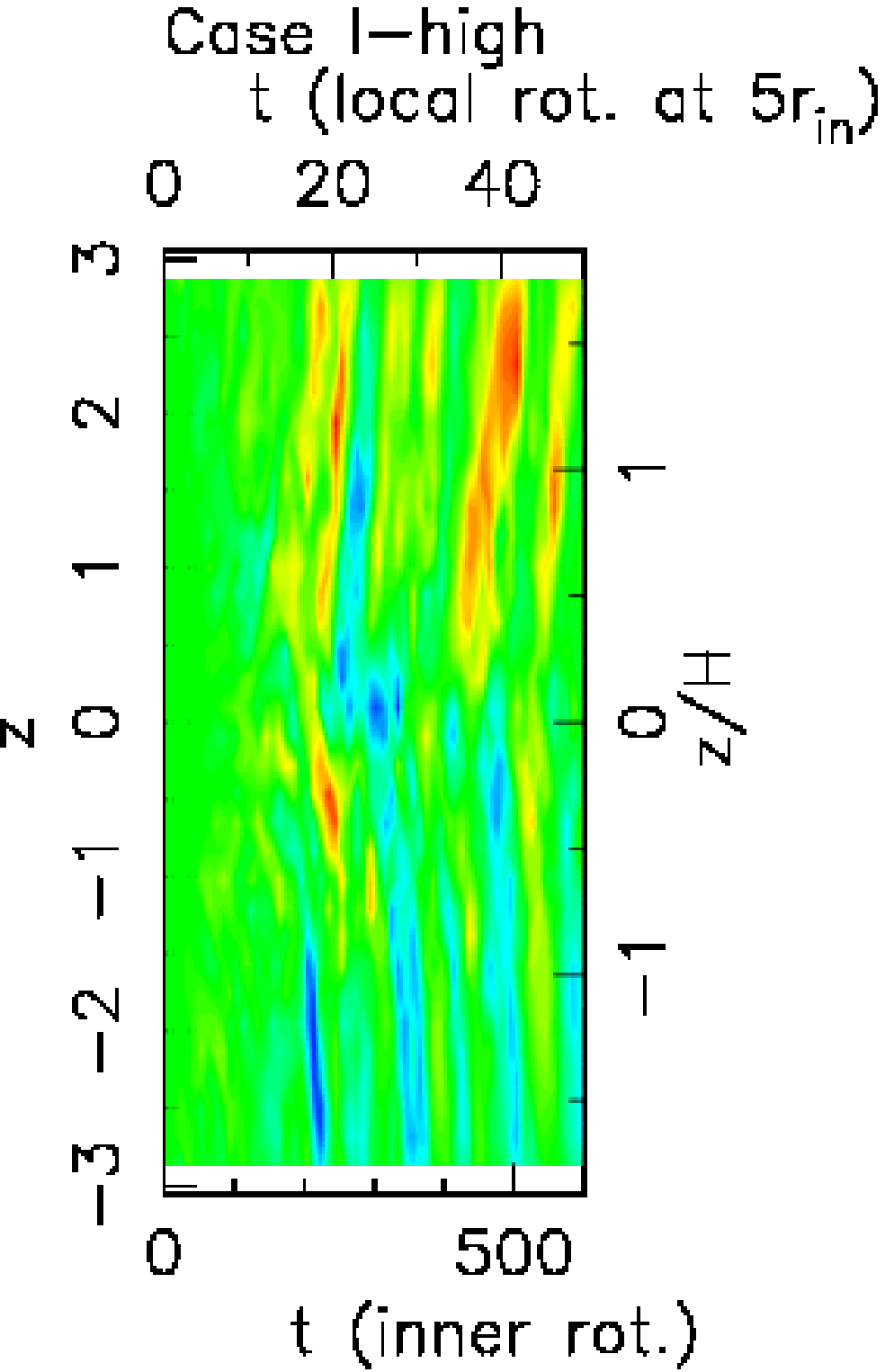}\\
\includegraphics[height=0.35\textheight,width=0.69\textwidth]{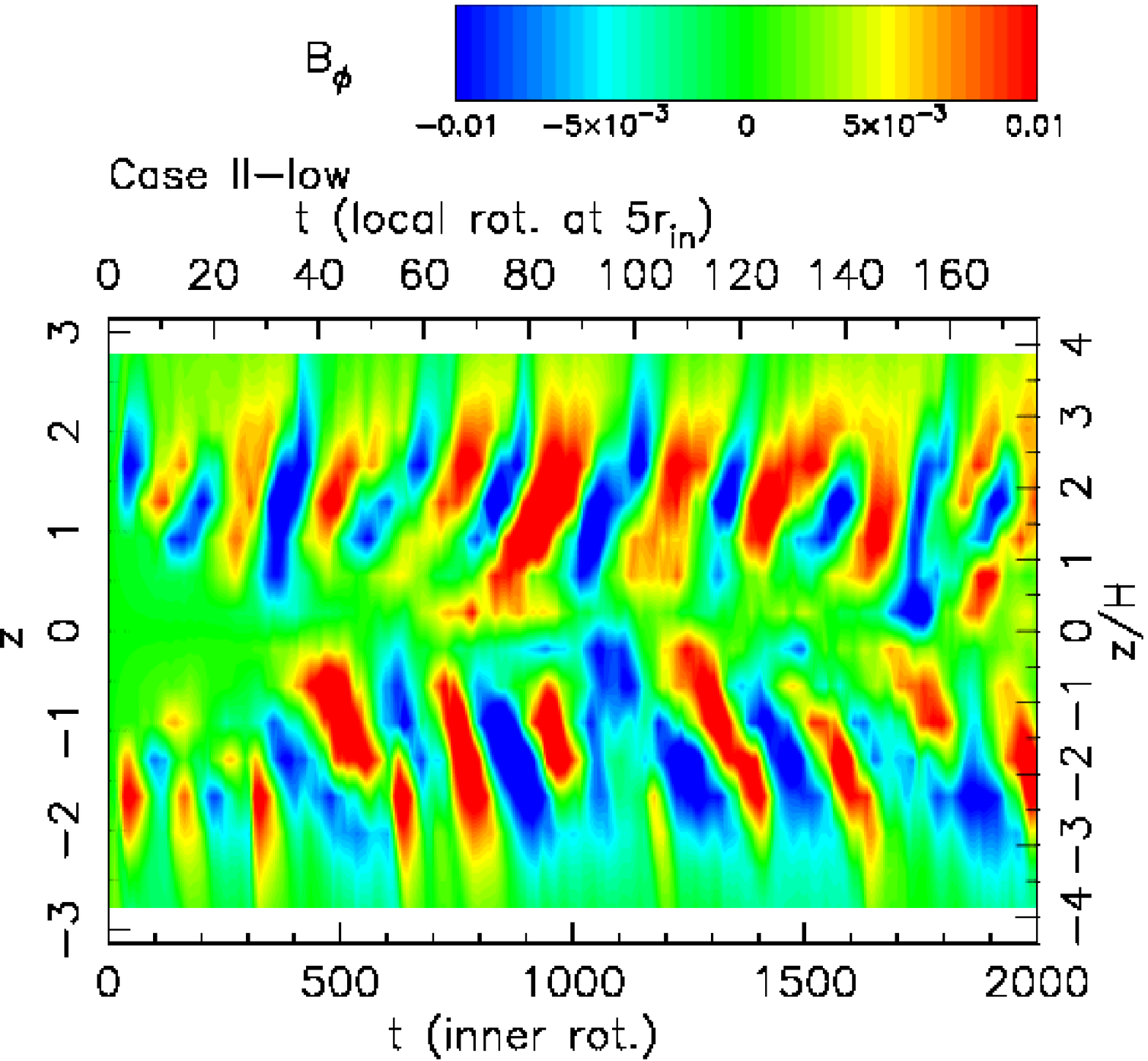}
\includegraphics[height=0.3\textheight,width=0.64\textwidth]{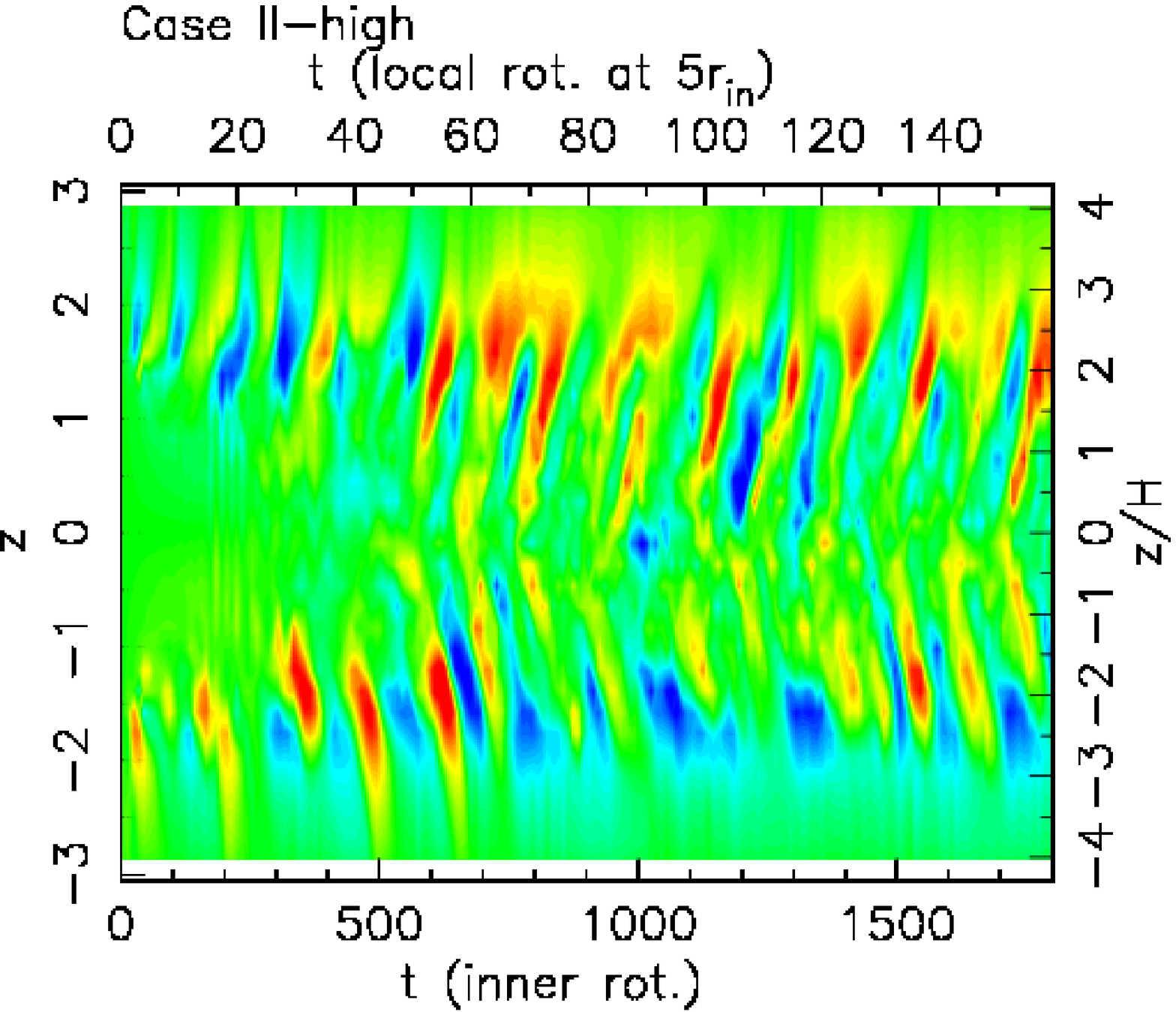}
\end{center}
\caption{$t-z$ diagrams of $\langle B_{\phi}\rangle_{\phi}(r_1,z)$ 
at $R=r_1(=5r_{\rm in})$ for Case I-low ({\it top-left}), Case I-high 
({\it top-right}), Case II-low ({\it Middle}), and Case II-high 
({\it Bottom}). On the 
horizontal axis are shown time in units of the inner rotation ({\it top} 
of panels) and time in units of the local rotation at $r=r_1$ ({\it bottom} 
of panels). On the vertical axis are shown $z$ ({\it left} to panels) 
and $z/H$ ({\it right} to panels). }
\label{fig:tzBphi}
\end{figure*}

Temporal oscillations of toroidal magnetic fields are a universal phenomenon 
in MRI-induced accretion disks \citep[e.g.,][]{dav10,flo11}.
Our global simulations show similar trends as illustrated in Figure 
\ref{fig:tzBphi} which displays the $t$--$z$ diagrams of $\langle B_{\phi} 
\rangle_{\phi}(r_1,z)$.
Cases II-high and II-low show more distinctive oscillating features than 
Cases I-high and I-low because Cases II-high and II-low cover the larger 
vertical region measured in scale height. This indicates that the magnetic 
fields at high altitudes ($z\gtrsim 2H$) are important in the flip-flops 
of the toroidal magnetic fields. 

Compared to results of local simulations \citep[e.g.,][]{dav10}, 
the quasi-periodic nature of oscillatory features is deformed in 
our global simulations, in a manner similar to the time-dependent properties 
of the disk winds (Figure \ref{fig:tzvz} in \S \ref{sec:dwd}). 
The four cases show that in the upper hemisphere ($z>0$) $\langle B_{\phi} 
\rangle$ tends to be positive (redder colors), while in the lower 
hemisphere ($z<0$) $\langle B_{\phi} \rangle$ tends to be negative 
(bluer colors). This is related to the layered advection of the vertical 
magnetic fields discussed in \S \ref{sec:rf}. $\langle B_z \rangle$ is 
advected inward in the surface regions. Focusing on a single field line, 
it rotates faster in the surface regions as a consequence of this inward 
advection. Thus, positive (negative) $\langle B_{\phi} \rangle$ is 
created near the upper (lower) surface. 
However, this might be affected by the boundary condition at the surfaces 
as discussed in \S \ref{sec:rf}. If a global poloidal magnetic field 
is anchored in coronal regions which are outside the simulation box, 
the layered advection of the vertical field  lines would eventually be
inhibited and the systematic generation of $\langle B_{\phi} \rangle$ would 
also be suppressed.   

\section{Discussion}

\subsection{Entire Region of Case II}
\label{sec:disII}
\begin{figure}[h]
\begin{center}
\includegraphics[height=0.36\textheight,width=0.5\textwidth]{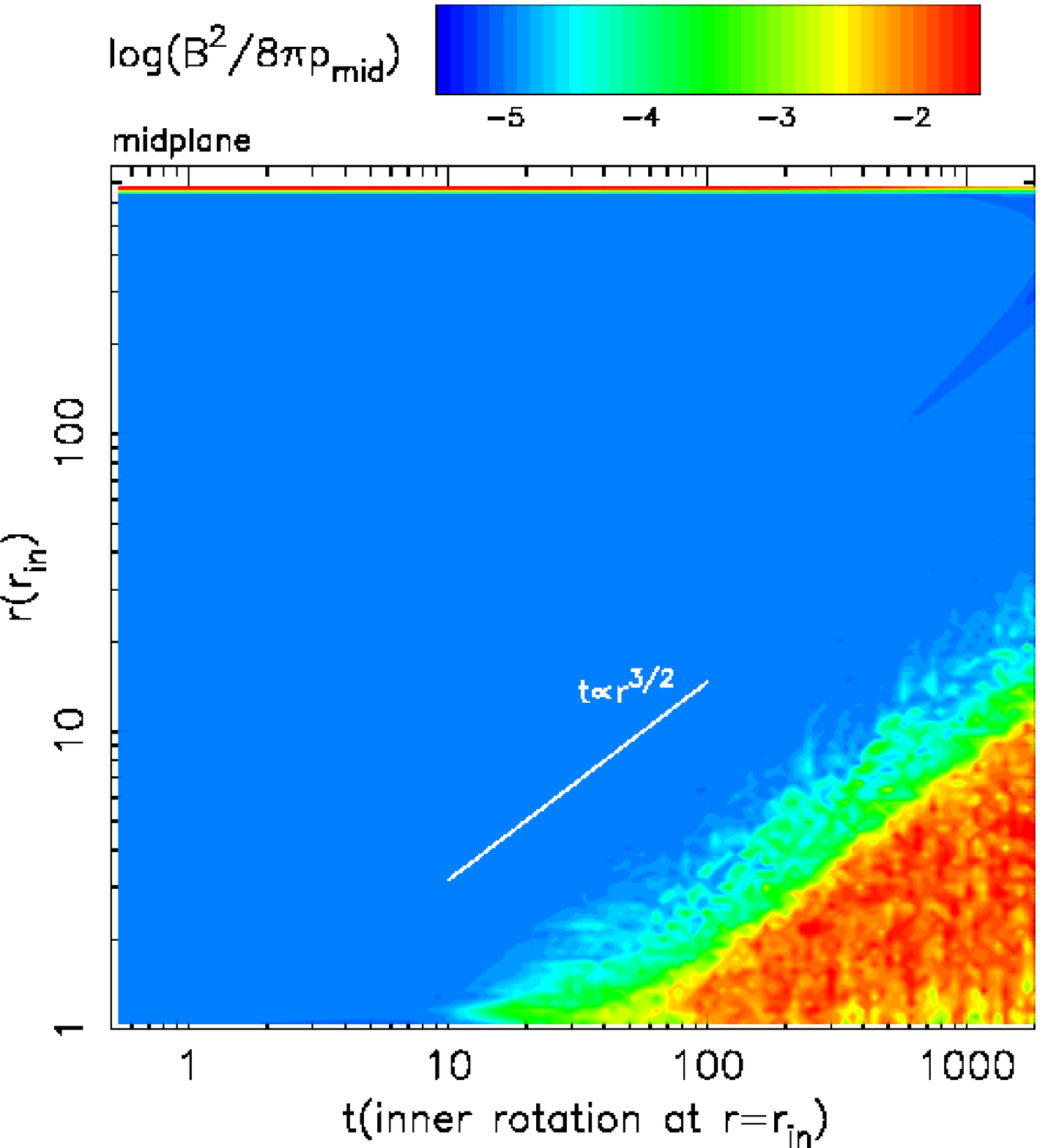}\\
\includegraphics[height=0.36\textheight,width=0.5\textwidth]{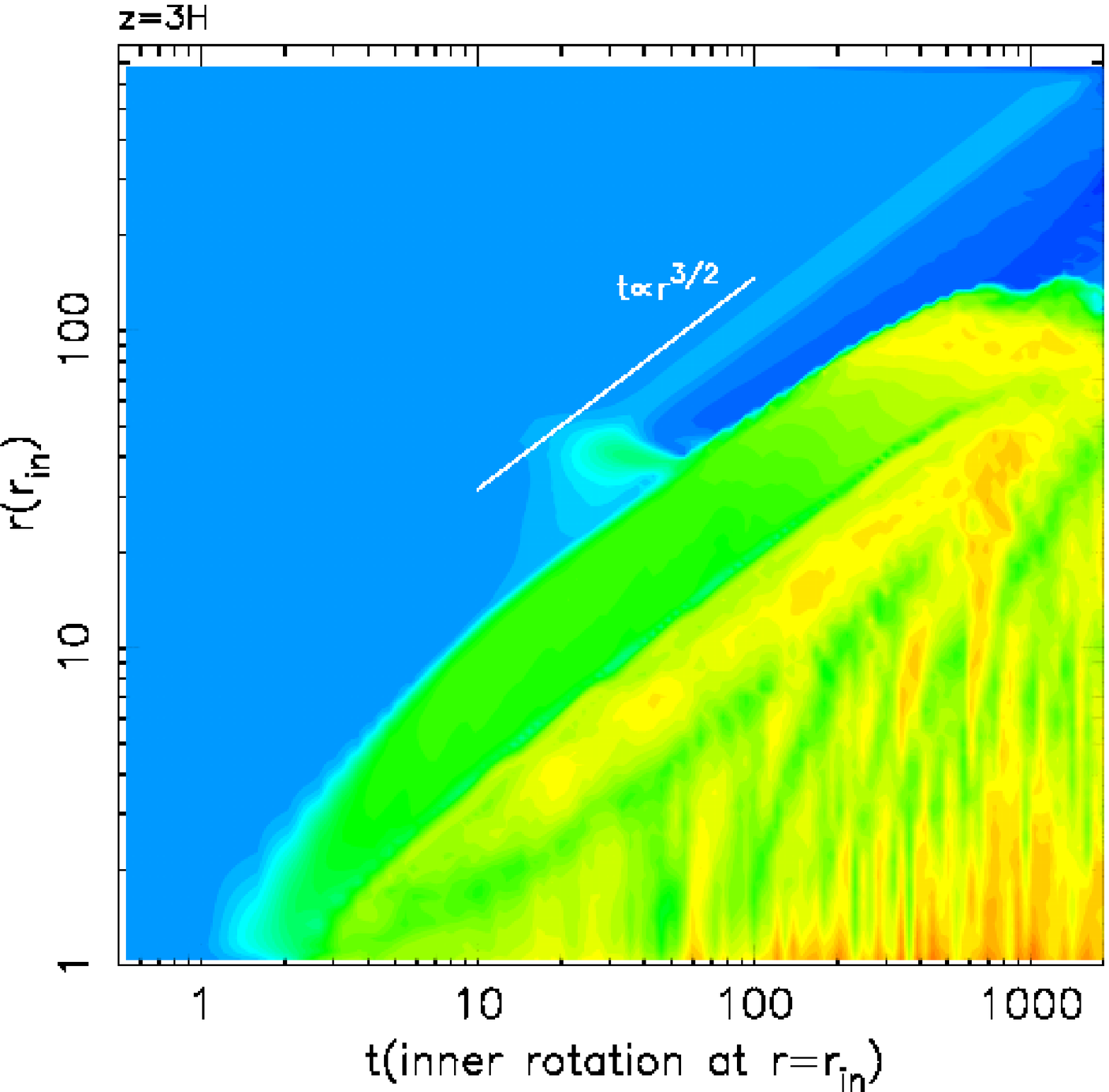}
\end{center}
\caption{$t$--$r$ diagrams of the magnetic energy normalized by the 
gas pressure at the midplane, $\langle B^2 \rangle_{\phi}(t,z,r)
/8\pi\langle p_{\rm mid} \rangle_{\phi}(t,z,r)$, at $z=0$ 
(midplane; {\it upper}) and at $z=3H$ ({\it lower}) of Case II-high. 
Both axes are in logarithmic scale. The white line shows the relation of 
the Keplerian rotation time, $t\propto r^{3/2}$.
}
\label{fig:trB2}
\end{figure}

In Case II we use the very large radial extent ($\sim 500 r_{\rm in}$) 
of the simulation box (Table \ref{tab:models}). 
However, since our purpose is to study the saturated state of the magnetic 
field, we have focused on the region in $r\le 10r_{\rm in}$ so far; in the 
outer region, the magnetic field is still in the growth phase. 
Here, we briefly introduce the evolution of the magnetic field in the entire 
radial extent of Case II-high. Figure \ref{fig:trB2} presents the evolution of 
the magnetic energy at different heights, $z=0$ (upper panel) and $z=3H$ 
(lower panel), normalized by the gas pressure at the midplane, 
\begin{equation}
\frac{\langle B^2\rangle_{\phi}(t,r,z)}
{8\pi\langle p_{\rm mid}\rangle_{\phi}(t,r)}
\end{equation}
in the time (horizontal axis) -- $r$ (vertical axis) diagram.

Both panels show that the amplification of the magnetic fields proceeds 
with time in proportion to the Keplerian rotation time, $t\propto r^{3/2}$. 
At the midplane the magnetic field is mainly amplified by the MRI. 
In the surface region, the growth of the magnetic field is roughly 
ten times faster, which is triggered by the vertical differential rotation.

\subsection{Implication for Observation of Spirals in Protoplanetary Disks}

In Figure \ref{fig:faceon} in \S \ref{sec:fov} we showed spiral structures 
in the face-on views of $\log (1/\beta)$. 
Although the disks are turbulent, pressure balance is 
supposed to be roughly satisfied between high-$\beta$ (weak $B$) region 
and low-$\beta$ (strong $B$) regions, where high-$\beta$ regions 
correspond to denser regions. Applying the result of Figure \ref{fig:faceon} 
to protoplanetary disks, denser high-$\beta$ regions are expected to have 
a larger number of dust grains than lower-$\beta$ regions. 
Thus, similar spiral structures are expected to be 
seen in scattered-light observations, {\it e.g.,} in the near-infrared 
wavelength. 
Recent near-infrared observations of protoplanetary disks found many
remarkable spiral structures. Many authors tend to interpret those
structures as being caused by unseen planets embedded in the disks 
\citep[e.g.,][]{has11,mut12,fuk13}.  However, we
should keep in mind that those structures can be naturally produced by
magnetic structures, just as shown in our simulations.  
The spiral structures seen in our simulations are typically 
short-lived with lifetimes of the order of rotation time, e.g., 1000 yr 
at 100 AU for a disk around a central star with the solar mass. However, such 
structures are ubiquitously created and can almost always observed be 
somewhere in the simulated disks as shown in the movies for Figure 
\ref{fig:faceon} ({\it online materials}; also available at 
www.ta.phys.nagoya-u.ac.jp/stakeru/research/glbdsk). 
For a detailed comparison with observations, we need to model the dust 
component in a reasonable way not only in the disk itself \citep{an06,nn06} 
but also in wind regions \citep[e.g.,][]{hei11} since dust grains are 
expected to remain at high altitudes above the disk \citep{suz10,mcj13}. 
A more detailed
comparison with more realistic simulations with radiative transfer taking 
into account dust might be interesting in future work.

\section{Summary}
\begin{figure}[h]
\begin{center}
\includegraphics[width=0.5\textwidth]{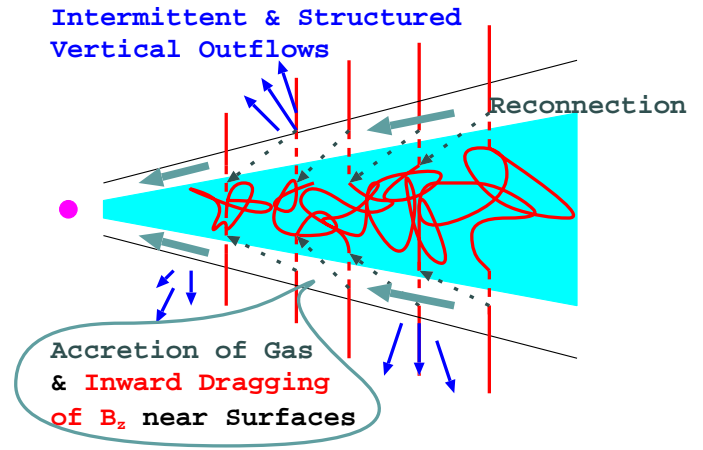}
\end{center}
\caption{Schematic summary of the simulation results. The inward 
accretion of the gas mainly takes place in the surface regions. The net $B_z$ 
(red lines) is also dragged inward in the surface regions, while in the midplane
region the turbulence magnetic field stochastically moves 
inward and outward; namely $B_z$ 
in the surface regions and $B_z$ in the midplane continuously reconnect owing 
to the numerical resistivity (\S \ref{sec:rf}). Vertical outflows 
intermittently stream out by the Poynting flux (blue arrows).}
\label{fig:simpic}
\end{figure}
We have performed 3D MHD simulations of global accretion disks 
threaded with weak vertical magnetic fields in the two types of 
the temperature profiles.  
In the simulations MHD turbulence is triggered and amplified by MRI, and 
in the saturated states the diagnostics for MRI, $-2\langle B_R B_{\phi}
\rangle/B^2$ and $\langle B_R^2\rangle/\langle B_{\phi}^2\rangle$, are 
well correlated with the numerical resolutions as discussed in previous 
works \citep{haw11,flo11}. 

In addition, the effect of the different 
temperature profiles also affects the results through the vertical 
differential rotation. In the cases with spatially constant temperature 
(Case I), the rotation frequency is constant along the initial vertical 
field lines, and the overall properties of the MHD turbulence are similar 
to those observed in local shearing box simulations. On the other 
hand, in the cases with the profile of $T\propto 1/r$ (Case II), the gas in 
the surface regions rotates slower than the gas near the midplane. 
As a result, coherent magnetic fields are amplified 
in the surface regions, which also contribute to the Maxwell stress there. 

This result indicates that thermal processes could play an 
important role in determining the saturation of the magnetic field and the 
properties of the turbulence in an accretion disk. 
In our simulations the temperature profiles are fixed, which implicitly 
assumes that external mechanisms such as irradiation from a central object 
regulate the temperatures. 
In other words, the thermodynamics determined by external mechanisms 
controls the dynamics of the disks.  
In reality, however, the temperature profile of a disk is also affected 
by internal mechanisms because the dissipation of the turbulence leads to 
the heating of a disk. A self-consistent treatment in determining the 
temperature distribution is necessary for our future studies.

The velocity fluctuations are dominated by the radial component and give 
$\approx 0.1-0.2$ of the sound speed around the midplane. 
The density fluctuations, $\delta \rho/\rho$ are also $\approx 0.1-0.2$. 
These values for the density and velocity perturbations are somewhat larger
than those obtained in global simulations without initial vertical magnetic 
fields \citep{flo11}. In the case with $T\propto 1/r$, the regions with 
large $\delta \rho / \rho$, which coincide with the regions 
with large $-2\langle B_R B_{\phi}\rangle/\langle B^2 \rangle$ (high 
activities of MRI), are radially localized and stay for long times, which 
may be analogous to zonal flows seen in local simulations \citep{joh09}.  
If applied to protoplanetary disks, such large density fluctuations 
are expected to influence the dynamics of dust particles. 

The azimuthal power spectra of the magnetic fields show quite shallow power-law 
indice with respect to mode $m$, probably because the energy injection 
by the MRI is from the small 
scale (large $m$). On the other hand, the azimuthal power spectra of the 
velocity and density show $\propto m^{-1 \sim -2}$. 

The onsets of intermittent and structured vertical disk winds are observed 
in the global simulations (Figures \ref{fig:edgeon}, \ref{fig:tzvz}, \& 
\ref{fig:simpic}), 
similar to that seen in the local simulations \citep{si09,suz10}. They are 
driven by the Poynting flux associated with the MHD turbulence. The magnetic 
pressure component gives the larger contribution than the magnetic tension 
component, which is in contrast to the local simulations in which both 
are contribute almost equally. 
Although the magnitude is smaller than that obtained in the local simulation, 
the injection regions of the magnetic tension form at $z\approx\pm 1.3H$. 
The acoustic waves, which are probably linked to the injection regions, are 
directed to the midplane. In protoplanetary disk conditions, these sound-like 
waves enhance the sedimentation of dust particles to the midplane. 

In both Cases I \& II, the mass accretions take place in the surface regions, 
because the $\alpha$ stresses are larger there. 
At the midplane the gas moves radially outward at a very slow speed.  
The velocity difference between the midplane and the surfaces may cause 
meridional circulation. 
Applied to protoplanetary disks, this causes the outward migration of 
solid particles at the midplane and possibly explains the observed 
crystalline dusts in the outer parts of disks. 
The radial motion of the vertical magnetic field lines also follows these 
tendencies of the gas, although the velocities of the magnetic fields and 
the gas are not the same because of the magnetic diffusion at the sub-grid 
scale. 
The magnetic flux is dragged inward in the surface regions, while at 
the midplane the turbulent magnetic flux moves stochastically outward 
and inward; the vertical magnetic fields continually reconnect at the 
sub-grid scales (Figure \ref{fig:simpic}).
However, we should carefully note that the observed layered accretion 
of the gas and the inward dragging of the vertical magnetic fields might 
be affected by the adopted boundary condition at the disk surfaces.

This work was supported in part by Grants-in-Aid for 
Scientific Research from the MEXT of Japan, 22864006 (TKS), 23244027, and
23103005 (SI). We thank the referee for many constructive 
comments. 
We are also grateful to Dr. Mario Flock, Dr. Xuening Bai, and 
Dr. Satoshi Okuzumi for many fruitful discussions. 
Numerical simulations in this work were carried out at the Yukawa 
Institute Computer Facility, SR16000 and at the Cray XT4 and XC30 
operated in CfCA, National Astrophysical Observatory of Japan.

\begin{appendix}
\section{Setting Initial $B_z$ in Spherical Coordinates}
In our simulations, we use the method of constrained transport 
\citep{eh88} in the spherical coordinates to handle the evolution of 
magnetic fields while keeping $\mbf{\nabla}\cdot\mbf{B}=0$. 
In order to set up the initial vertical magnetic fields (Equation 
\ref{eq:Bzinit}) which exactly satisfy $\mbf{\nabla}\cdot\mbf{B}=0$ in all 
the cells, we use the vector potential, $\mbf{A}$. 
To achieve the initial profile, $B_z\propto 
R^{-\mu/2}=(r\sin\theta)^{-\mu/2}$  for $\mu\ne 4$, we set 
\begin{equation}
(A_r,A_{\theta},A_{\phi}) = \left(0,0,\frac{1}{2-\mu/2}B_{z,{\rm in}}
\left(\frac{r\sin \theta}{r_{\rm in}}\right)^{1-\mu/2}\right). 
\label{eq:vecpot}
\end{equation}
Then, we can set up the initial $\mbf{B}$ from 
$\mbf{B}=\mbf{\nabla}\times\mbf{A}$, or in the explicit form, 
\begin{equation}
(B_r,B_{\theta},B_{\phi}) = \left(\frac{1}{r\sin\theta}\frac{\partial}{\partial 
\theta}(\sin\theta A_{\phi}),-\frac{1}{r}\frac{\partial}{\partial r}
(r A_{\phi}),0\right). 
\label{eq:A2B}
\end{equation}
In our simulations as well as other simulations adopting the constrained 
transport method, $\mbf{B}$ are located at the face-centered positions and 
$\mbf{A}$ are at the sides of each grid cell, and the discretization of 
Equation (\ref{eq:A2B}) is straightforward. 

Substituting Equation (\ref{eq:vecpot}) into Equation (\ref{eq:A2B}) for 
confirmation, we can recover the required result, 
\begin{equation}
(B_r,B_{\theta},B_{\phi}) = (B_z\cos\theta,-B_z\sin\theta,0),
\end{equation}
where $B_z=B_{z,{\rm in}}\left(\frac{r \sin\theta}{r_{\rm in}}\right)^{-\mu/2}$.

\end{appendix}


\begin{thebibliography}{86}
\expandafter\ifx\csname natexlab\endcsname\relax\def\natexlab#1{#1}\fi

\bibitem[{{Aikawa} \& {Nomura}(2006)}]{an06}
{Aikawa}, Y., \& {Nomura}, H. 2006, \apj, 642, 1152

\bibitem[{{Bai} \& {Stone}(2013{\natexlab{a}})}]{bs13a}
{Bai}, X.-N., \& {Stone}, J.~M. 2013{\natexlab{a}}, \apj, 767, 30

\bibitem[{{Bai} \& {Stone}(2013{\natexlab{b}})}]{bs13b}
---. 2013{\natexlab{b}}, \apj, 769, 76

\bibitem[{{Balbus} \& {Hawley}(1991)}]{bh91}
{Balbus}, S.~A., \& {Hawley}, J.~F. 1991, \apj, 376, 214

\bibitem[{{Balbus} \& {Hawley}(1998)}]{bh98}
---. 1998, Reviews of Modern Physics, 70, 1

\bibitem[{{Beckwith} {et~al.}(2009){Beckwith}, {Hawley}, \& {Krolik}}]{bec09}
{Beckwith}, K., {Hawley}, J.~F., \& {Krolik}, J.~H. 2009, \apj, 707, 428

\bibitem[{{Blandford} \& {Payne}(1982)}]{bp82}
{Blandford}, R.~D., \& {Payne}, D.~G. 1982, \mnras, 199, 883

\bibitem[{{Bouwman} {et~al.}(2008){Bouwman}, {Henning}, {Hillenbrand}, {Meyer},
  {Pascucci}, {Carpenter}, {Hines}, {Kim}, {Silverstone}, {Hollenbach}, \&
  {Wolf}}]{bou08}
{Bouwman}, J., {Henning}, T., {Hillenbrand}, L.~A., {et~al.} 2008, \apj, 683,
  479

\bibitem[{{Brandenburg} {et~al.}(1995){Brandenburg}, {Nordlund}, {Stein}, \&
  {Torkelsson}}]{bra95}
{Brandenburg}, A., {Nordlund}, A., {Stein}, R.~F., \& {Torkelsson}, U. 1995,
  \apj, 446, 741

\bibitem[{{Burgers}(1939)}]{bur39}
{Burgers}, J.~M. 1939, Verhand. Kon.Neder. Akad. Wetenschappen, Afd.,
  Natuurkunde, Eerste Sectie, 17, 1

\bibitem[{{Chandrasekhar}(1961)}]{cha61}
{Chandrasekhar}, S. 1961, {Hydrodynamic and hydromagnetic stability} (Oxford:
  Clarendon)

\bibitem[{{Cho} \& {Lazarian}(2003)}]{cl03}
{Cho}, J., \& {Lazarian}, A. 2003, \mnras, 345, 325

\bibitem[{{Clarke}(1996)}]{cl96}
{Clarke}, D.~A. 1996, \apj, 457, 291

\bibitem[{{Davis} {et~al.}(2010){Davis}, {Stone}, \& {Pessah}}]{dav10}
{Davis}, S.~W., {Stone}, J.~M., \& {Pessah}, M.~E. 2010, \apj, 713, 52

\bibitem[{{Evans} \& {Hawley}(1988)}]{eh88}
{Evans}, C.~R., \& {Hawley}, J.~F. 1988, \apj, 332, 659

\bibitem[{{Flock} {et~al.}(2012){Flock}, {Dzyurkevich}, {Klahr}, {Turner}, \&
  {Henning}}]{flo12}
{Flock}, M., {Dzyurkevich}, N., {Klahr}, H., {Turner}, N., \& {Henning}, T.
  2012, \apj, 744, 144

\bibitem[{{Flock} {et~al.}(2011){Flock}, {Dzyurkevich}, {Klahr}, {Turner}, \&
  {Henning}}]{flo11}
{Flock}, M., {Dzyurkevich}, N., {Klahr}, H., {Turner}, N.~J., \& {Henning}, T.
  2011, \apj, 735, 122

\bibitem[{{Fricke}(1968)}]{fri68}
{Fricke}, K. 1968, \zap, 68, 317

\bibitem[{{Fromang} {et~al.}(2013){Fromang}, {Latter}, {Lesur}, \&
  {Ogilvie}}]{fro13}
{Fromang}, S., {Latter}, H., {Lesur}, G., \& {Ogilvie}, G.~I. 2013, \aap, 552,
  A71

\bibitem[{{Fromang} {et~al.}(2011){Fromang}, {Lyra}, \& {Masset}}]{fro11}
{Fromang}, S., {Lyra}, W., \& {Masset}, F. 2011, \aap, 534, A107

\bibitem[{{Fromang} \& {Nelson}(2006)}]{fn06}
{Fromang}, S., \& {Nelson}, R.~P. 2006, \aap, 457, 343

\bibitem[{{Fukagawa} {et~al.}(2013){Fukagawa}, {Tsukagoshi}, {Momose}, {Saigo},
  {Ohashi}, {Kitamura}, {Inutsuka}, {Muto}, {Nomura}, {Takeuchi}, {Kobayashi},
  {Hanawa}, {Akiyama}, {Honda}, {Fujiwara}, {Kataoka}, {Takahashi}, \&
  {Shibai}}]{fuk13}
{Fukagawa}, M., {Tsukagoshi}, T., {Momose}, M., {et~al.} 2013, \pasj, 65, L14

\bibitem[{{Gammie}(1996)}]{gam96}
{Gammie}, C.~F. 1996, \apj, 457, 355

\bibitem[{{Goldreich} \& {Schubert}(1967)}]{gs67}
{Goldreich}, P., \& {Schubert}, G. 1967, \apj, 150, 571

\bibitem[{{Goldreich} \& {Sridhar}(1995)}]{gs95}
{Goldreich}, P., \& {Sridhar}, S. 1995, \apj, 438, 763

\bibitem[{{Guan} \& {Gammie}(2011)}]{gg11}
{Guan}, X., \& {Gammie}, C.~F. 2011, \apj, 728, 130

\bibitem[{{Hashimoto} {et~al.}(2011){Hashimoto}, {Tamura}, {Muto}, {Kudo},
  {Fukagawa}, {Fukue}, {Goto}, {Grady}, {Henning}, {Hodapp}, {Honda},
  {Inutsuka}, {Kokubo}, {Knapp}, {McElwain}, {Momose}, {Ohashi}, {Okamoto},
  {Takami}, {Turner}, {Wisniewski}, {Janson}, {Abe}, {Brandner}, {Carson},
  {Egner}, {Feldt}, {Golota}, {Guyon}, {Hayano}, {Hayashi}, {Hayashi}, {Ishii},
  {Kandori}, {Kusakabe}, {Matsuo}, {Mayama}, {Miyama}, {Morino}, {Moro-Martin},
  {Nishimura}, {Pyo}, {Suto}, {Suzuki}, {Takato}, {Terada}, {Thalmann},
  {Tomono}, {Watanabe}, {Yamada}, {Takami}, \& {Usuda}}]{has11}
{Hashimoto}, J., {Tamura}, M., {Muto}, T., {et~al.} 2011, \apjl, 729, L17

\bibitem[{{Hawley}(2000)}]{haw00}
{Hawley}, J.~F. 2000, \apj, 528, 462

\bibitem[{{Hawley} {et~al.}(1995){Hawley}, {Gammie}, \& {Balbus}}]{hgb95}
{Hawley}, J.~F., {Gammie}, C.~F., \& {Balbus}, S.~A. 1995, \apj, 440, 742

\bibitem[{{Hawley} {et~al.}(2011){Hawley}, {Guan}, \& {Krolik}}]{haw11}
{Hawley}, J.~F., {Guan}, X., \& {Krolik}, J.~H. 2011, \apj, 738, 84

\bibitem[{{Hawley} {et~al.}(2013){Hawley}, {Richers}, {Guan}, \&
  {Krolik}}]{haw13}
{Hawley}, J.~F., {Richers}, S.~A., {Guan}, X., \& {Krolik}, J.~H. 2013, \apj,
  772, 102

\bibitem[{{Heinzeller} {et~al.}(2011){Heinzeller}, {Nomura}, {Walsh}, \&
  {Millar}}]{hei11}
{Heinzeller}, D., {Nomura}, H., {Walsh}, C., \& {Millar}, T.~J. 2011, \apj,
  731, 115

\bibitem[{{Hirose} {et~al.}(2009){Hirose}, {Blaes}, \& {Krolik}}]{hir09}
{Hirose}, S., {Blaes}, O., \& {Krolik}, J.~H. 2009, \apj, 704, 781

\bibitem[{{Hirose} {et~al.}(2006){Hirose}, {Krolik}, \& {Stone}}]{hir06}
{Hirose}, S., {Krolik}, J.~H., \& {Stone}, J.~M. 2006, \apj, 640, 901

\bibitem[{{Io} \& {Suzuki}(2014)}]{is13}
{Io}, Y., \& {Suzuki}, T.~K. 2014, \apj, 780, 46

\bibitem[{{Iwasaki} \& {Inutsuka}(2011)}]{ii11}
{Iwasaki}, K., \& {Inutsuka}, S.-I. 2011, \mnras, 418, 1668

\bibitem[{{Jacquet}(2013)}]{jac13}
{Jacquet}, E. 2013, \aap, 551, A75

\bibitem[{{Johansen} {et~al.}(2009){Johansen}, {Youdin}, \& {Klahr}}]{joh09}
{Johansen}, A., {Youdin}, A., \& {Klahr}, H. 2009, \apj, 697, 1269

\bibitem[{{Kato} {et~al.}(2004){Kato}, {Hayashi}, \& {Matsumoto}}]{kat04}
{Kato}, Y., {Hayashi}, M.~R., \& {Matsumoto}, R. 2004, \apj, 600, 338

\bibitem[{{Keller} \& {Gail}(2004)}]{kg04}
{Keller}, C., \& {Gail}, H.-P. 2004, \aap, 415, 1177

\bibitem[{{Kolmogorov}(1941)}]{k41}
{Kolmogorov}, A. 1941, Akademiia Nauk SSSR Doklady, 30, 301

\bibitem[{{Kozlowski} {et~al.}(1978){Kozlowski}, {Jaroszynski}, \&
  {Abramowicz}}]{koz78}
{Kozlowski}, M., {Jaroszynski}, M., \& {Abramowicz}, M.~A. 1978, \aap, 63, 209

\bibitem[{{Kudoh} {et~al.}(1998){Kudoh}, {Matsumoto}, \& {Shibata}}]{kud98}
{Kudoh}, T., {Matsumoto}, R., \& {Shibata}, K. 1998, \apj, 508, 186

\bibitem[{{Lesur} {et~al.}(2013){Lesur}, {Ferreira}, \& {Ogilvie}}]{les13}
{Lesur}, G., {Ferreira}, J., \& {Ogilvie}, G.~I. 2013, \aap, 550, A61

\bibitem[{{Lubow} {et~al.}(1994){Lubow}, {Papaloizou}, \& {Pringle}}]{lub94}
{Lubow}, S.~H., {Papaloizou}, J.~C.~B., \& {Pringle}, J.~E. 1994, \mnras, 267,
  235

\bibitem[{{Lynden-Bell} \& {Pringle}(1974)}]{lp74}
{Lynden-Bell}, D., \& {Pringle}, J.~E. 1974, \mnras, 168, 603

\bibitem[{{Machida} {et~al.}(2000){Machida}, {Hayashi}, \& {Matsumoto}}]{mac00}
{Machida}, M., {Hayashi}, M.~R., \& {Matsumoto}, R. 2000, \apjl, 532, L67

\bibitem[{{Machida} \& {Matsumoto}(2003)}]{mm03}
{Machida}, M., \& {Matsumoto}, R. 2003, \apj, 585, 429

\bibitem[{{Machida} {et~al.}(2013){Machida}, {Nakamura}, {Kudoh}, {Akahori},
  {Sofue}, \& {Matsumoto}}]{mac13}
{Machida}, M., {Nakamura}, K.~E., {Kudoh}, T., {et~al.} 2013, \apj, 764, 81

\bibitem[{{Maeder}(1999)}]{mae99}
{Maeder}, A. 1999, \aap, 347, 185

\bibitem[{{Matsumoto} \& {Tajima}(1995)}]{mt95}
{Matsumoto}, R., \& {Tajima}, T. 1995, \apj, 445, 767

\bibitem[{{McJunkin} {et~al.}(2014){McJunkin}, {France}, {Schneider},
  {Herczeg}, {Brown}, {Hillenbrand}, {Schindhelm}, \& {Edwards}}]{mcj13}
{McJunkin}, M., {France}, K., {Schneider}, P.~C., {et~al.} 2014, \apj, 780, 150

\bibitem[{{Mo{\'s}cibrodzka} \& {Proga}(2009)}]{mp09}
{Mo{\'s}cibrodzka}, M., \& {Proga}, D. 2009, \mnras, 397, 2087

\bibitem[{{Muto} {et~al.}(2012){Muto}, {Grady}, {Hashimoto}, {Fukagawa},
  {Hornbeck}, {Sitko}, {Russell}, {Werren}, {Cur{\'e}}, {Currie}, {Ohashi},
  {Okamoto}, {Momose}, {Honda}, {Inutsuka}, {Takeuchi}, {Dong}, {Abe},
  {Brandner}, {Brandt}, {Carson}, {Egner}, {Feldt}, {Fukue}, {Goto}, {Guyon},
  {Hayano}, {Hayashi}, {Hayashi}, {Henning}, {Hodapp}, {Ishii}, {Iye},
  {Janson}, {Kandori}, {Knapp}, {Kudo}, {Kusakabe}, {Kuzuhara}, {Matsuo},
  {Mayama}, {McElwain}, {Miyama}, {Morino}, {Moro-Martin}, {Nishimura}, {Pyo},
  {Serabyn}, {Suto}, {Suzuki}, {Takami}, {Takato}, {Terada}, {Thalmann},
  {Tomono}, {Turner}, {Watanabe}, {Wisniewski}, {Yamada}, {Takami}, {Usuda}, \&
  {Tamura}}]{mut12}
{Muto}, T., {Grady}, C.~A., {Hashimoto}, J., {et~al.} 2012, \apjl, 748, L22

\bibitem[{{Nelson} {et~al.}(2013){Nelson}, {Gressel}, \& {Umurhan}}]{nel13}
{Nelson}, R.~P., {Gressel}, O., \& {Umurhan}, O.~M. 2013, \mnras, 435, 2610

\bibitem[{{Nelson} \& {Papaloizou}(2004)}]{np04}
{Nelson}, R.~P., \& {Papaloizou}, J.~C.~B. 2004, \mnras, 350, 849

\bibitem[{{Nishikori} {et~al.}(2006){Nishikori}, {Machida}, \&
  {Matsumoto}}]{nis06}
{Nishikori}, H., {Machida}, M., \& {Matsumoto}, R. 2006, \apj, 641, 862

\bibitem[{{Noble} {et~al.}(2010){Noble}, {Krolik}, \& {Hawley}}]{nob10}
{Noble}, S.~C., {Krolik}, J.~H., \& {Hawley}, J.~F. 2010, \apj, 711, 959

\bibitem[{{Nomura} \& {Nakagawa}(2006)}]{nn06}
{Nomura}, H., \& {Nakagawa}, Y. 2006, \apj, 640, 1099

\bibitem[{{Ohsuga} {et~al.}(2009){Ohsuga}, {Mineshige}, {Mori}, \&
  {Kato}}]{ohs09}
{Ohsuga}, K., {Mineshige}, S., {Mori}, M., \& {Kato}, Y. 2009, \pasj, 61, L7

\bibitem[{{Okuzumi} \& {Hirose}(2011)}]{oh11}
{Okuzumi}, S., \& {Hirose}, S. 2011, \apj, 742, 65

\bibitem[{{Okuzumi} \& {Hirose}(2012)}]{oh12}
---. 2012, \apjl, 753, L8

\bibitem[{{Okuzumi} \& {Ormel}(2013)}]{oo13a}
{Okuzumi}, S., \& {Ormel}, C.~W. 2013, \apj, 771, 43

\bibitem[{{Ormel} \& {Okuzumi}(2013)}]{oo13b}
{Ormel}, C.~W., \& {Okuzumi}, S. 2013, \apj, 771, 44

\bibitem[{{Papaloizou} \& {Nelson}(2003)}]{pn03}
{Papaloizou}, J.~C.~B., \& {Nelson}, R.~P. 2003, \mnras, 339, 983

\bibitem[{{Parkin} \& {Bicknell}(2013{\natexlab{a}})}]{pb13a}
{Parkin}, E.~R., \& {Bicknell}, G.~V. 2013{\natexlab{a}}, \apj, 763, 99

\bibitem[{{Parkin} \& {Bicknell}(2013{\natexlab{b}})}]{pb13b}
---. 2013{\natexlab{b}}, \mnras, 435, 2281

\bibitem[{{Pessah} {et~al.}(2007){Pessah}, {Chan}, \& {Psaltis}}]{pes07}
{Pessah}, M.~E., {Chan}, C.-k., \& {Psaltis}, D. 2007, \apjl, 668, L51

\bibitem[{{Proga} \& {Begelman}(2003)}]{pb03}
{Proga}, D., \& {Begelman}, M.~C. 2003, \apj, 592, 767

\bibitem[{{Rothstein} \& {Lovelace}(2008)}]{rl08}
{Rothstein}, D.~M., \& {Lovelace}, R.~V.~E. 2008, \apj, 677, 1221

\bibitem[{{Sano} {et~al.}(1999){Sano}, {Inutsuka}, \& {Miyama}}]{san99}
{Sano}, T., {Inutsuka}, S., \& {Miyama}, S.~M. 1999, in Astrophysics and Space
  Science Library, Vol. 240, Numerical Astrophysics, ed. S.~M. {Miyama},
  K.~{Tomisaka}, \& T.~{Hanawa} (Boston, MA: Kluwer), 383

\bibitem[{{Sano} {et~al.}(2004){Sano}, {Inutsuka}, {Turner}, \&
  {Stone}}]{san04}
{Sano}, T., {Inutsuka}, S.-i., {Turner}, N.~J., \& {Stone}, J.~M. 2004, \apj,
  605, 321

\bibitem[{{Shakura} \& {Sunyaev}(1973)}]{ss73}
{Shakura}, N.~I., \& {Sunyaev}, R.~A. 1973, \aap, 24, 337

\bibitem[{{Shi} {et~al.}(2010){Shi}, {Krolik}, \& {Hirose}}]{shi10}
{Shi}, J., {Krolik}, J.~H., \& {Hirose}, S. 2010, \apj, 708, 1716

\bibitem[{{Simon} {et~al.}(2011){Simon}, {Hawley}, \& {Beckwith}}]{sim11}
{Simon}, J.~B., {Hawley}, J.~F., \& {Beckwith}, K. 2011, \apj, 730, 94

\bibitem[{{Stone} {et~al.}(1996){Stone}, {Hawley}, {Gammie}, \&
  {Balbus}}]{sto96}
{Stone}, J.~M., {Hawley}, J.~F., {Gammie}, C.~F., \& {Balbus}, S.~A. 1996,
  \apj, 463, 656

\bibitem[{{Stone} \& {Norman}(1992)}]{sn92}
{Stone}, J.~M., \& {Norman}, M.~L. 1992, \apjs, 80, 791

\bibitem[{{Stone} \& {Pringle}(2001)}]{sp01}
{Stone}, J.~M., \& {Pringle}, J.~E. 2001, \mnras, 322, 461

\bibitem[{{Suzuki} \& {Inutsuka}(2006)}]{si06}
{Suzuki}, T.~K., \& {Inutsuka}, S.-i. 2006, Journal of Geophysical Research
  (Space Physics), 111, 6101

\bibitem[{{Suzuki} \& {Inutsuka}(2009)}]{si09}
---. 2009, \apjl, 691, L49

\bibitem[{{Suzuki} {et~al.}(2010){Suzuki}, {Muto}, \& {Inutsuka}}]{suz10}
{Suzuki}, T.~K., {Muto}, T., \& {Inutsuka}, S.-i. 2010, \apj, 718, 1289

\bibitem[{{Takeuchi} \& {Lin}(2002)}]{tl02}
{Takeuchi}, T., \& {Lin}, D.~N.~C. 2002, \apj, 581, 1344

\bibitem[{{Turner} {et~al.}(2003){Turner}, {Stone}, {Krolik}, \&
  {Sano}}]{tur03}
{Turner}, N.~J., {Stone}, J.~M., {Krolik}, J.~H., \& {Sano}, T. 2003, \apj,
  593, 992

\bibitem[{{Velikhov}(1959)}]{vel59}
{Velikhov}, E.~P. 1959, Zh. Eksp. Teor. Fiz., 36, 1398

\bibitem[{{von Zeipel}(1924)}]{vz24}
{von Zeipel}, H. 1924, \mnras, 84, 665

\bibitem[{{Weber} \& {Davis}(1967)}]{wd67}
{Weber}, E.~J., \& {Davis}, Jr., L. 1967, \apj, 148, 217

\end{thebibliography}

\end{CJK*}

\end{document}